\documentclass[a4paper,10pt]{article}
\usepackage{amsmath,amsthm,amssymb,mathtools}
\usepackage{thmtools} 
\usepackage{color}
\PassOptionsToPackage{hyphens}{url}\usepackage[colorlinks,citecolor=blue,urlcolor=magenta]{hyperref}
\usepackage[authoryear,round]{natbib}
\usepackage{doi}
\usepackage{booktabs,paralist}
\usepackage{longtable}
\usepackage{subcaption}
\usepackage[flushleft]{threeparttablex}
\usepackage{comment}
\usepackage[export]{adjustbox} 

\usepackage{epstopdf}
\usepackage{xpatch}

\usepackage{bbm}
\usepackage{bm} 

\usepackage{accents}
\newcommand{\ubar}[1]{\underaccent{\bar}{#1}}
\usepackage[ruled,vlined]{algorithm2e}
\usepackage{setspace}
\usepackage{geometry}
\usepackage{titling} 

\newtheorem{theorem}{Theorem}[section]

\newtheorem{proposition}[theorem]{Proposition}
\newtheorem{lemma}[theorem]{Lemma}
\theoremstyle{remark}

\newtheorem{Remark}{Remark}
\theoremstyle{plain}
\newtheorem{assumption}[theorem]{Assumption}

\newtheoremstyle{case}{}{}{}{}{}{:}{ }{}
\theoremstyle{case}

\makeatletter
\newcounter{proofpart}
\xpretocmd{\proof}{\setcounter{proofpart}{0}}{}{}
\newcommand{\proofpart}[1]{%
	\par
	\addvspace{\medskipamount}%
	\stepcounter{proofpart}%
	\noindent\emph{Part \theproofpart: #1}\par\nobreak\smallskip
	\@afterheading
}
\makeatother

\numberwithin{equation}{section}

\renewcommand{\hat}{\widehat} 
\renewcommand{\Pr}{\operatorname{P}}
\newcommand{\lr}[1]{\left[#1 \right]}
\newcommand{\uexp}[1]{\mathbb{E}\lr{#1}}
\newcommand{\cexp}[1]{\mathbb{E}_t\lr{#1}}

\defcitealias{LIS2020}{LIS}

\makeatletter
\newcommand*{\rom}[1]{\expandafter\@slowromancap\romannumeral #1@}
\makeatother

\DeclareMathOperator{\Var}{Var}

\DeclareMathOperator{\cov}{\mathbb{COV}_t}
\newcommand{\unccov}{\mathbb{COV}}
\newcommand{\vartilde}{\tilde{\mathbb{VAR}}_t}

\theoremstyle{definition}
\newtheorem{example}{Example}[section]

\newcommand{\qcexp}[1]{\tilde{\mathbb{E}}_t\lr{#1}}
\newcommand{\scexp}[1]{\mathbb{E}^*_t\lr{#1}}

\newcommand{\covhat}[2]{\widetilde{\cov}\lr{#1,#2}}

\newcommand{\ind}[1]{\mathbbm{1}\left(#1\right)}
\newcommand{\abs}[1]{\left| #1 \right|}

\newcommand{\risky}{R_{t+1}}

\newcommand{\ofree}{R_{f,t\to N}}
  
\newcommand{\osdf}{M_{t \to N}}

\newcommand{\cprob}{\mathbb{P}_t}

\newcommand{\texpneut}{\tilde{\mathbb{E}}_t}
\newcommand{\ttexp}[1]{\mathbb{E}_t \lr{#1}}

\newcommand\norm[1]{\left\lVert#1\right\rVert}
\newcommand{\TJ}[1]{{\color{magenta} TJ : #1}}
\newcommand{\bigoh}[1]{\mathcal{O}\left(#1\right)}
\newcommand{\chabi}[1]{\tilde{\mathbb{M}}_{t \to N}^{(#1)}}
\newcommand{\trunc}[2]{\tilde{\mathbb{M}}_{t \to N }^{(#1)}[#2]}

\newcommand{\mrkt}{R_{m,t+1}}
\newcommand{\pde}[2]{\frac{\partial #1}{\partial #2}}

\newcommand{\pdfmrk}{\tilde{f}_{t}}
\newcommand{\lrb}{\mathrm{CLB}_{t,\tau}}
 
\newcommand{\lrbdp}{\mathrm{LB}_{t,\tau}} 
 
\DeclareMathOperator*{\argmin}{arg\,min}
\newcommand{\lro}[1]{\left(#1\right)}
\newcommand{\pareto}[2]{\mathbf{Par}\left(#1,#2\right)} 
\newcommand{\unif}[2]{\textbf{Unif}\left[#1,#2\right]}

\renewcommand{\tilde}{\widetilde}

\newcommand{\uncfree}{R_f}

\newcommand{\uncmrkt}{R_m}
\newcommand{\uncsdf}{M} 
\newcommand{\uncrisky}{R}
\newcommand{\cquant}{\tilde{Q}_{t,\tau}}
\newcommand{\cquantstar}{\tilde{Q}_{t,\tau^*}}
\newcommand{\iid}{\textsc{iid}\xspace}
\newcommand{\cpquant}{Q_{t,\tau}}

\newcommand{\qhatln}{\hat{Q}_{t,\tau}^{\text{logn}}}

\newcommand{\diff}{\mathop{}\!\mathrm{d}} 
\newcommand{\func}{\widehat{F}}
\newcommand{\funct}{\widehat{\tilde{F}}}
\newcommand{\qunct}{\widehat{\tilde{Q}}}
\newcommand{\ordunc}{\widehat{\phi}}
\newcommand{\quantunc}{\widehat{\theta}}

\newcommand{\uncfreehat}{\hat{R}_f}
\newcommand{\omrkt}{R_{m,t\to N}}
\newcommand{\texptilde}[1]{\tilde{\mathbb{E}}_t\lro{#1}}

\newcommand{\LTtau}{L_{T,\tau}}
\newcommand{\Ltau}{L_\tau}

\makeatletter
\newcommand*\rel@kern[1]{\kern#1\dimexpr\macc@kerna}
\newcommand*\widebar[1]{%
	\begingroup
	\def\mathaccent##1##2{%
		\rel@kern{0.8}%
		\overline{\rel@kern{-0.8}\macc@nucleus\rel@kern{0.2}}%
		\rel@kern{-0.2}%
	}%
	\macc@depth\@ne
	\let\math@bgroup\@empty \let\math@egroup\macc@set@skewchar
	\mathsurround\z@ \frozen@everymath{\mathgroup\macc@group\relax}%
	\macc@set@skewchar\relax
	\let\mathaccentV\macc@nested@a
	\macc@nested@a\relax111{#1}%
	\endgroup
}
\makeatother

\begin{document}
	
\begin{titlepage}
	\title{A Tale of Two Tails: A Model-free Approach to Estimating Disaster Risk Premia and Testing Asset Pricing Models}
	\author{Tjeerd de Vries\thanks{Department of Economics, University of California San Diego. Email: \href{mailto:tjdevrie@ucsd.edu}{tjdevrie@ucsd.edu}. I am particularly indebted to my advisors, Allan Timmermann and Alexis Toda, for their support and feedback throughout the writing process. I would also	like to thank James Hamilton, Xinwei Ma, Yixiao Sun, Rossen Valkanov, Brendan Beare, Caio Almeida, Ali Uppal, Rob Engle (discussant), Andrew Lo (discussant), Lerby Ergun, Casper de Vries, Jens Jackwerth and seminar participants of the 2022 SoFiE conference, 2023 JOIM conference and ESI Brown Bag Seminar for useful feedback. All errors are my own.}}
	\date{\today}
	\maketitle

	\begin{abstract}
		
	I introduce a model-free methodology to assess the impact of disaster risk on the market return. Using S\&P500 returns and the risk-neutral quantile function derived from option prices, I employ quantile regression to estimate local differences between the conditional physical and risk-neutral distributions. The results indicate substantial disparities primarily in the left-tail, reflecting the influence of disaster risk on the equity premium. These differences vary over time and persist beyond crisis periods. On average, the bottom 5\% of returns contribute to 17\% of the equity premium, shedding light on the Peso problem. I also find that disaster risk increases the stochastic discount factor's volatility. Using a lower bound observed from option prices on the left-tail difference between the physical and risk-neutral quantile functions, I obtain similar results, reinforcing the robustness of my findings.

		\noindent \\
		\vspace{0in}
		\vspace{0in}
		
		\bigskip
	\end{abstract}
	\setcounter{page}{0}
	\thispagestyle{empty}
\end{titlepage}
\pagebreak \newpage


\section{Introduction}

Disaster risk has emerged as a pervasive and influential concept in asset pricing, offering a prominent explanation of the equity premium puzzle, as well as other asset pricing puzzles.\footnote{See, for example, \citet{rietz1988equity}, \citet{barro2006rare,Barro2009},  \citet{Drechsler2011}, \citet{gabaix2012variable}, \citet{wachter2013can}, \citet{Constantinides2017}, \citet{Isore2017}, \citet{Farhi2018}, \citet{Seo2019} and \citet{Schreindorfer2020}.} Little is known, however, about the quantitative properties of disaster risk and evidence for it is often inferred indirectly, such as from the historically high equity premium. Nevertheless, a high equity premium does not necessarily arise due to disaster risk, and the literature has yet to reach an unambiguous conclusion regarding its ability to explain asset pricing puzzles (see, e.g., \citet{Julliard2012}). \citet{ross2015recovery} refers to disaster risk as dark matter and eloquently summarizes the concept as follows: ``It is unseen and not directly observable but it exerts a force that can change over time and that can profoundly influence markets''. \\

In this paper, I propose a model-free methodology to measure and track through time disaster risk in S\&P500 returns. My results unequivocally show that disaster risk is pervasive and is the primary determinant of the equity premium. In establishing these results, I confront two critical challenges that have hindered inference about disaster risk. Firstly, to estimate disaster risk in a model-free manner, the literature often estimates the stochastic discount factor (SDF), defined as the ratio of risk-neutral to physical density. Disaster risk is then defined by the SDF taking large values in the left-tail of the return distribution. However, this approach faces scrutiny due to the potential for erratic results when estimating the density ratio in the tails, thereby complicating robust inference. Secondly, it is crucial to account for changing conditioning information. Typically,  estimation of the physical density involves pooling historical returns, while the risk-neutral density relies on forward-looking option prices. This disparity in information sets can lead to inconsistent estimates of conditional disaster risk.\\

To address these challenges, I consider an approach that avoids the need for density estimation. Starting from the absence of arbitrage opportunities,  a risk-neutral distribution exists that can be identified from option prices \citep{breeden1978prices}. However, the conditional physical distribution, which describes the actual evolution of the market return, remains unobserved. To proceed, I use quantile regression (QR) to estimate 
\begin{equation}\label{eq:intro}
	\underbrace{\cpquant(\omrkt)}_{\text{Unobserved}}  = \beta_0(\tau) + \beta_1(\tau)  \underbrace{\cquant(\omrkt)}_{\text{Observed}} \qquad \tau \in  (0,1),
\end{equation}
where $\cpquant$ and $\cquant$ represent the physical and risk-neutral $\tau$-quantiles, respectively, of the market return $\omrkt$, from period $t$ to $t+N$. The parameters in \eqref{eq:intro} can be estimated using quantile regression, with the observed time series of returns, $\{\omrkt\}_{t=1}^T$, as the dependent variable and $\{\cquant\}_{t=1}^T$ as the regressor.  Importantly, both $\omrkt$ and $\cquant$ are conditioned on the same information set. In general, quantile regression estimates the best linear approximation to the physical quantile function, from which $\omrkt$ is drawn. Hence, any deviation from the risk-neutral benchmark, $[\beta_0(\tau),\beta_1(\tau)] = [0,1]$, signifies a local difference between the physical and risk-neutral measures at the $\tau$-quantile. Since the equity premium is determined by these differences, it is  natural to define the \emph{disaster risk premium} as the difference between $\cpquant$ and $\cquant$ in the left-tail, i.e.\ for values of $\tau$ close to zero.\\

Based on the QR estimates, two key findings emerge: \begin{inparaenum}[(i)] 
	\item the risk-neutral benchmark cannot be rejected in the right-tail ($\tau \ge 0.7$) but it is rejected in the left-tail ($\tau \le 0.3$); and 
	\item the in-sample and out-of-sample explanatory power of the risk-neutral quantile is significantly higher in the right-tail compared to the left-tail.
\end{inparaenum}
Both findings suggest that disaster risk is the main driver of the equity premium.\\

 Building on these results, I estimate the conditional Lorenz curve and Gini coefficient associated with the equity premium. These statistics summarize how much the conditional equity premium is driven by the lowest returns, akin to its interpretation of wealth inequality in labor economics. I find that the Lorenz curve is always concave, and the Gini coefficients are far above zero in every time period, thus showing that disaster risk is a pervasive feature of the data. On average, I find that the bottom 5\% of returns contribute to 17\% of the total equity premium.\\

While this result demonstrates that disaster risk is an important driver of expected returns, it also adds nuance to the degree of disaster risk necessary to explain the equity premium. In particular, previous papers attribute about 90\% of the equity premium to the lowest 5\% of returns \citep{Barro2009,backus2011disasters,Beason2022}. The results differ since I account for conditioning information, whereas unconditional estimates of disaster risk tend to overestimate this risk, since the physical distribution acquires fatter tails when averaging out state variables.\\

Comparing the physical and risk-neutral quantile functions over time also sheds light on the role of risk aversion and forward-looking beliefs in jointly determining disaster risk premia. Particularly during crises, the value of an insurance contract that hedges against disaster risk increases, resulting in a downward movement in the left-tail of the risk-neutral quantile function. Simultaneously, investors often revise their beliefs about the likelihood of another disaster, frequently assigning a higher probability to such an event. This effect drives down the left-tail of the physical quantile function, creating an ambiguous overall impact on disaster risk premia. However, the quantile regression estimates indicate that the risk-neutral quantile function decreases proportionally more, highlighting the greater influence of risk aversion in determining disaster risk premia.\\

Given that the equity premium is primarily driven by disaster risk premia, the discussion above implies that the left-tail of the risk-neutral quantile function can predict the equity premium. An OLS regression of the equity premium against the 5\% risk-neutral quantile shows preliminary evidence of forecasting ability, especially out-of-sample. In line with theoretical expectations, a decline in the 5\% risk-neutral quantile is associated with a substantial increase in the equity premium. Notably, during the 2008 financial crisis and the 2020 Covid-19 crisis, monthly estimates of the equity premium reached peaks of around 5\%.\\



Besides the equity premium puzzle, the QR estimates shed light on the role of first-order stochastic dominance and the pricing kernel puzzle.   Specifically, I find that $\cquant < \cpquant$ holds across most of the distribution, except in the far right-tail, where $\cquant > \cpquant$ frequently occurs. This violation of stochastic dominance  raises questions in asset pricing models using the expected utility framework, as it suggests that a representative investor exhibits negative risk aversion. Furthermore, I show that a violation of stochastic dominance implies that the pricing kernel is not monotonic, thereby confirming the pricing kernel puzzle while accounting for conditioning information, and without the need to estimate a density ratio.\\

To further understand the influence of disaster risk on the pricing kernel, I introduce a \emph{distribution bound} on the SDF volatility that is closely related to the \citet{hansen1991implications} bound.  The distribution bound summarizes the risk-return trade-off of an asset paying out one dollar when the market return falls below a certain threshold. I show that disaster risk makes the risk-return trade off highly favorably by going short in an asset paying one dollar in case of a disaster. The price of such an asset is high because investors are willing to pay a significant premium to insure against disaster risk, but the risk is limited since the actual probability of a disaster is comparatively low. The Sharpe ratio associated to this investment therefore dominates the Sharpe ratio of a direct investment in the market portfolio. Specifically, in the data, the Sharpe ratio on selling an asset that pays out one dollar if the return falls below the 5th percentile is 30\% in monthly units, while the Sharpe ratio of investing in the market portfolio is only 13\%. I also show that models which do not embed a source of disaster risk, such as conditional lognormal models, cannot rationalize this finding.\\

 I conclude by proposing a model-free lower bound on disaster risk premia to assess the robustness of my earlier findings. This lower bound is observed from option prices and is inspired by recent bounds on the equity premium \citep{martin2017expected,chabi2020conditional}. Using quantile regression, I show that the lower bound explains a substantial proportion of the fluctuation in disaster risk premia over time. Moreover, the lower bound relaxes the assumption of a time-homogeneous relation between the physical and risk-neutral quantile functions.  Empirically, the lower bound closely aligns with the disaster risk estimates derived from quantile regression, further strengthening the robustness of my earlier findings.\\

\subsection{Related Literature} 
My approach, which uses quantile regression to measure local dispersion between the physical and risk-neutral distribution, is related to a larger body of literature that estimates the pricing kernel from returns and option data \citep{ait2000nonparametric,jackwerth2000recovering,rosenberg2002empirical,beare2016empirical,linn2018pricing,cuesdeanu2018pricing}. However, estimating the pricing kernel from returns and options can be challenging, especially in the tails of the distribution, where the ratio of densities that defines the pricing kernel can become unstable. In addition, using historical returns to estimate the physical density can lead to inconsistent results \citep{linn2018pricing}. \citet{Beason2022} apply a similar methodology to decompose the unconditional equity premium.\\

In contrast, QR can be used to draw inference on the pricing kernel indirectly, by leveraging  the \emph{observed} realized return and risk-neutral distribution, which avoids the estimation of a density ratio. Furthermore, QR can account for changes in the shape and scale of the underlying SDF over time due to changing conditional information, while the approach of \citet{cuesdeanu2018pricing} renders an estimate of the SDF that only allows the normalizing constant to be time-varying, since the shape and scale are time invariant (see Section \ref{sec:logn}).\\

The QR approach also lays out a new framework to analyze disaster risk in a model-free fashion, defining this risk as the conditional quantile difference between the physical and risk-neutral measure in the left-tail. Since my results indicate that these differences are substantial, I offer new empirical evidence supporting disaster risk as a key explanation of the equity premium, as in the models of \citet{rietz1988equity} and \citet{barro2006rare,Barro2009}, or more recently \citet{Schreindorfer2020} and \citet{Ai2021}. Furthermore, my analysis adds nuance to the understanding of the equity premium's drivers. While \citet{Beason2022} assert that the worst 5\% of returns account for 91.5\% of the equity premium, my methodology reveals that these returns contribute only 17\% to the equity premium. The results differ since I account for conditioning information, and my time series of returns includes the Covid-19 crisis.\\

Complementary to the QR estimates, I derive a nonparametric bound on the SDF volatility closely related to the bound of  \citet{hansen1991implications}. They argue that the SDF is necessarily volatile and use this observation to screen asset pricing models. Several papers have built on this insight using higher-order moment bounds  \citep{snow1991diagnosing,almeida2012assessing,liu2020index} and entropy bounds \citep{stutzer1995bayesian,bansal1997growth,alvarez2005using,backus2014sources}.  These bounds all provide a measure of how much the risk-neutral distribution differs from the physical distribution. Unlike the distribution bound, all of these measures are global in that they rely on averages over the entire state space. The distribution bound in this paper is a function rather than a single statistic and can be considered an intermediate approach between a single bound and a complete estimate of the SDF.\\

This paper is also related to the growing literature on using options to estimate forward-looking equity premiums \citep{martin2017expected,martin2019expected,chabi2020conditional}. However, unlike those papers that focus on the conditional expectation of excess returns, this paper uses option data to predict conditional return quantiles. The relationship between option prices and expected market return shocks has been extensively studied in the literature \citep{bates1991crash,bates2000post,bates2008market,coval2001expected,bollerslev2011tails,backus2011disasters,ross2015recovery}. Similar to \citet{bollerslev2011tails}, this paper obtains a nonparametric measure of disaster risk. However, my approach differs in that it only uses risk-neutral information and is motivated by the interplay between the physical and risk-neutral quantile functions.\\

 This approach also complements the recovery literature  as it derives forward-looking approximations to the left- and right-tail of the physical distribution using option data \citep{ross2015recovery,borovivcka2016misspecified,qin2017long,Bakshi2018,qin2018long,Schneider2019,jackwerth2020does}. The time variation in the approximation for the left-tail quantile documented in this paper is consistent with the time-varying disaster risk models of \citet{gabaix2012variable}, \citet{wachter2013can}, \citet{Constantinides2017}, \citet{Isore2017},  \citet{Farhi2018} and \citet{Seo2019}.\\ 


Finally, the QR approach is related to conditional mean regressions that are common in the equity premium literature. The performance evaluation of conditional expected return predictors is well established in the literature, with important contributions from \citet{campbell2008predicting} and \citet{welch2008comprehensive}. To evaluate the performance of the QR approach, I draw on earlier work of \citet{koenker1999goodness} and extend the evaluation toolkit to the quantile setting, specifically focusing on out-of-sample performance. This paper thus complements the literature on conditional return prediction by extending it to the entire distribution.\\

The rest of this paper is organized as follows.  Section \ref{sec:qr_reg}  presents the main empirical results from the quantile regressions and its consequences for the equity premium and SDF are discussed in  Section \ref{sec:impli_sdf}. Section \ref{sec:Robust_QR} provides further evidence on the robustness of QR to estimate disaster risk relative to extant approaches. Section \ref{sec:HJbound} introduces the distribution bound, discusses its use in asset pricing models, and presents estimates of the distribution bound from empirical data. Building on the results of Sections \ref{sec:qr_reg} and \ref{sec:impli_sdf}, Section \ref{sec:dark} establishes a model-free lower bound on disaster risk premia. Finally, Section \ref{sec:conclusion} concludes.

\section{Empirical Estimates of Quantile Difference}\label{sec:qr_reg}
This section documents empirical estimates of the conditional difference between the physical and risk-neutral quantile functions. I first discuss the notation and then consider an example to clarify the idea and motivate the methodology.

\subsection{Notation}\label{sec:notation}
Let $\omrkt$ denote the market return from period $t$ to $t+N$, where $N$ can be measured in days or years, depending on the context. The risk-free rate over the same period is denoted by  $\ofree$, which is assumed to be known at time $t$. In the absence of arbitrage, there exists a positive random variable $\osdf$ such that, conditional on all time $t$ information
\begin{equation}\label{eq:sdf_def}
	\ttexp{\osdf \omrkt  } = 1.
\end{equation}
The random variable $\osdf$ is referred to as the stochastic discount factor (SDF) and the expectation in \eqref{eq:sdf_def} is calculated under the \emph{physical} probability measure $\mathbb{P}_t$, which is the actual distribution of the market return, i.e.\ $\omrkt \sim \mathbb{P}_t$. The SDF can potentially depend on many state variables, but  these are suppressed from the notation for brevity. It is convenient to restate \eqref{eq:sdf_def} in terms of risk-neutral probabilities:
\begin{equation*}
	\texptilde{\omrkt} = 1/\ttexp{\osdf} = \ofree,
\end{equation*}
where the expectation is calculated under the \emph{risk-neutral} measure $\tilde{\mathbb{P}}_t$ induced by $\osdf$. Finally, $F_t(x) \coloneqq \mathbb{P}_t(\omrkt \le x )$ denotes the physical CDF of the market return conditional on all information available at time $t$, $f_t(\cdot)$ denotes the conditional probability density function (PDF) and $\cpquant$ denotes the conditional $\tau$-quantile. As before, a tilde superscript refers to the risk-neutral measure, so that 
\begin{equation*}
	\tilde{F}_t(\cquant) = \tilde{\mathbb{P}}_t\lro{\omrkt \le \cquant} = \tau, \qquad \forall \tau \in (0,1).
\end{equation*}
The physical and risk-neutral quantiles depend on the underlying random variable $\omrkt$ (i.e., $\cquant \coloneqq \cquant(\omrkt)$), but I typically omit this dependence as the underlying random variable always refers to the market return. \\

To clarify my approach of using quantiles to analyze disaster risk, I consider the following asset pricing model that will be used several times in the paper. 

\renewcommand\thmcontinues[1]{Continued}

\begin{figure}[!htb]
	\centering
	\begin{subfigure}[b]{0.49\textwidth}
		\includegraphics[width=\textwidth]{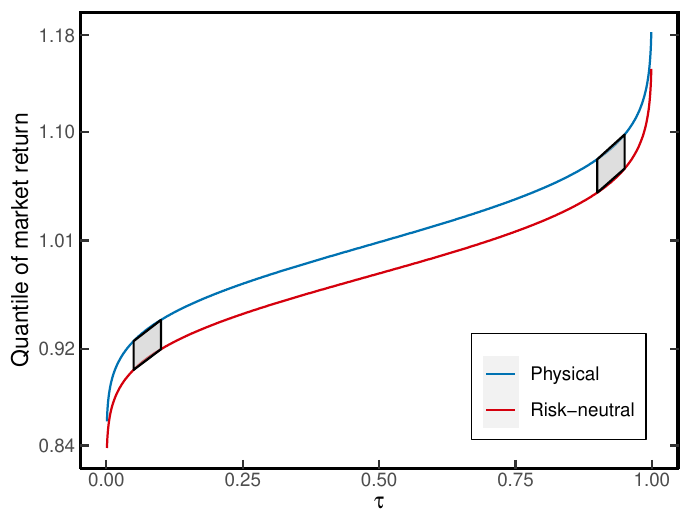}
		\caption{Without jumps}
		\label{fig:q_noJump}
	\end{subfigure}
	\begin{subfigure}[b]{0.49\textwidth}
		\includegraphics[width=\textwidth]{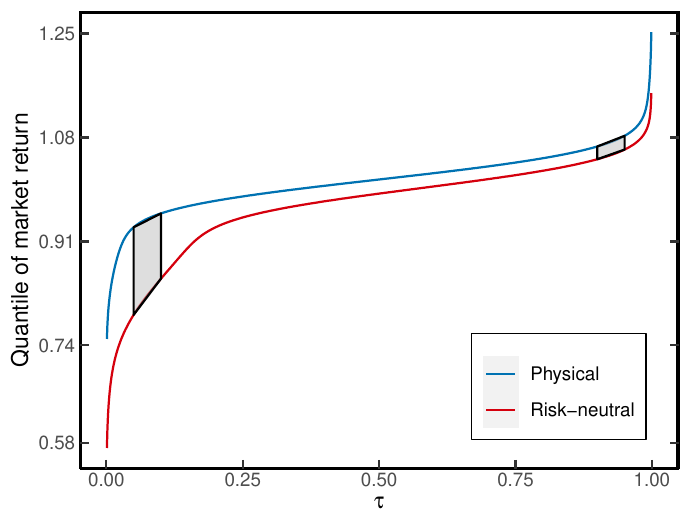}
		\caption{With jumps}
		\label{fig:q_jump}
	\end{subfigure}
	\caption{\textbf{Effect of jumps on physical and risk-neutral quantile functions.} \footnotesize The left panel displays the physical and risk-neutral quantile functions without jumps ($\omega = 0$), while the right panel illustrates the quantile functions with jumps ($\omega = 1.4$). In both cases, the mean of the disaster shock ($\theta$) is set to $-0.0074$. The market return is defined as a levered claim on the consumption asset. The trapezoids represent the difference in quantile functions at the 10th and 90th percentiles.}
\end{figure}

\begin{example}[Disaster risk]{\protect \label{ex:disaster_risk_jumps}}
	Consider the disaster risk model analyzed in \citet{backus2011disasters}. The SDF process is given by
	\begin{align*}
		\log \osdf  = \log \beta - \gamma \log G_{t \to N},
	\end{align*}
	where $\beta$ is a time discount factor, $\gamma$ is the coefficient of relative risk aversion, and	$G_{t \to N} = C_{t+N}/C_t$ is consumption growth in period $t+N$. Consumption growth follows a two-component structure:
	\begin{equation*}
		\log G_{t \to N}= z_{1,t+N} + z_{2,t+N}, \quad z_{1,t+N}  \sim \mathcal{N}(\mu,\sigma^2),
	\end{equation*}
	and $z_{2,t+N}$ is a Poisson mixture of normals to capture jumps representing rare shocks to consumption growth that are large in magnitude. The number of jumps, $j$, take on nonnegative integer values with probability $e^{-\omega} \omega^j/j!$, and conditional on $j$, the jump term is normal: $z_{2,t+N}|j \sim \mathcal{N}(j \theta,j \delta^2)$. \citet{backus2011disasters} show that the risk-neutral distribution of consumption growth in a representative agent model is again a normal mixture with parameters:
	\begin{equation}\label{eq:change_of_meas}
		\tilde{\mu} = \mu - \gamma \sigma^2, \quad \tilde{\omega} = \omega e^{-\gamma \theta + (\gamma \delta)^2}, \quad \tilde{\theta} = \theta - \gamma \delta^2.
	\end{equation}
	 In this setup, risk aversion amplifies the jump frequency ($\tilde{\omega} > \omega$ if $\theta < 0$) as well as the jump size ($\tilde{\theta} < \theta$). If the model is calibrated such that $\theta \ll 0$, then $z_{2,t+N}$ can be interpreted as a disaster shock if a jump takes place $(j \ge 1)$.\\

	 Figures \ref{fig:q_noJump} and \ref{fig:q_jump} illustrate the impact of jumps on the physical and risk-neutral quantile functions.
	 Specifically, in the absence of jumps, the market return follows a lognormal distribution and Figure \ref{fig:q_noJump} shows that the difference between the physical and risk-neutral quantile functions is approximately equal in both tails. However, when jumps are introduced, this difference is almost entirely concentrated in the left-tail. This result is driven by the impact of jumps on the risk-neutral distribution, and the requirement that $\theta<0$ is crucial to drive a wedge between the physical and risk-neutral measures in the left-tail (see \eqref{eq:change_of_meas}). The question is whether these distinct shape restrictions on the physical and risk-neutral distribution are supported by the data. \\
	 
	
	
\end{example}

\subsection{Methodology and Econometric Model}\label{sec:methodology}
Building on the discussion in Example {\protect \ref{ex:disaster_risk_jumps}}, it is of interest to estimate the quantile difference between the physical and risk-neutral measures.  The disaster risk model predicts that these differences are
significant in the left tail while negligible in the right tail. Therefore, I refer to $\cpquant - \cquant$ in the left-tail as \emph{disaster risk premia}.\\

While the conditional risk-neutral distribution and its quantile function can be inferred from option prices without specific modeling assumptions \citep{breeden1978prices}, the same cannot be said for the physical distribution, unless strong assumptions are made about the martingale component of the SDF \citep{ross2015recovery, borovivcka2016misspecified}. The information available about the conditional physical distribution is limited to a single realization of the market return, as $\omrkt$ follows $\mathbb{P}_t$ conditional on time $t$. Consequently, the primary challenge in measuring disaster risk premia lies in the unobservable nature of $\cpquant$, which has made model-free inference challenging thus far. 


\subsubsection{Risk-Neutral Quantile Regression}\label{sec:rn_qr_reg}
In order to overcome this difficulty, I assume the following model for the physical quantile function
\begin{equation}\label{eq:tail}
	\underbrace{\cpquant(\omrkt)}_{\text{Unobserved}} = \beta_0(\tau) + \beta_1(\tau) \underbrace{\cquant(\omrkt)}_{\text{Observed}}, \qquad \forall \tau \in (0,1).
\end{equation}
If the world is risk-neutral, $[\beta_0(\tau),\beta_1(\tau)] = [0,1]$ for all $\tau$. Departures from risk-neutrality at a specific percentile $\tau$ are reflected by point estimates of $[\beta_0(\tau),\beta_1(\tau)]$ that are far from the $[0,1]$ benchmark. Given a sample of $T$ observations $\{\omrkt,\cquant \}_{t=1}^T$, the unknown parameters in \eqref{eq:tail} can be estimated by  quantile regression \citep{koenker1978regression}:
\begin{equation}\label{eq:qr_est}
	[\hat{\beta}_0(\tau),\hat{\beta}_1(\tau)] = \argmin_{(\beta_0,\beta_1)  \in \mathbb{R}^2} \sum_{t=1}^{T} \rho_\tau(\omrkt - \beta_0 - \beta_1 \cquant),
\end{equation}
where $\rho_\tau(\cdot)$ is the check function from quantile regression
\begin{equation*}
	\rho_\tau(x) = 
	\begin{cases}
		\tau x, & \text{if } x \ge 0\\
		(\tau-1)x & \text{if } x < 0.
	\end{cases}
\end{equation*}\\

 Even if the world is not risk-neutral, the model in \eqref{eq:tail} can still be correctly specified, as is the case for conditional lognormal models (see Section \ref{sec:logn}).  When the model is misspecified, the estimation in \eqref{eq:qr_est} remains meaningful as QR finds the best linear approximation to the conditional quantile function \citep{angrist2006quantile}.\footnote{This result is analogous to OLS, which finds the best linear approximation to the conditional \emph{expectation} function, even if the model is misspecified.} Since the risk-neutral quantile itself is a highly non-linear transformation of state variables, the model can accommodate non-linear dependence between the physical quantile function and state variables driving the economy.  The benefit of using the risk-neutral quantile function as a regressor is that it does not require the econometrician  to take a stand on the state variables driving the physical distribution. Furthermore, both $\omrkt$ and $\cquant$ are conditioned on the same information set, thus avoiding the mismatched information critique of \citet{linn2018pricing}. \\

 In addition, theory often suggests tantalizing links between the tails of the physical and risk-neutral distribution. Table \ref{tab:correlation_Q} presents correlations between $\cpquant$ and $\cquant$ for both left and right tails in different asset pricing models. In most models, these correlations are nearly one, indicating a strong positive relation that can be modeled by \eqref{eq:tail}. Only for $\tau = 0.3$, the correlation is notably lower at 41\% in the \citet{campbell1999force} model and -67\% in the \citet{Drechsler2011} model.\\

  In Appendix \ref{app:oos_linear}, I consider non-linear specifications as alternatives to the linear model in \eqref{eq:tail}. Broadly speaking, I find that the linear model outperforms all non-linear models when predicting the physical quantile function out-of-sample. Based on this evidence, and the close linear approximation suggested by asset pricing models, I use the linear specification throughout most of the paper.\\

\begin{table}[!htb]
\captionsetup{width=12cm}
\centering
\caption{\textbf{Tail correlations (in \%) of physical and risk-neutral quantile function in asset pricing models}}
\label{tab:correlation_Q}
\begin{adjustbox}{max width=\textwidth}
\begin{threeparttable}
	\begin{tabular}{lcccccccc}
		\toprule
		\midrule
		Percentile & 0.05 & 0.1 & 0.2 & 0.3 & 0.7 & 0.8 & 0.9 & 0.95 \\ 
		& & & & & &  & & \\
		\underline{Lognormal} & & & & & &  & & \\
		\citet{campbell1999force} & 96.94 & 94.49 & 83.61 & 40.88 & 86.51 & 94.25 & 97.27 & 98.26 \\
		\citet{bansal2004risks} & 99.97 & 99.97 & 99.98 & 99.98 & 99.99 & 99.99 & 99.99 & 100.00 \\
		\underline{Disaster} & & & & & &  & & \\
		\citet{Drechsler2011} & 99.90 & 99.44 & 94.67 & -67.16 & 96.88 & 98.75 & 99.45 & 99.67 \\
		\citet{wachter2013can} & 95.40 & 99.63 & 99.57 & 98.98 & 99.71 & 99.88 & 99.94 & 99.97 \\
		\citet{Constantinides2017} & 99.86 & 99.72 & 99.21 & 97.58 & 85.96 & 94.68 & 97.32 & 97.90 \\ 
		\bottomrule
	\end{tabular}%
	\begin{tablenotes}
		\footnotesize
		\item \textit{Note}: This table reports the correlation between $\cpquant$ and $\cquant$ in conditional lognormal models and models that embed a source of conditional disaster risk. The correlations at various percentiles are obtained by simulating $10^6$ draws of the ergodic distribution of states in each model. 
	\end{tablenotes}
\end{threeparttable}
\end{adjustbox}
\end{table}

\begin{Remark}\label{remark:sdf_scale}
	An alternative to QR is nonparametric estimation of the SDF as proposed by \citet{ait2000nonparametric}, \citet{jackwerth2000recovering} and \citet{rosenberg2002empirical}. This method can infer the quantile difference from the estimated SDF but relies on pooled historical returns, which can be problematic for forward-looking distribution estimation \citep{linn2018pricing}. More recently,  \citet{linn2018pricing} and \citet{cuesdeanu2018pricing} proposed an estimator of the SDF that accounts for forward-looking information. However, this method presents challenges such as non-convex optimization, the objective function might be undefined due to the small number of existing risk-neutral moments (see Figure \ref{fig:lee_bound}), ambiguity in basis function selection, and the inability to account for shape changes in the SDF leading to incorrect conditional inference. QR, on the other hand, avoids these issues, as shown in more detail in Section  \ref{sec:logn}.

\end{Remark}

\subsubsection{Measures of Fit}
Based on the quantile regression \eqref{eq:qr_est}, I consider two measures of fit  to evaluate how well the risk-neutral quantile locally approximates the physical distribution. The first in-sample measure, $R^{1}(\tau)$, is defined as\footnote{It is well known that $b_0$ in the denominator of \eqref{eq:R1tau} equals the in-sample $\tau$-quantile.}
\begin{equation}\label{eq:R1tau}
	R^1(\tau) \coloneqq 1 - \frac{\min_{b_0,b_1} \sum_{t=1}^T \rho_\tau(\omrkt - b_0 - b_1 \cquant)  }{\min_{b_0} \sum_{t=1}^T \rho_\tau(\omrkt - b_0)}.
\end{equation}
This measure of fit was proposed by \citet{koenker1999goodness} and is a clean substitute for the OLS $R^2$. I also consider an out-of-sample measure of fit 
\begin{equation}\label{eq:Roos_rn}
	R_{oos}^1(\tau) \coloneqq 1 - \frac{\sum_{t=w}^T \rho_\tau(\omrkt - \cquant)}{\sum_{t=w}^T \rho_\tau(\omrkt - \widebar{Q}_{t,\tau})},
\end{equation}
where $\widebar{Q}_{t,\tau}$ is the historical rolling quantile of the market return from time $t-w+1$ to $t$, and $w$ is the rolling window length.  Notice that \eqref{eq:Roos_rn} is a genuine out-of-sample metric since no parameter estimation is used. In the equity premium literature, \citet{campbell2008predicting} stress the importance of out-of-sample predictability; \eqref{eq:Roos_rn}  is analogous to their out-of-sample $R^2$.

\subsection{Data and Estimation}\label{sec:data}
To estimate the quantile regression in \eqref{eq:qr_est}, I require data on the market return and the risk-neutral distribution over time.  I use overlapping returns on the S\&P500 index from WRDS over the period 2003--2021 to represent the market return. I calculate the market return over a horizon of 30-, 60-, and 90-days. Second, over the same horizon, I use put and call option prices on the S\&P500 on each day $t$ from OptionMetrics to estimate the risk-neutral quantile function based on the \citet{breeden1978prices} formula:
\begin{equation}\label{eq:breeden}
	\tilde{F}_t\lro{\frac{K}{S_t}} = \ofree \pde{}{K} \mathrm{Put}_t(K),
\end{equation}
where $\mathrm{Put}_t(K)$ denotes the time $t$ price of a European put option on the S\&P500 index with stock price $S_t$,  strike price $K$ and expiration date $t+N$. This formula is model-free and only requires a no-arbitrage assumption.  Due to the lack of a continuum of option prices, interpolation of different maturity options and missing data for option prices far in-- and out-of-the money, it is a nontrivial exercise to obtain accurate estimates of $\tilde{F}_t$ (and hence $\cquant$) from \eqref{eq:breeden}. A detailed description of my approach that overcomes these issues is described in Appendix \ref{app:risk_neutral_quantile_function}, which is based on \citet{filipovic2013density}.\footnote{This approach uses a kernel density and adds several correction terms to approximate the risk-neutral density. I follow \citet{barletta2018analyzing} and use a principal components step to avoid overfitting in the tails.} Finally, I obtain the risk-free rate from Kenneth French's website.\footnote{See \url{http://mba.tuck.dartmouth.edu/pages/faculty/ken.french/data_library.html\#Research}}\\

Table \ref{tab:only.rn.quantile} shows the QR estimates of \eqref{eq:qr_est}. The point estimates are close to the $[0,1]$ benchmark in the right-tail ($\tau \ge 0.7$), but not in the left-tail ($\tau \le 0.3$). Additionally, the joint restriction that $[\beta_0(\tau),\beta_1(\tau)] = [0,1]$ is  rejected for all $\tau \le 0.2$, at all horizons. In contrast, the null hypothesis is never rejected for $\tau \ge 0.8$. The fact that the risk-neutral distribution provides a good approximation of the physical distribution in the right-tail is confirmed by the measures of fit, $R^1(\tau)$ and $R_{oos}^1(\tau)$, which are  also shown in Table  \ref{tab:only.rn.quantile}. Specifically, both in- and out-of-sample, the risk-neutral quantile fits the physical distribution much better in the right-tail.\\

\begin{Remark}
 The standard errors for the quantile regression in Table \ref{tab:only.rn.quantile} are obtained by the smooth extended tapered block bootstrap (SETBB) of \citet{gregory2018smooth}, which is robust to heteroscedasticity and weak dependence.\footnote{It may seem counterintuitive that the standard errors decrease in the tails, which are generally harder to estimate. However, since the regressor $\cquant$ changes with $\tau$, there is an opposing effect that can cause the standard errors to decrease in the tails. This happens if $\cquant$ is more variable in the tails, akin to the intuition in OLS that more variability in the regressor decreases the standard error. In the data, $\cquant$ is much more variable in the tails.} This robustness is important in the estimation, since I use overlapping returns which creates time dependence in the error term, akin to the overlapping observation problem in OLS \citep{hansen1980forward}. SETBB also renders an estimate of the covariance matrix between $\hat{\beta}_0(\tau)$ and $\hat{\beta}_1(\tau)$, which can be used to test joint restrictions on the coefficients.\footnote{I use the $\texttt{QregBB}$ function from the $R$-package $\texttt{QregBB}$, available on the author's Github  page: \url{https://rdrr.io/github/gregorkb/QregBB/man/QregBB.html}. The only user required input for this method is the block length in the bootstrap procedure.}\\
\end{Remark}

\begin{table}[!htb]
\captionsetup{width=12cm}	
\centering
\caption{\textbf{Risk-neutral quantile regression}}
\label{tab:only.rn.quantile}
\begin{adjustbox}{max width=0.9\textwidth}
\begin{threeparttable}
\begin{tabular}{llccccccc}
\toprule
\midrule
Horizon& $\tau$ & $\hat{\beta}_0(\tau)$ & $\hat{\beta}_1(\tau)$ &$\underset{(p\text{-value})}{\text{Wald test}}$  & $R^1(\tau)$[\%] & $R_{oos}^1(\tau)$[\%] & $\widebar{\mathrm{Hit}}$[\%] & $\hat{Q}_{t,\tau} > \cquant$[\%] \\ 
\cmidrule(lr){1-1}
\cmidrule(lr){2-9}
30 days\textsuperscript{*} & $ 0.05 $ & $\underset{( 0.208 )}{\text{ 0.43 }}$ & $\underset{( 0.223 )}{\text{ 0.56 }}$ & 0.00 & 6.28 & 6.11 & $\underset{( 0.676 )}{\text{ -2.67 }}$ & 99.88 \\
& $ 0.1 $ & $\underset{( 0.201 )}{\text{ 0.45 }}$ & $\underset{( 0.209 )}{\text{ 0.54 }}$ & 0.01 & 3.45 & 1.01 & $\underset{( 1.089 )}{\text{ -3.56 }}$ & 98.52 \\
& $ 0.2 $ & $\underset{( 0.284 )}{\text{ 0.69 }}$ & $\underset{( 0.290 )}{\text{ 0.30 }}$ & 0.02 & 0.55 & 0.89 & $\underset{( 1.719 )}{\text{ -3.73 }}$ & 90.98 \\
& $ 0.3 $ & $\underset{( 0.357 )}{\text{  1.02   }}$ & $\underset{( 0.360 )}{\text{ -0.02 }}$ & 0.00 & 0.00 & 2.49 & $\underset{( 2.147 )}{\text{ -5.51 }}$ & 99.58 \\
& $ 0.4 $ & $\underset{( 0.237 )}{\text{  1.17   }}$ & $\underset{( 0.237 )}{\text{ -0.16 }}$ & 0.00 & 0.03 & 1.75 & $\underset{( 2.357 )}{\text{ -7.32 }}$ & 97.25 \\
& $ 0.6 $ & $\underset{( 0.216 )}{\text{ -0.45 }}$ & $\underset{( 0.213 )}{\text{  1.44   }}$ & 0.00 & 4.62 & 4.19 & $\underset{( 2.468 )}{\text{ -8.05 }}$ & 99.93 \\
& $ 0.7 $ & $\underset{( 0.162 )}{\text{ -0.18 }}$ & $\underset{( 0.159 )}{\text{  1.18   }}$ & 0.03 & 7.79 & 7.47 & $\underset{( 2.220 )}{\text{ -5.84 }}$ & 99.95 \\
& $ 0.8 $ & $\underset{( 0.141 )}{\text{ -0.09 }}$ & $\underset{( 0.137 )}{\text{  1.09   }}$ & 0.19 & 12.44 & 12.50 & $\underset{( 1.886 )}{\text{ -3.24 }}$ & 99.95 \\
& $ 0.9 $ & $\underset{( 0.113 )}{\text{ 0.03 }}$ & $\underset{( 0.108 )}{\text{ 0.97 }}$ & 0.96 & 20.41 & 21.88 & $\underset{( 1.235 )}{\text{ -0.04 }}$ & 55.85 \\
\textsuperscript{*}{\footnotesize (Obs. 4333)}& 0.95 & $\underset{( 0.120 )}{\text{ 0.12 }}$ & $\underset{( 0.114 )}{\text{ 0.89 }}$ & 0.57 & 27.07 & 31.31 & $\underset{( 0.863 )}{\text{ 0.27 }}$ & 22.41 \\
\cmidrule(lr){1-1}
\cmidrule(lr){2-9}
{60 days}\textsuperscript{**} &0.05 & $\underset{( 0.303 )}{\text{ 0.45 }}$ & $\underset{( 0.343 )}{\text{ 0.54 }}$ & 0.00 & 3.12 & 13.14 & $\underset{( 0.875 )}{\text{ -3.33 }}$ & 100.00 \\
& $ 0.1 $ & $\underset{( 0.263 )}{\text{ 0.58 }}$ & $\underset{( 0.283 )}{\text{ 0.41 }}$ & 0.00 & 1.79 & 3.50 & $\underset{( 1.320 )}{\text{ -5.57 }}$ & 100.00 \\
& $ 0.2 $ & $\underset{( 0.336 )}{\text{ 0.78 }}$ & $\underset{( 0.345 )}{\text{ 0.21 }}$ & 0.01 & 0.38 & -0.03 & $\underset{( 2.351 )}{\text{ -6.60 }}$ & 99.95 \\
& $ 0.3 $ & $\underset{( 0.434 )}{\text{ 0.93 }}$ & $\underset{( 0.438 )}{\text{ 0.07 }}$ & 0.00 & 0.01 & -0.12 & $\underset{( 3.012 )}{\text{ -7.81 }}$ & 99.47 \\
& $ 0.4 $ & $\underset{( 0.325 )}{\text{ 0.36 }}$ & $\underset{( 0.323 )}{\text{ 0.65 }}$ & 0.02 & 0.25 & 2.34 & $\underset{( 3.439 )}{\text{ -8.48 }}$ & 99.79 \\
& $ 0.6 $ & $\underset{( 0.342 )}{\text{ -0.65 }}$ & $\underset{( 0.333 )}{\text{  1.64   }}$ & 0.02 & 5.57 & 4.60 & $\underset{( 3.465 )}{\text{ -7.68 }}$ & 99.77 \\
& $ 0.7 $ & $\underset{( 0.266 )}{\text{ -0.31 }}$ & $\underset{( 0.256 )}{\text{  1.30   }}$ & 0.05 & 8.41 & 7.65 & $\underset{( 3.260 )}{\text{ -7.34 }}$ & 99.91 \\
& $ 0.8 $ & $\underset{( 0.183 )}{\text{ -0.08 }}$ & $\underset{( 0.174 )}{\text{  1.08   }}$ & 0.07 & 12.70 & 12.23 & $\underset{( 2.683 )}{\text{ -5.53 }}$ & 100.00 \\
& $ 0.9 $ & $\underset{( 0.147 )}{\text{ 0.04 }}$ & $\underset{( 0.138 )}{\text{ 0.96 }}$ & 0.58 & 21.66 & 22.79 & $\underset{( 1.707 )}{\text{ -1.94 }}$ & 92.86 \\
\textsuperscript{**}{\footnotesize (Obs. 4312)}& 0.95 & $\underset{( 0.135 )}{\text{ 0.04 }}$ & $\underset{( 0.126 )}{\text{ 0.96 }}$ & 0.90 & 31.07 & 34.19 & $\underset{( 1.046 )}{\text{ 0.43 }}$ & 13.73 \\
\cmidrule(lr){1-1}
\cmidrule(lr){2-9}
{90 days}\textsuperscript{***} & 0.05 & $\underset{( 0.405 )}{\text{ 0.60 }}$ & $\underset{( 0.478 )}{\text{ 0.37 }}$ & 0.01 & 2.90 & 15.63 & $\underset{( 1.102 )}{\text{ -2.95 }}$ & 100.00 \\
& $ 0.1 $ & $\underset{( 0.321 )}{\text{ 0.59 }}$ & $\underset{( 0.356 )}{\text{ 0.40 }}$ & 0.00 & 3.46 & 3.84 & $\underset{( 1.495 )}{\text{ -6.36 }}$ & 100.00 \\
& $ 0.2 $ & $\underset{( 0.516 )}{\text{ 0.57 }}$ & $\underset{( 0.534 )}{\text{ 0.43 }}$ & 0.03 & 0.83 & 1.93 & $\underset{( 2.896 )}{\text{ -7.53 }}$ & 100.00 \\
& $ 0.3 $ & $\underset{( 0.637 )}{\text{ 0.62 }}$ & $\underset{( 0.643 )}{\text{ 0.39 }}$ & 0.04 & 0.17 & -0.52 & $\underset{( 3.668 )}{\text{ -8.42 }}$ & 99.84 \\
& $ 0.4 $ & $\underset{( 0.468 )}{\text{ 0.42 }}$ & $\underset{( 0.463 )}{\text{ 0.60 }}$ & 0.02 & 0.22 & -1.76 & $\underset{( 4.199 )}{\text{ -9.52 }}$ & 99.77 \\
& $ 0.6 $ & $\underset{( 0.426 )}{\text{ -0.84 }}$ & $\underset{( 0.413 )}{\text{  1.82   }}$ & 0.01 & 6.37 & 3.81 & $\underset{( 4.542 )}{\text{ -11.60 }}$ & 99.98 \\
& $ 0.7 $ & $\underset{( 0.307 )}{\text{ -0.46 }}$ & $\underset{( 0.293 )}{\text{  1.45   }}$ & 0.02 & 10.45 & 8.87 & $\underset{( 4.056 )}{\text{ -9.43 }}$ & 100.00 \\
& $ 0.8 $ & $\underset{( 0.204 )}{\text{ -0.23 }}$ & $\underset{( 0.192 )}{\text{  1.23   }}$ & 0.10 & 15.47 & 16.54 & $\underset{( 3.189 )}{\text{ -6.66 }}$ & 100.00 \\
& $ 0.9 $ & $\underset{( 0.170 )}{\text{ -0.02 }}$ & $\underset{( 0.157 )}{\text{  1.02   }}$ & 0.79 & 23.18 & 27.92 & $\underset{( 1.971 )}{\text{ -1.12 }}$ & 100.00 \\
\textsuperscript{***}{\footnotesize (Obs. 4291)}& 0.95 & $\underset{( 0.153 )}{\text{ 0.08 }}$ & $\underset{( 0.139 )}{\text{ 0.93 }}$ & 0.86 & 32.14 & 39.88 & $\underset{( 1.366 )}{\text{ -0.06 }}$ & 52.37 \\
\bottomrule
\end{tabular}%
\begin{tablenotes}
\footnotesize
\item\textit{Note}: This table reports the QR estimates of \eqref{eq:qr_est} over the sample period 2003--2021 at different horizons, using overlapping returns.	Standard errors are shown in parentheses and based on SETBB with a block length equal to the prediction horizon.  \emph{Wald test} denotes the $p$-value of the joint restriction $[\beta_0(\tau),\beta_1(\tau)] =[0,1]$.  $R^1(\tau)$ denotes the goodness of fit measure \eqref{eq:R1tau}. $R_{oos}^1(\tau)$ is the out-of-sample goodness of fit \eqref{eq:Roos_rn}, using a rolling window of size 10 times the prediction horizon.  \emph{$\widebar{\mathrm{Hit}}$} refers to the sample expectation defined in \eqref{eq:hit} and standard errors are reported in parentheses, which are obtained by stationary bootstrap based on 10,000 bootstrap samples. The last column indicates the time series average of the event that $\hat{Q}_{t,\tau} > \cquant$, where $\hat{Q}_{t,\tau}  = \hat{\beta}_0(\tau) + \hat{\beta}_1(\tau) \cquant$.
\end{tablenotes}
\end{threeparttable}
\end{adjustbox}
\end{table}

\section{Equity Premium Puzzle and SDF Implications}\label{sec:impli_sdf}
Building on the estimates in Table  \ref{tab:only.rn.quantile}, this section shows that the conditional equity premium is driven by disaster risk, and that disaster risk is a pervasive feature of the data, which poses a new challenge to asset pricing models.  I further comment on two implications of Table \ref{tab:only.rn.quantile} that relate to properties of the SDF that have previously received attention in the literature.

\subsection{Equity Premium Puzzle}\label{sec:eq_prem_puzzle}
The results in Table \ref{tab:only.rn.quantile} show that the physical distribution is close to risk-neutral in the right-tail, but not in the left-tail. Investors in the market portfolio thus get compensated for bearing downside risk, but not upside risk. This result has important repercussions for explanations of the equity premium puzzle. To see this, consider the following decomposition of the equity premium\footnote{See Appendix \ref{subsec:tail_risk} for a derivation.}  
\begin{align}
	\cexp{\omrkt} - \ofree &=  \int_0^1 \lro{ \cpquant - \cquant } \diff \tau \nonumber \\
	&= \underbrace{\int_0^{\ubar{\tau}} \lro{ \cpquant - \cquant }\diff \tau}_{\text{disaster risk} } + \int_{\ubar{\tau}}^1 \lro{ \cpquant - \cquant } \diff \tau, \label{eq:eqDecomp}
\end{align}
where $\ubar{\tau}$ is a percentile close to zero. The first term on the right-hand side aggregates the local difference between the risk-neutral and physical quantiles in the left-tail, which I define as the contribution of disaster risk. The results in Table \ref{tab:only.rn.quantile} show that these differences are the primary determinant for the equity premium, as in the right-tail we have $\cpquant\approx \cquant$. The latter finding is consistent with the modeling assumption in (time-varying) disaster risk models  that shocks to the market return are negative conditional on a disaster occurring (see, e.g., the condition $\theta < 0$ in Example {\protect \ref{ex:disaster_risk_jumps}}). Hence, an asset pricing model seeking to explain the (conditional) equity premium of the market return must embed a source of disaster risk.\\

To illustrate the pervasiveness  of disaster risk in the data, I consider the Lorenz curve associated with the conditional equity premium
\begin{equation*}
L_t(x) \coloneqq \frac{\int_0^x  \lro{\cpquant - \cquant} \diff \tau  }{ \cexp{\omrkt} - \ofree  }  \overset{\eqref{eq:eqDecomp}}{=}  \frac{\int_0^x  \lro{\cpquant - \cquant} \diff \tau  }{\int_0^1  \lro{\cpquant - \cquant} \diff \tau  } ,  \quad 0 \le x \le 1.
\end{equation*}
The Lorenz curve summarizes the proportion of the equity premium contributed by the bottom $x$\% of returns, akin to its interpretation in labor economics to summarize wealth inequality. Since $\cpquant$ is unobserved, I use instead the inferred value, $\hat{Q}_{t,\tau} = \hat{\beta}_0(\tau) + \hat{\beta}_1(\tau) \cquant$, with the estimated parameters coming from the QR estimates in \eqref{eq:qr_est}.\\

\begin{figure}[!htb]
	\centering
	\begin{subfigure}[b]{0.49\textwidth}
		\centering
		\includegraphics[width=\textwidth]{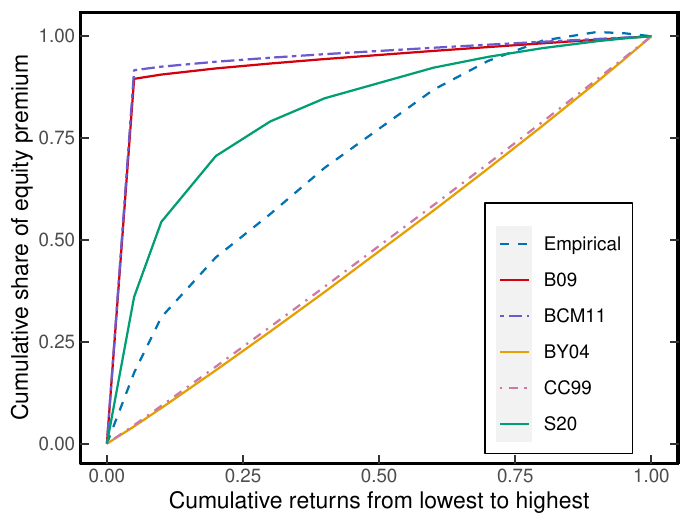}
		\caption{Average Lorenz curve}
		\label{fig:lorenz}
	\end{subfigure}
	\hfill
	\begin{subfigure}[b]{0.49\textwidth}
		\centering
		\includegraphics[width=\textwidth]{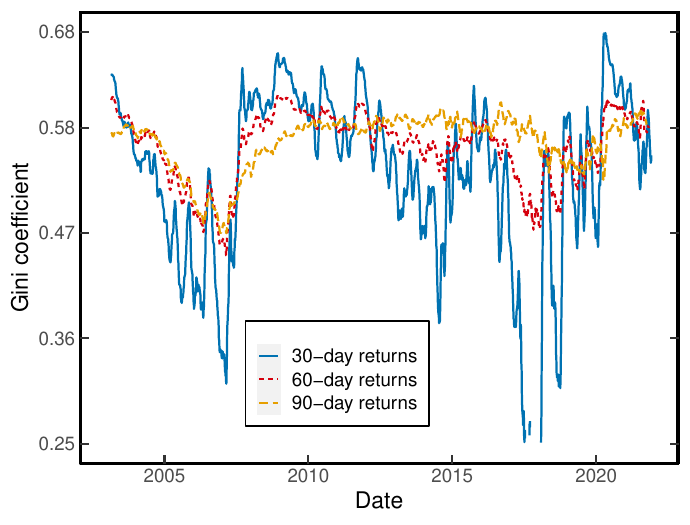}
		\caption{Conditional Gini coefficient}
		\label{fig:gini}
	\end{subfigure}
	\begin{subfigure}[b]{0.49\textwidth}
		\centering
		\includegraphics[width=\textwidth]{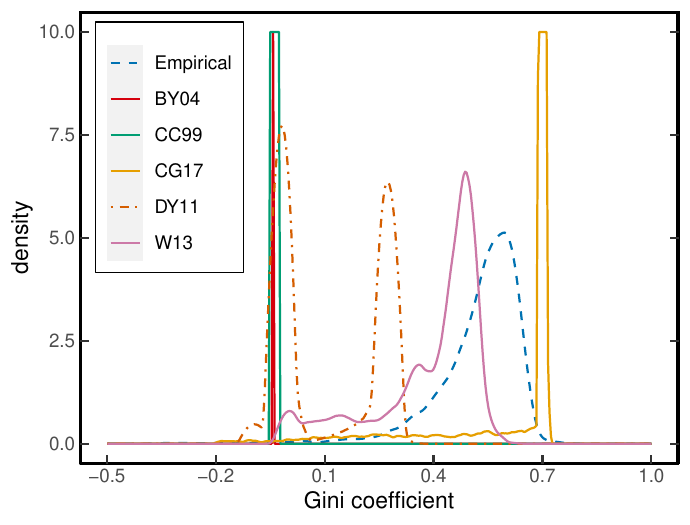}
		\caption{Ergodic distribution of Gini coefficients}
		\label{fig:gini_ergodic}
\end{subfigure}
\caption{\textbf{Lorenz curve and Gini coefficient of the conditional equity premium.}   \footnotesize This figure presents the Lorenz curve and Gini coefficient associated with the conditional equity premium in both empirical data and asset pricing models. Panel (a) displays the time-averaged Lorenz curve estimated from 30-day returns (Empirical) alongside Lorenz curves implied by the unconditional asset pricing models of \citet{Barro2009} (B09), \citet{backus2011disasters} (BCM11), and \citet{Schreindorfer2020} (S20), as well as the average Lorenz curve from conditional asset pricing models by \citet{campbell1999force} (CC99) and \citet{bansal2004risks} (BY04). Panel (b) depicts the estimated Gini coefficient over time for different return horizons and is smoothed using a 30-day rolling window. Panel (c) shows the ergodic distribution of Gini coefficients estimated from 30-day returns (Empirical) and those implied by the conditional asset pricing models of CC99, BY04, \citet{Drechsler2011} (DY11),  \citet{wachter2013can} (W13), and \citet{Constantinides2017} (CG17). Model parameters are calibrated on a monthly frequency, and the ergodic distribution is derived from 10,000 state draws.}
\label{fig:Lorenz_Gini_all}
\end{figure}

Figure \ref{fig:lorenz} shows the average Lorenz curve in the data, together with the Lorenz curve implied by various asset pricing models.\footnote{I thank \citet{Beason2022} for making the code to simulate from these models publicly available.} In the data, the Lorenz curve is quite concave, thus showing that the majority of the equity premium is contributed by the left-tail. At the same time, my estimation adds nuance to the degree of disaster risk influencing the equity premium. Specifically, while the disaster risk models of \citet{Barro2009} and \citet{backus2011disasters} attribute approximately 90\% of the equity premium to the lowest 5\% of returns, empirical estimates suggest this proportion is only around 17\%. These findings also deviate substantially from the nonparametric estimates of \citet{Beason2022}, who report that 91.5\% of the equity premium is driven by the bottom 5\% of returns. Our results differ because I account for conditioning information, while \citet{Beason2022} employ an unconditional approach. Using unconditional averages can inflate the tails of the physical distribution \citep{ChabiYo2008}, leading to an overestimation of disaster risk. \\

 On the other hand, the models of \citet{campbell1999force} and \citet{bansal2004risks} are even more misspecified since the Lorenz curve in these models is slightly convex, thus attributing more than 50\% of the equity premium to upside returns. The model of \citet{Schreindorfer2020} matches the Lorenz curve best, even though it also overestimates the contribution of disaster risk to the equity premium.\\

I also consider the Gini coefficient derived from the Lorenz curve
\begin{equation*}
G_t = 2 \int_0^1 L_t(\tau) \diff \tau -1. 
\end{equation*}
By construction, the Gini coefficient is between -1 and 1, and a value closer to 1 indicates that a bigger proportion of the equity premium is coming from the left-tail.  In contrast, a value of 0 suggests that the equity premium is evenly distributed across the return distribution, while negative values imply that the right-tails contribute more to the equity premium than the left-tails.  Figure \ref{fig:gini} shows the time series of conditional Gini coefficients for various return horizons. For 30-day returns, the Gini coefficient mostly hovers between 0.33 and 0.68. At longer horizons, the Gini coefficients exhibit less variability and typically range between 0.47 to 0.6. These coefficients are also countercyclical, peaking during periods associated with economic downturns, such as the 2008 financial crisis and the Covid-19 crisis. Overall, the Gini coefficients consistently exhibit strong positive values, highlighting the pervasiveness of conditional disaster risk in the data, which extends beyond crisis periods.\\

Finally, I analyze the ergodic distribution of Gini coefficients in time-varying asset pricing models and compare it to the distribution implied by the data.\footnote{In the model, I obtain the distribution of Gini coefficients from the state distribution. In the data, I rely on the time series average. If the data are generated by the model and the system is ergodic, Birkhoff's theorem implies that the state and time averages are equal almost everywhere.} Figure \ref{fig:gini_ergodic} displays these distributions and shows that many asset pricing models have difficulty in matching the empirical distribution. The conditional lognormal models of \citet{campbell1999force} and \citet{bansal2004risks} imply negative Gini coefficients, with minimal variation among different states, contrary to what the data indicate. 
The models of \citet{Drechsler2011}, \citet{wachter2013can}, \citet{Constantinides2017} all incorporate a source of disaster risk, but they also have difficulty to match the empirics.  In particular, the models of \citet{Drechsler2011} and \citet{wachter2013can} embed too little disaster risk, while the model of \citet{Constantinides2017} overestimates the impact of disaster risk.

\subsection{Driver of Disaster Risk Premia: Insurance or Beliefs?}
Disaster risk premia have two components: an insurance effect and a forward-looking beliefs effect (under rational expectations). To see this, consider a short position in a derivative security that pays one dollar if the market return is below a threshold, denoted by $x$, in the left tail. The return on such an investment can be expressed as
\begin{equation*}
 \underbrace{\texpneut\lr{\ind{\omrkt \le x}}}_{\text{price of insurance}} - \underbrace{\cexp{\ind{\omrkt \le x}}}_{\text{forward looking belief}} = \tilde{F_t}(x) - F_t(x).
\end{equation*}

During a crisis, the price of this security tends to rise. This effect can occur in the disaster risk model (Example {\protect \ref{ex:disaster_risk_jumps}}), if risk aversion increases when a disaster hits, leading to an increase in $\tilde{F}_t(x)$ and a subsequent decrease in $\cquant$. Simultaneously, investors may believe that the actual probability of a disaster increases during a crisis. This belief drives up $F_t(x)$ and, consequently, pushes down $\cpquant$.\\

\begin{figure}[!htb]
	\centering
	\begin{subfigure}[b]{0.49\textwidth}
		\centering
		\includegraphics[width=\textwidth]{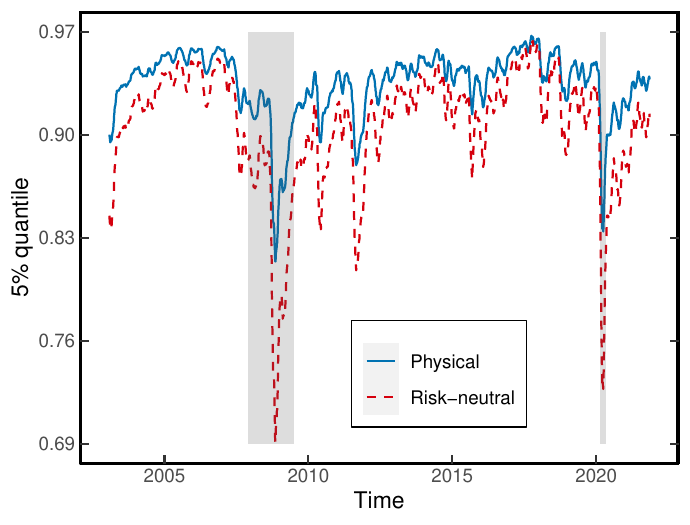}
		\caption{5\% quantile}
		\label{fig:q_qtilde}
	\end{subfigure}
	\hfill
	\begin{subfigure}[b]{0.49\textwidth}
		\centering
		\includegraphics[width=\textwidth]{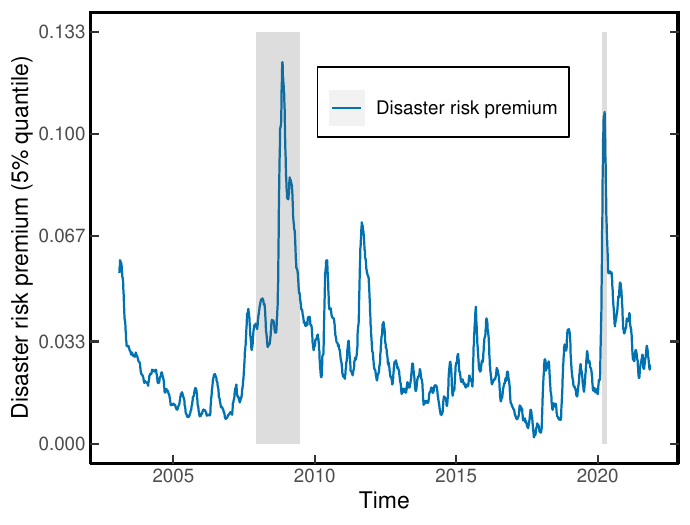}
		\caption{Disaster risk premium at 5\% quantile}
		\label{fig:dp}
	\end{subfigure}
	\caption{\textbf{Disaster risk premia for 30-day returns at the 5th percentile}. \footnotesize Panel (a) shows the physical and risk-neutral quantile functions over time at $\tau = 0.05$. The physical quantile function is estimated from the quantile regression in \eqref{eq:qr_est}. Panel (b) shows the associated disaster risk premium, $\cpquant - \cquant$. Both panels are smoothed using a 30-day moving window. The two shaded bars denote the Great Recession period (Dec 2007 -- June 2009) and Covid-19 crisis (Feb 2020 -- April 2020). }
	\label{fig:dp_time}
\end{figure}

Building on this discussion, it is not immediately clear what the net effect is on disaster risk premia ($\cpquant - \cquant$), as both $\cpquant$ and $\cquant$ tend to decrease during periods of heightened market uncertainty. Figure \ref{fig:q_qtilde} illustrates this effect for 30-day returns and $\tau = 0.05$. Notably, during the global financial crisis and Covid-19 crisis, both the physical and risk-neutral quantile functions exhibit significant drops.\\

However, the downward spikes in the risk-neutral quantile function are more pronounced, as it decreases to 63\% in these periods. In contrast, the physical quantile function only drops to 78\%, suggesting that a monthly loss of 22\% or more had a 5\% probability. To put this in perspective, this probability is 14 times higher than the estimate obtained from historical monthly S\&P500 returns (from 1926 to 2021). This calculation shows that historical estimates can diverge significantly from forward-looking beliefs. Furthermore, the time fluctuations in the physical quantile function lend empirical support to the notion of time-varying disaster risk, as proposed in various models such as \citet{gabaix2012variable}, \citet{wachter2013can}, \citet{Constantinides2017}, \citet{Isore2017}, \citet{Farhi2018} and \citet{Seo2019}.\\


To shed light on the net effect on disaster risk premia during crises, Figure \ref{fig:dp} displays the evolution of disaster risk premia over time. The most significant change occurs during the peak of the global financial crisis and the Covid-19 crisis. In these turbulent periods, disaster risk premia consistently rise, suggesting that the insurance effect is more substantial than the forward-looking beliefs effect.\footnote{I find similar results for 60- and 90-day returns.} Because of these large increases, disaster risk is a more important driver of the equity premium, which clarifies the countercyclical Gini coefficients in Figure \ref{fig:gini}.\\


\subsection{Predicting the Equity Premium}
The previous results establish that, in times of heightened market uncertainty, the equity premium is driven more by disaster risk. This observation suggests a strong link between $\cexp{\omrkt} - \ofree$ and the tail of the risk-neutral distribution, which motivates the predictive regression
\begin{equation}\label{eq:ols}
\omrkt - \ofree = \beta_0 + \beta_1 \cquant + \varepsilon_{t \to T},
\end{equation}
where $\cquant$ is evaluated at $\tau = 0.05$.\\

Table \ref{tab:ols_reg} shows the results at several return horizons. In all cases, the coefficient is negative, consistent with previous findings that the equity premium increases under market uncertainty. Following \citet{welch2008comprehensive}, the table also reports the out-of-sample $R^2$, denoted by $R_{oos}^2$, which compares the predictions of \eqref{eq:ols} to a rolling average of excess returns. Precisely, I estimate \eqref{eq:ols} using the sub-sample covering 2003--2012, and fix the estimated parameters to predict excess returns over the out-of-sample period 2013--2021. Encouragingly,  $R_{oos}^2$ is always positive and statistically significant according to the  \citet{Diebold1995} test, thus suggesting that the left-tail of the risk-neutral quantile function outperforms the  historical mean benchmark. These values are also substantially higher compared to the $R_{oos}^2$ reported by \citet{welch2008comprehensive} using various valuation ratios, or \citet{martin2017expected} using SVIX.\footnote{The latter is not directly comparable however, since SVIX does not require parameter estimation.}\\

\begin{table}[!htb]
	\captionsetup{width=12cm}	
	\centering
	\caption{\textbf{OLS estimates of conditional equity premium}}
	\label{tab:ols_reg}
	\begin{tabular}{lccccc}
		\toprule
		\midrule
		{Horizon} & $\hat{\beta}_0$ & $\hat{\beta}_1$ & $R^2$[\%] &$R_{oos}^2$[\%] & $p$-value DM \\ 
		\cmidrule(lr){2-6}
	30 days & $\underset{( 0.088 )}{\text{  0.13   }}$ & $\underset{( 0.095 )}{\text{ -0.14 }}$ & 10.44 & 1.82 & 0.00 \\
	60 days & $\underset{( 0.120 )}{\text{  0.17   }}$ & $\underset{( 0.136 )}{\text{ -0.18 }}$ & 18.16 & 3.30 & 0.00 \\
	90 days & $\underset{( 0.150 )}{\text{  0.20   }}$ & $\underset{( 0.178 )}{\text{ -0.22 }}$ & 26.67 & 4.20 & 0.00 \\
	 &  & & & & \\ 
	\cmidrule(lr){2-3}
	\cmidrule(lr){4-6}
	Sample & \multicolumn{2}{c}{2003--2021 }  & \multicolumn{3}{c}{2013--2021 } \\
		\bottomrule
	\end{tabular}%
	\caption*{\textit{Note}:  \footnotesize This table reports the OLS estimates of \eqref{eq:ols} for 30-, 60- and 90-day returns.  Standard errors are shown in parentheses and calculated using stationary bootstrap, with an average block length equal to the return horizon. $R_{oos}^2$ denotes the out-of-sample $R^2$ using the historical rolling mean of excess returns. The window length is equal to 5 years. \emph{$p$-value DM} denotes the $p$-value of the \citet{Diebold1995} test that  the risk-neutral quantile exhibits equal out-of-sample forecasting accuracy as the rolling mean. The  ``\emph{Sample}'' row indicates the specific time periods used for estimation.} 
\end{table}

\begin{figure}[!htb]
	\centering
	\begin{subfigure}[b]{0.49\textwidth}
		\centering
		\includegraphics[width=\textwidth]{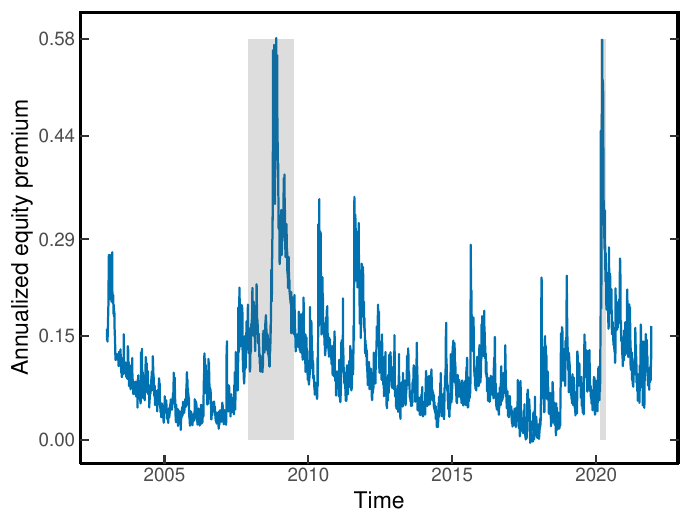}
		\caption{30-day returns}
		\label{fig:eq_30}
	\end{subfigure}
	\begin{subfigure}[b]{0.49\textwidth}
		\centering
		\includegraphics[width=\textwidth]{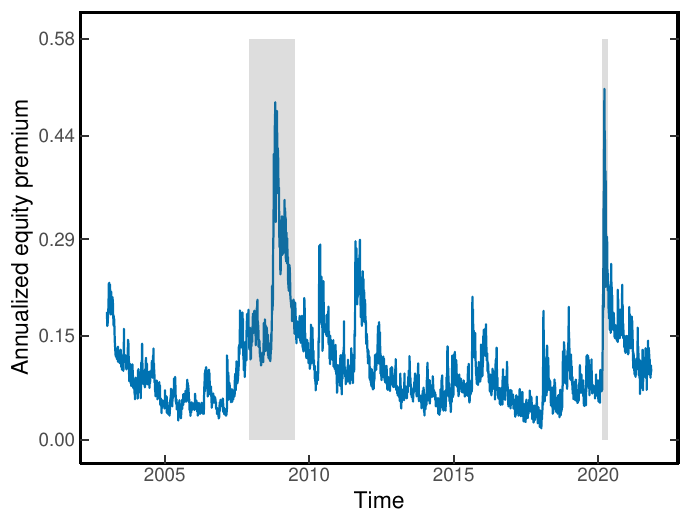}
		\caption{60-day returns}
		\label{fig:eq_60}
	\end{subfigure}
	\caption{\textbf{Estimated equity premium.} \footnotesize This figure shows the estimated equity premium based on \eqref{eq:ols} for 30-day returns (Panel \ref{fig:eq_30}) and 60-day returns (Panel \ref{fig:eq_60}). In both cases, the equity premium is converted to annual units. The two shaded bars signify the Great Recession period (Dec 2007 – June 2009) and Covid-19 crisis (Feb 2020 – April 2020).}
	\label{fig:ols}
\end{figure}

Figure \ref{fig:ols} shows the estimated equity premium over time for 30- and 60-day returns. The panels are annualized to make them comparable. Both panels display considerable variation in the equity premium over time and large values during the global financial crisis and Covid-19 crisis.  In these periods, Figure \ref{fig:eq_30} suggests that the annualized equity premium peaks at 58\%, which is substantial relative to more conventional estimates based on dividend-price ratios. On the other hand, the estimates around the 2008 financial crisis are in line with \citet[Figure \rom{4}]{martin2017expected}.

\subsection{Pricing Kernel Monotonicity and Stochastic Dominance} 
Besides the equity premium puzzle, the QR estimates in Table \ref{tab:only.rn.quantile} also provide insights into other asset pricing anomalies, such as pricing kernel monotonicity.  Pricing kernel monotonicity refers to the property that $\osdf(\omrkt) \coloneqq \uexp{\osdf | \omrkt}$ is a decreasing function of the market return. Asset pricing models that link the SDF to the marginal rate of substitution imply that the pricing kernel is indeed  a decreasing function. Empirically, there is suggestive evidence that the pricing kernel is not monotonic, which is puzzling as it contradicts that a representative investor is risk-averse (see \citet{ait1998nonparametric}, \citet{jackwerth2000recovering}, \citet{rosenberg2002empirical}, \citet{bakshi2010returns}, \citet{beare2016empirical} and \citet{cuesdeanu2018pricing}). However, a formal statistical test that can detect violations of monotonicity is challenging as one needs uniform confidence bands for the estimated SDF, which requires tools from empirical process theory (see, e.g., \citet{beare2016empirical}).\\

I consider a different approach based on stochastic dominance. Proposition \ref{prop:monFOSD} in the Appendix shows that pricing kernel monotonicity implies that the physical distribution is first-order stochastic dominant (FOSD) over the risk-neutral distribution, i.e., $F_t(x) \le \tilde{F}_t(x)$ for all $x$. The latter condition can be rephrased as $F_t(\cquant) \le \tau$ for all $\tau \in (0,1)$. A violation of stochastic dominance, and hence pricing kernel monotonicity, is thus implied if there is statistical evidence that $F_t(\cquant) > \tau$ for a single $\tau$. To investigate this possibility, let\footnote{The $\mathrm{Hit}_{t\to N}$ function was first introduced by \citet{engle2004caviar} in a different context.} 
\begin{align}\label{eq:hit}
	\mathrm{Hit}_{t\to N} &= \ind{\omrkt < \cquant} - \tau, \nonumber \\ 
	\widebar{\mathrm{Hit}} &= \frac{1}{T} \sum_{t=1}^{T} \mathrm{Hit}_{t\to N}. 
\end{align}
Hence,  $\widebar{\mathrm{Hit}}$ provides an estimate of $\mathbb{E}(F_t(\cquant) - \tau)$ which ought to be negative for all $\tau$ under FOSD.\footnote{$\widebar{\mathrm{Hit}}$ also yields another measure of the difference between $F_t$ and $\tilde{F}_t$. Consistent with the quantile regression estimates, the Hit statistic shows that $F_t$ and $\tilde{F}_t$ are similar in the right-tail, but different in the left-tail.} The ``$\widebar{\mathrm{Hit}}$'' column in Table  \ref{tab:only.rn.quantile} reports the value of \eqref{eq:hit}, which is positive for $\tau = 0.95$ at the 30- and 60-day horizon. However, these estimates are not significant at the conventional levels and a violation of FOSD cannot be concluded. \\

Since $F_t(x) \le \tilde{F}_t(x)$ if and only if $\cpquant > \cquant$, it follows that violations of stochastic dominance can also be identified directly from the quantile function. Based on the QR estimates \eqref{eq:qr_est}, consider the predicted quantile function $\hat{Q}_{t,\tau} = \hat{\beta}_0(\tau) + \hat{\beta}_1(\tau) \cquant$. The last column in Table \ref{tab:only.rn.quantile} displays the time series average of instances where $\hat{Q}_{t,\tau} > \cquant$. Broadly speaking, for all horizons, violations of stochastic dominance are infrequent, except far in the right-tail. At $\tau = 0.95$, stochastic dominance is frequently violated, consistent with a non-monotonic pricing kernel.\footnote{The most significant violations occur during two major financial crises: the 2008 financial crisis and the 2020 Covid-19 crisis.} In representative agent models, this result is puzzling as it contradicts the assumption of decreasing marginal utility of wealth (see Proposition \ref{prop:sufFOSD} in the Appendix).

\subsection{Belief Recovery}  A recent literature asks to what extent Arrow prices can be used to learn about the underlying probability distribution of the data, or the subjective probabilities used by investors. Since Arrow prices are confounded by risk aversion, it is impossible to identify the underlying probabilities from Arrow prices alone, unless one imposes additional restrictions \citep{ross2015recovery,borovivcka2016misspecified,Bakshi2018,qin2018long,jackwerth2020does}. For example, \citet{ross2015recovery} uses the Perron-Frobenius theorem to recover investors' beliefs, which agrees with the underlying 
physical measure under rational expectations.\\

Complementary to this insight, the QR estimates in Table \ref{tab:only.rn.quantile} show that the right-tail of the physical distribution can approximately be recovered from the right-tail of the risk-neutral distribution, which aligns with the investor's belief under rational expectations. In contrast, the left-tail of the physical distribution cannot be recovered even though the risk-neutral quantile serves as a conservative lower bound.   In Section \ref{sec:dark}, I propose a more stringent lower bound to recover the left-tail of the physical distribution as well from option data.

\section{QR and Robust Estimation of Disaster Risk}\label{sec:Robust_QR}
Section \ref{sec:eq_prem_puzzle} demonstrated that the conditional lognormal assumption is inconsistent with the observed disaster risk premia in the market. At the same time, Figure \ref{fig:lorenz} showed that disaster risk models tend to overestimate the magnitude of disaster risk in the data. These conclusions heavily rely on the accuracy of QR in providing estimates of the physical quantile function.\\

In this section, I compare QR to nonparametric SDF methods for estimating disaster risk. Foreshadowing the results, I show that QR is more robust and argue that the SDF approach tends to overestimate disaster risk. These results help explain the current disagreement about the extent of disaster risk in the data, and provide further support for QR to estimate this risk.\\

\subsection{QR in the Conditional Lognormal Model}\label{sec:logn}
To convey the intuition, it is convenient to work with a discretized version of the \citet{black1973pricing} model. There is a riskless asset that offers a certain return, $\ofree \equiv R_f = e^{r_{f} N}$, and a risky asset with return
\begin{equation}\label{eq:logn}
\omrkt = \exp([\mu_t - \frac{1}{2}\sigma_t^2]N + \sigma_t \sqrt{N} Z_{t+N}),
\end{equation}
where $\mu_t$ represents the conditional mean return, $\sigma_t$ is the conditional volatility, and  $Z_{t+N}$ is a random shock that follows a standard normal distribution. In this setup, $\osdf \coloneqq \exp(-[r_{f} + \xi_t^2/2]N - \xi_t \sqrt{N} Z_{t+N})$ is a valid SDF with conditional Sharpe ratio
\begin{equation}\label{eq:cond_sharpe}
	\xi_t = \frac{\mu_t - r_{f}}{\sigma_t}.
\end{equation}

Hence, under risk-neutral measure, the conditional distribution of $\omrkt$ is given by
\begin{equation}\label{eq:r_logn}
\log \tilde{R}_{m,t \to N} \sim \mathcal{N}\lro{ (r_{f} - \frac{1}{2}\sigma_t^2)N, \sigma_t^2 N}.
\end{equation}
Notice that $\sigma_t$ is implicitly observed from the risk-neutral distribution, but $\mu_t$ is unobserved with mean $\mu \coloneqq \uexp{\mu_t}$ and  variance  $\sigma_\mu^2 \coloneqq \Var(\mu_t) < \infty$.  The following result characterizes the limiting behavior of the QR estimates \eqref{eq:qr_est} in the lognormal model when the variance of the equity premium is small. A convenient way to model this is by means of a drifting sequence $\sigma_\mu^T \to 0$ as $T \to \infty$, which captures the intuition that the volatility of the equity premium is much smaller than the return volatility.

\begin{proposition}[QR in Lognormal Model]\label{lemma:bs}
	In the lognormal model described above with return observations $\{\omrkt\}_{t=1}^T$ and risk-neutral quantile functions $\{\cquant\}_{t=1}^T$, the following hold.
	\begin{enumerate}[(i)]	
		\item 	Suppose that conditional on time $t$, $\mu_t$ follows a normal distribution  $\mu_t \sim \mathcal{N}(\mu,\sigma_\mu^2)$,  independent of $\sigma_t$.	Let $\cpquant(\sigma_t,\sigma_\mu)$ denote the physical quantile function of $\omrkt$ conditional on $\sigma_t$ only. Then, for all $\tau \in \mathcal{I} \coloneqq$ a closed subset of $[\varepsilon,1-\varepsilon]$ for $0 < \varepsilon < 1$, the physical quantile function satisfies \label{item:logn1}
		\begin{align*}
		\cpquant(\sigma_t,\sigma_\mu) &= \exp\lr{(\mu - \frac{1}{2}\sigma_t^2)N + \lro{\sqrt{\sigma_{\mu}^2 N^2 + \sigma_t^2 N}} \Phi^{-1} (\tau)}\\
		&= \cquant e^{(\mu-r_f)N}\lro{1 + \bigoh{\sigma_\mu N}},
		\end{align*}
	where $\Phi^{-1} (\tau)$ denotes the quantile function of the standard normal distribution.
	\item Consider a drifting sequence for $\sigma_\mu$, denoted by $\sigma_\mu^T \to 0$ as $T \to \infty$. Then, under Assumption \ref{ass:sig_small} in the Appendix, the estimated parameters in the quantile regression \label{item:logn2}
	\begin{equation*}
	\lr{\hat{\beta}_0(\sigma_\mu^T;\tau),\hat{\beta}_1(\sigma_\mu^T;\tau)} = \argmin_{(\beta_0,\beta_1)  \in \mathbb{R}^2} \sum_{t=1}^{T} \rho_\tau(\omrkt - \beta_0 - \beta_1 \cquant),
	\end{equation*}
	satisfy
	\begin{equation}\label{eq:claim}
		\lr{\hat{\beta}_0(\sigma_\mu^T;\tau),\hat{\beta}_1(\sigma_\mu^T;\tau)} = \lr{0,e^{(\mu-r_f)N}} + o_p(1).
	\end{equation}
Furthermore, the quantile forecast based on the QR estimates satisfies 
	\begin{equation}\label{eq:q_forecast}
\hat{\beta}_0(\sigma_\mu^T;\tau) +  \hat{\beta}_1(\sigma_\mu^T;\tau)   \cquant 	= \cpquant + o_p(1).
	\end{equation}
\end{enumerate}
\end{proposition}
\begin{proof}
	See Appendix \ref{app:proof_logn}.
\end{proof}
Proposition \ref{lemma:bs}(\ref{item:logn1}) shows that the risk-neutral quantile function is a good predictor of $\cpquant(\sigma_t;\sigma_{\mu})$ when $\sigma_\mu$ is small, and the difference between the two functions is governed by the  unconditional equity premium $e^{(\mu-r_f)N}$. In this case, Proposition \ref{lemma:bs}(\ref{item:logn2}) suggests that the QR estimates are almost constant across $\tau$ and close to $[0,e^{(\mu-r_f)N}]$. This result obtains without assuming that $\mu_t$ follows a normal distribution. The wedge between $\cpquant(\sigma_t; \sigma_{\mu})$ and $\cquant$ not explained by the equity premium can be attributed to uncertainty about $\mu_t$, which increases the variance of the physical distribution.  The assumption that $\sigma_\mu$ is small relative to $\sigma_t$ accords with empirical findings of \citet[Table \rom{1}]{martin2017expected}, who finds that $2.4\% \le \sigma_{\mu} \le 4.6\%$, whereas $\sigma_t$ hovers around 20\%. Unreported simulations show that the approximation in \eqref{eq:claim} obtains closely when the model is calibrated to match these stylized facts.  As a result, the physical quantile forecast based on the QR estimates in \eqref{eq:q_forecast} is also highly accurate. \\

\subsection{QR versus Nonparametric SDF Estimation}
Because of the availability of closed-form expressions in the lognormal model, it is instructive to compare the QR approach to alternative methods for estimating the physical distribution. Since the SDF represents the Radon–Nikodym derivative of the risk-neutral and physical measures, it is possible to obtain the physical quantile function from the estimated SDF. There is a substantial literature on how to estimate the SDF in a forward-looking manner (see Remark \ref{remark:sdf_scale}). For this comparison, I consider the state-of-the-art SDF estimator proposed by \citet{cuesdeanu2018pricing} (CJ).\\

 After some algebra, the SDF in the Black-Scholes model can be expressed as a function of the market return:
\begin{equation}\label{eq:bs_sdf}
\osdf = 	\exp\lro{-\frac{N}{2} \lr{\mu_t + r_f  + \frac{r_f^2 - \mu_t^2}{\sigma_t^2}}} (\omrkt)^{-\xi_t/\sigma_t},
\end{equation}
where $\xi_t$ is the conditional Sharpe ratio \eqref{eq:cond_sharpe}. CJ project the unobserved SDF in \eqref{eq:bs_sdf} on the market return and estimate an SDF of the form
\begin{equation*}
\hat{M}_{t \to N} = C_t g(\omrkt),
\end{equation*}
where $C_t$ is a time-varying constant, and $g(\cdot)$ is an unknown function that can be estimated by choosing a sieve basis. Since $g(\cdot)$ is \emph{time-homogeneous}, it is evident that changes in the shape of the true SDF in \eqref{eq:bs_sdf} are not captured by the estimated SDF. Specifically, in times when the Sharpe ratio is high, the physical and risk-neutral measures exhibit more distinct differences, as the true SDF becomes steeper. Because the estimated SDF does not account for these shape changes, it leads to a severe underestimation of the physical quantile function in the left-tail. Proposition \ref{lemma:bs} demonstrates that the QR approach does not suffer from this limitation.\\

\begin{figure}[!htb]
	\centering
	\begin{subfigure}[b]{0.49\textwidth}
		\centering
		\includegraphics[width=\textwidth]{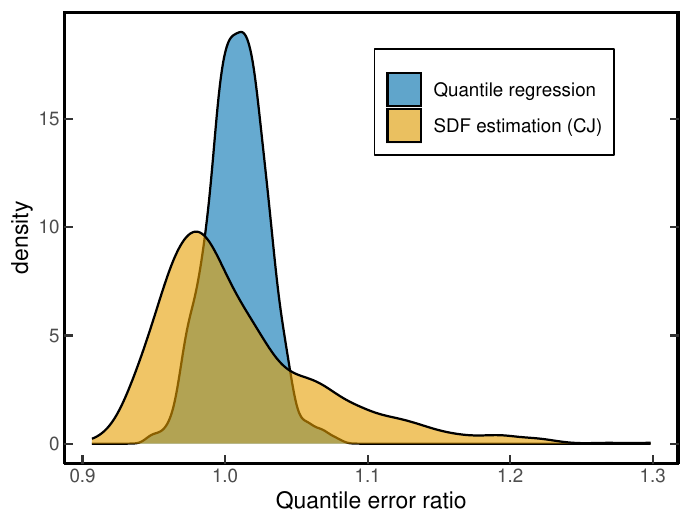}
		\caption{Black-Scholes}
		\label{fig:q_bs1}
	\end{subfigure}
	\hfill
	\begin{subfigure}[b]{0.49\textwidth}
		\centering
		\includegraphics[width=\textwidth]{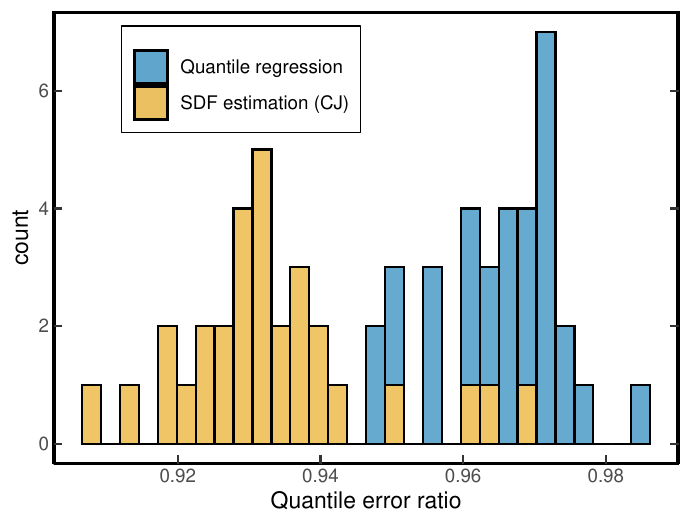}
		\caption{Black-Scholes, largest risk-adjustment}
		\label{fig:q_bs2}
	\end{subfigure}
	\begin{subfigure}[b]{0.49\textwidth}
		\centering
		\includegraphics[width=\textwidth]{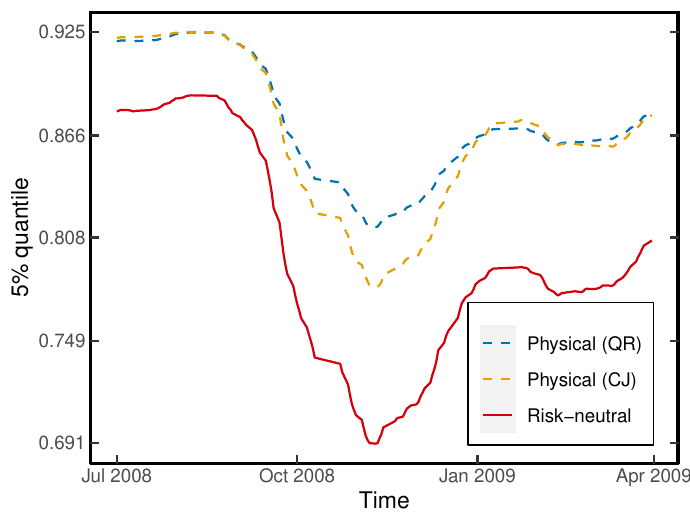}
		\caption{2008 financial crisis}
		\label{fig:q_cj1}
	\end{subfigure}
	\hfill
	\begin{subfigure}[b]{0.49\textwidth}
		\centering
		\includegraphics[width=\textwidth]{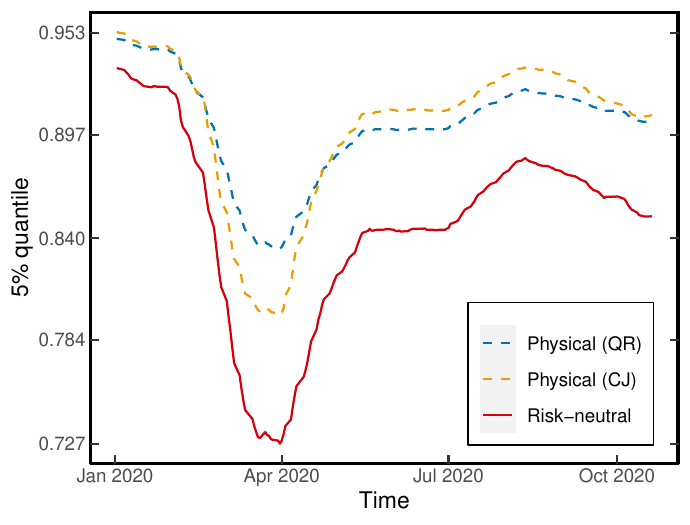}
		\caption{Covid-19 crisis}
		\label{fig:q_cj2}
	\end{subfigure}
	\caption{\textbf{Disaster risk premia at the 5th percentile}. \footnotesize Panel (a) shows the quantile error ratio, $\hat{Q}_{t,\tau}/\cpquant$, in a conditional \citet{black1973pricing} model for $\tau = 0.05$, where $\hat{Q}_{t,\tau}$ is the predicted physical quantile based on QR or the SDF estimate of CJ. Volatility is generated according to an $\mathrm{AR}(1)$-model with mean value 0.2, standard deviation 0.03 and a persistency of 0.9. The mean of the physical distribution follows $\mu_t \sim \mathcal{N}(0.07,0.02^2)$, the risk-free rate equals $r_f = 0.01$, the time horizon is one-year, and the number of observations is 1,000. Panel (b) shows the histogram of error ratios conditioned the 30 events for which $\cpquant - \cquant$ is maximal. The bottom panels illustrate the difference between the predicted physical quantile obtained from QR, and the predicted quantile coming from the SDF estimate of CJ, during the global financial crisis and the Covid-19 crisis. Both estimates are based on 30-day returns, using the data from Section \ref{sec:data}. The bottom panels are smoothed using a 30-day rolling window.}
	\label{fig:qp_bs}
\end{figure}

To illustrate this discussion, I simulate returns from the lognormal model and estimate the physical quantile function at the 5th percentile using QR and the SDF estimate of CJ. Since the conditional (physical) quantile function is known analytically in the lognormal model, I evaluate the forecast accuracy using the quantile error ratio, $\hat{Q}_{t,\tau}/\cpquant$, where $\hat{Q}_{t,\tau}$ is the predicted physical quantile based on QR or the SDF estimate. Panel \ref{fig:q_bs1} displays the empirical density of error ratios obtained by simulating 1,000 returns. In line with Proposition \ref{lemma:bs}(\ref{item:logn2}), the error ratio corresponding to QR is symmetric and closely centered around one. In contrast, when the physical quantile is inferred from the estimated SDF, the error density is biased and exhibits fat tails since the estimated SDF cannot change shape. Consequently, in periods of high disaster risk premia, the CJ method severely \emph{underestimates} $\cpquant$.\\

 Panel \ref{fig:q_bs2} presents the histogram of error ratios conditioned on the 30 largest values of $\cpquant - \cquant$, clearly illustrating the downward bias in the SDF method. On average, the predicted physical quantile is 7\% lower than its actual value when disaster risk premia are high.  The QR approach is less affected by this bias because it can capture changes in the shape of the SDF. The computational benefits of QR are also notable, as the computation of the physical quantile forecast takes less than a second. On the other hand, the SDF method requires more than 20 minutes to complete the same task.\footnote{Moreover, the optimization problem required to implement the sieve estimation did not converge, as the maximum number of iterations were exceeded. This problem occurs due to the large number of parameters to estimate, and because the optimization problem is not convex (see Remark \ref{remark:sdf_scale}).}\\


The bottom panels of Figure \ref{fig:qp_bs} further illustrate the difference between QR and CJ using the 30-day return data from Section \ref{sec:data}, particularly during the 2008 financial crisis and the Covid-19 crisis. At the height of both crises, both methods predict increases in disaster risk premia as $\hat{Q}_{t,\tau} - \cquant$ rises significantly. As mentioned earlier, the SDF approach implies that disaster risk premia increase less relative to the QR approach, as the shape of the SDF remains constant over time. However, it is worth noting that while the QR approach performs well when returns are conditional lognormal, Appendix \ref{app:add_logn} demonstrates that the quantile forecasts based on QR contradict \eqref{eq:q_forecast}, casting further doubt on the validity of the lognormal assumption in the data.


\section{Disaster Risk and SDF Volatility}\label{sec:HJbound}
Section \ref{sec:eq_prem_puzzle} demonstrated that the physical and risk-neutral distributions locally differ most in the left-tail. In this section, I show that these local differences imply that the SDF must be highly volatile; an observation that is closely related to the \citet{hansen1991implications} bound. Furthermore, I use this insight to argue that the left-tail of the physical distribution cannot be too predictable, which clarifies the low explanatory power in Table \ref{tab:only.rn.quantile}.

\subsection{A Bound on the SDF Volatility}
For ease of notation, I define $\phi_t(\tau) \coloneqq F_t(\cquant)$, which can be interpreted as the ordinal dominance curve of the measures $\mathbb{P}_t$ and $\tilde{\mathbb{P}}_t$ \citep{hsieh1996nonparametric}. Furthermore, let 
\begin{equation*}
\aleph_t^+ := \{\osdf: \osdf \ge 0 \text{ and } \cexp{\osdf \omrkt} = 1 \},
\end{equation*}
which is the space of all nonnegative conditional SDFs. The volatility bound on the SDF can now be stated as follows.
\begin{proposition}[Distribution bound]\label{prop:dist_bound}
	Assume no-arbitrage, then for any $\osdf \in \aleph_t^+$, we have
	\begin{equation}\label{eq:quantile_bound}
		\frac{\sigma_t(\osdf)}{\cexp{\osdf}} \ge \frac{\abs{\tau - \phi_t(\tau)}}{\sqrt{\phi_t(\tau) (1- \phi_t(\tau))}} \qquad \forall \tau \in (0,1).
	\end{equation}
	If a risk-free asset exists, then  $\cexp{\osdf} = 1/\ofree$ and \eqref{eq:quantile_bound} simplifies to
	\begin{equation*}\label{eq:quantile_bound_RiskFree}
		\sigma_t(\osdf) \ge \frac{1}{\ofree} \frac{\abs{\tau - \phi_t(\tau)}}{\sqrt{\phi_t(\tau) (1- \phi_t(\tau))}} \qquad \forall \tau \in (0,1).
	\end{equation*} 
The bound can be further rewritten in terms of the conditional CDFs only
\begin{equation}\label{eq:CDF_rep}
		\sigma_t(\osdf) 	=  \frac{1}{\ofree} \frac{\abs{\tilde{F}_t(x) - F_t(x)}}{F_t(x)(1-F_t(x))} \qquad \forall x > 0.
\end{equation}
\end{proposition}
\begin{proof}
	See Appendix \ref{app:proof_qbound}. 
\end{proof}
If $\mathbb{P}_t = \tilde{\mathbb{P}}_t$, agents are risk-neutral and the dominance curve evaluates to $\phi_t(\tau) = \tau$. In that case the distribution bound degenerates to zero. Proposition \ref{prop:dist_bound} makes precise the sense in which any local difference between the physical and risk-neutral distribution leads to a volatile SDF. Compare this to the classical \citet{hansen1991implications} (HJ) bound:
\begin{equation}\label{eq:HJbound}
	\sigma_t(\osdf) \ge \frac{1}{\ofree} \frac{\abs{\cexp{\omrkt} - \ofree}}{\sigma_t(\omrkt)}.
\end{equation}
The lower bound in \eqref{eq:HJbound} shows that any excess return leads to a volatile SDF. Essentially, \eqref{eq:HJbound} uses three sources of information: \begin{inparaenum}[(i)]
	\item the mean of the physical distribution
	\item the mean of the risk-neutral distribution
	\item the variance of the physical distribution.
\end{inparaenum} The lower bound in \eqref{eq:HJbound} is also a global measure of distance between $\mathbb{P}_t$ and $\tilde{\mathbb{P}}_t$, since the mean and volatility are averages across the whole distribution.\\

In contrast, the bound in \eqref{eq:CDF_rep} compares the physical and risk-neutral distribution at every point $x$, which is a \emph{local} measure of distance between $\mathbb{P}_t$ and $\tilde{\mathbb{P}}_t$. To clarify this local interpretation, consider the following decomposition of the equity premium
\begin{align}\label{eq:hoeff}
	\frac{\cexp{\omrkt} - \ofree}{\ofree} &= -\cov[\omrkt,\osdf] \nonumber \\
	&= \int_{0}^{\infty} \cov[\ind{\omrkt \le x},\osdf] \diff x,
\end{align}
where the first equation follows since the SDF prices the market return \eqref{eq:sdf_def}, and the second equation is a consequence of Hoeffding's identity (see Lemma \ref{lemma:hoeff}.). Equation \eqref{eq:hoeff} shows that $\cov[\ind{\omrkt \le x},\osdf]$ locally measures the dependence between the SDF and market return. In other words, it quantifies how the SDF's variability relates to the market return's variability at different quantiles.\\

To explain the equity premium and disaster risk premia, the SDF must exhibit sufficient variability. Since the distribution bound can be derived from applying the Cauchy-Schwarz inequality to $\cov[\ind{\omrkt \le x},\osdf]$, it is expected to yield sharper bounds on the SDF volatility than the HJ bound if, for example, there is high tail dependence between the SDF and market return such as in the disaster risk model.\footnote{See \citet[Chapter 7.2.4]{mcneil2015quantitative} for a formal definition of tail dependence.}\\



\subsection{Quantile Predictability in the Left-Tail}\label{sec:q_predict}
The bound presented in Proposition \ref{prop:dist_bound} sheds light on the seemingly ``low'' explanatory power observed in the left-tail quantile regressions in Table \ref{tab:only.rn.quantile}. For tractability, it is more convenient to show this for CDFs instead of quantile functions, but the intuition remains the same. Specifically, suppose one could predict $F_t(x)$ at some $x$ in the left-tail, then this prediction can be exploited by going short in an asset that pays $\ind{\omrkt \le x}$. The profit and risk associated to this investment are, respectively
\begin{align}
\texpneut\lr{\ind{\omrkt \le x}} - \cexp{\ind{\omrkt \le x}}, \label{eq:eq_premium} \\
\sigma_t(\ind{\omrkt \le x}) = \sqrt{F_t(x) (1-F_t(x))}. \nonumber
\end{align}

Although such binary state payoffs do not exist in reality, they can be replicated closely by a portfolio of put options. In consequence, high predictability of $F_t(x)$ in the left-tail would render too good a Sharpe ratio; a near-arbitrage opportunity. Following the reasoning in \citet[Chapter 5]{ross2005neoclassical}, a crude upper bound on the SDF volatility imposes limitations on the degree of predictability in the left-tail by the distribution bound in Proposition \ref{prop:dist_bound}. This argument breaks down in the right-tail since \eqref{eq:eq_premium} is roughly zero, and high predictability would not imply counterfactually high SDF volatility.

\subsection{Distribution Bound in Asset Pricing Models}
The estimated Gini coefficients in Section \ref{sec:eq_prem_puzzle} demonstrate that conditional disaster risk is a pervasive feature of the data. This section complements those findings using the unconditional version of the distribution bound in Proposition \ref{prop:dist_bound}:
\begin{equation}\label{eq:dist_bound_unc}
\frac{\sigma(M)}{\uexp{M}} \ge \frac{\tau - \phi(\tau)}{\sqrt{\phi(\tau)(1-\phi(\tau))}},
\end{equation}
where $\sigma(M)$ represents the unconditional SDF volatility, and $\phi(\tau) = F(\tilde{Q}_\tau)$. In this context, $F(\cdot)$ denotes the unconditional physical CDF, and $\tilde{Q}_\tau$ is the unconditional risk-neutral quantile function of the market return. The main benefit of using unconditional distributions is that they can be estimated without estimating the risk-neutral quantile regressions. Moreover, the bound in \eqref{eq:dist_bound_unc} only requires the estimation of distribution functions, whereas existing approaches typically use unconditional density functions to emphasize disaster risk (see, e.g.\ \citet{Beason2022}).\\

The subsequent examples demonstrate that the HJ bound is always stronger than the distribution bound in models that do not embed a source of disaster risk. In contrast, models that incorporate disaster risk can generate distribution bounds that exceed the HJ bound in the left-tail. Since I use unconditional distributions, the time subscripts will be omitted from the notation.

\begin{example}[CAPM]
The Capital Asset Pricing Model (CAPM) specifies the SDF as
\begin{equation*}
	\uncsdf = \alpha - \beta \uncmrkt,
\end{equation*}
where $\uncmrkt$ denotes the return on the market portfolio. In this case $M \notin \aleph^+$, since the SDF can become negative. However, this probability is very small over short time horizons or we can think of $M$ as an approximation to $M^* \coloneqq \max(0,M) \in \aleph^+$. Since the HJ bound is derived by applying the Cauchy-Schwarz inequality to $\unccov(\uncmrkt,\uncsdf)$, the inequality binds if $\uncsdf$ is a linear combination of $\uncmrkt$. Hence, under CAPM, the HJ bound is strictly stronger than the distribution bound regardless of the distribution of $\uncmrkt$. 
\end{example}

\begin{example}[Joint normality]\label{ex:joint_norm}
	Suppose that $\uncsdf$ and $\uncmrkt$ are jointly normally distributed and denote the mean and variance of $\uncmrkt$ by $\mu_R$ and $\sigma_R^2$ respectively. The normality assumption	violates no-arbitrage since $M$ can be negative, but could be defended as an approximation over short time horizons when the variance is small (see Example \ref{ex:joint_logn}). In Appendix \ref{app:joint_norm}, I prove that 
	\begin{equation}\label{eq:stein}
		\abs{\unccov\left(\ind{\uncmrkt \le \tilde{Q}_\tau},\uncsdf \right)} = f_R(\tilde{Q}_\tau) \abs{\unccov(\uncmrkt,\uncsdf)},
	\end{equation}
	where $f_R(\cdot)$ is the marginal density of $\uncmrkt$.\footnote{Notice that this is the marginal density under physical measure $\mathbb{P}$.} This identity gives an explicit expression for the weighting factor in Hoeffding's identity \eqref{eq:hoeff}. In Appendix \ref{app:joint_norm}, I also derive an explicit expression for the relative efficiency between the distribution  and HJ bound 
	\begin{align}\label{eq:rel_eff}
		\frac{\text{HJ bound}}{\text{distribution bound}}  = \frac{\sqrt{\phi(\tau)(1-\phi(\tau))}}{\sigma_R f_R(\tilde{Q}_\tau)}.
	\end{align}
	To see that the HJ bound is always stronger than the distribution bound, minimize \eqref{eq:rel_eff} with respect to $\tau$. Appendix \ref{app:min_qbound_normal} shows that the minimizer $\tau^*$ satisfies $\tilde{Q}_{\tau^*} = \mu_R$. For this choice, $\phi(\tau^*) = \mathbb{P}(\uncmrkt \le \tilde{Q}_{\tau^*}) = 1/2$ and $f_\uncrisky(\tilde{Q}_{\tau^*}) = 1/\sqrt{2 \pi \sigma_{R}^2}$. Therefore, \eqref{eq:rel_eff} can be bounded by
	\begin{equation*}
		\frac{\sqrt{\phi(\tau)(1-\phi(\tau))}}{\sigma_R f(\tilde{Q}_\tau)} \ge \frac{\sqrt{2 \pi}}{2} \approx 1.25.
	\end{equation*}
	Hence, the HJ bound is always stronger in a model where the SDF and return are jointly normal. 
\end{example}

\begin{example}[Joint lognormality]\label{ex:joint_logn}
	Let $Z_R$ and $Z_M$ be standard normal random variables with correlation $\rho$ and consider the specification
	\begin{align*}
		\uncmrkt &= e^{(\mu_R - \frac{\sigma_R^2}{2})\lambda + \sigma_R \sqrt{\lambda}Z_R} \\
		\uncsdf &= e^{-(r_f + \frac{\sigma_M^2}{2})\lambda + \sigma_M \sqrt{\lambda}Z_M},
	\end{align*}
	where $\lambda$ governs the time scale. Simple algebra shows that the no-arbitrage condition, $\uexp{ \uncsdf \uncmrkt} = 1$, is satisfied when $\mu_R - r_f = - \rho \sigma_R \sigma_M$. It is difficult to find an analytical solution for the relative efficiency between the HJ and distribution bound in this case, but linearization leads to a closed form expression which is quite accurate in simulations. The details are described in Appendix \ref{app:lognormal}, where I show that 
	\begin{equation}\label{eq:rel_eff_lognormal}
		\min_{\tau \in (0,1)} \frac{\text{HJ bound}}{\text{distribution bound}}  \approx  \frac{1}{2} \sqrt{\frac{2 \pi \sigma_R^2 \lambda}{\exp(\sigma_R^2 \lambda)-1}}.
	\end{equation}
	This expression is independent of $\mu_R$. An application of l'H\^ospital's rule reveals that the relative efficiency converges to $\sqrt{2 \pi}/2$ if $\lambda \to 0^+$.\footnote{This is the same relative efficiency in Example \ref{ex:joint_norm}, which is unsurprising as the linearization becomes exact in the limit as $\lambda \to 0^+$.} The ratio in \eqref{eq:rel_eff_lognormal} is less than 1 if $\sigma_R \ge 0.92$ and $\lambda = 1$. Since the annualized market return volatility is about 16\%,  the HJ bound is stronger than the distribution bound under any reasonable parameterization if the SDF and market return are lognormal. 
\end{example}

\begin{example}[continues=ex:disaster_risk_jumps]\label{ex:disaster}
	The disaster risk model discussed in Section \ref{sec:notation}  is calibrated according to the results in \citet[Table \rom{2}]{backus2011disasters}. The market return in this model is considered as a levered claim on consumption growth, i.e.\ an asset that pays dividends proportional to $G_{t\to N}^\lambda$. Here $\lambda$ governs the variability of the claim to equity. I convert the model implied volatility bounds to monthly units, to facilitate the comparison with the empirical bounds obtained in Section \ref{sec:empirical}.\\ 
	
	The distribution bound, HJ bound and SDF volatility are depicted in Panels \ref{fig:lrr_local} (without jumps) and \ref{fig:dis_loc}  (with jumps). Consistent with Example \ref{ex:joint_logn}, the distribution bound in the model without jumps never exceeds the HJ bound  because both the market return and SDF follow lognormal distributions. The distribution bound with jumps has a sharp peak at $\tau = 0.037$, after which it steadily decreases. Interestingly, there is a range of $\tau$ values for which the distribution bound is stronger than the HJ bound.\footnote{In Appendix \ref{app:pareto}, I show that the distribution bound can also exceed the HJ bound when returns follow the Pareto distribution.} This result can  be understood from the physical and risk-neutral quantile functions in Figure \ref{fig:q_jump}. The risk-neutral quantile function displays a heavy left-tail, owing to the implied disaster risk embedded in the SDF. Consequently, it is extremely profitable to sell digital put options which pay out in case of a disaster. These put options must have high Sharpe ratios as their prices are high (insurance against disaster risk), but the actual probability of a disaster event occurring is low enough that the risk associated with selling such insurance is limited.

	\begin{figure}[!htb]
		\centering
		\begin{subfigure}[b]{0.49\textwidth}
			\includegraphics[width=\textwidth]{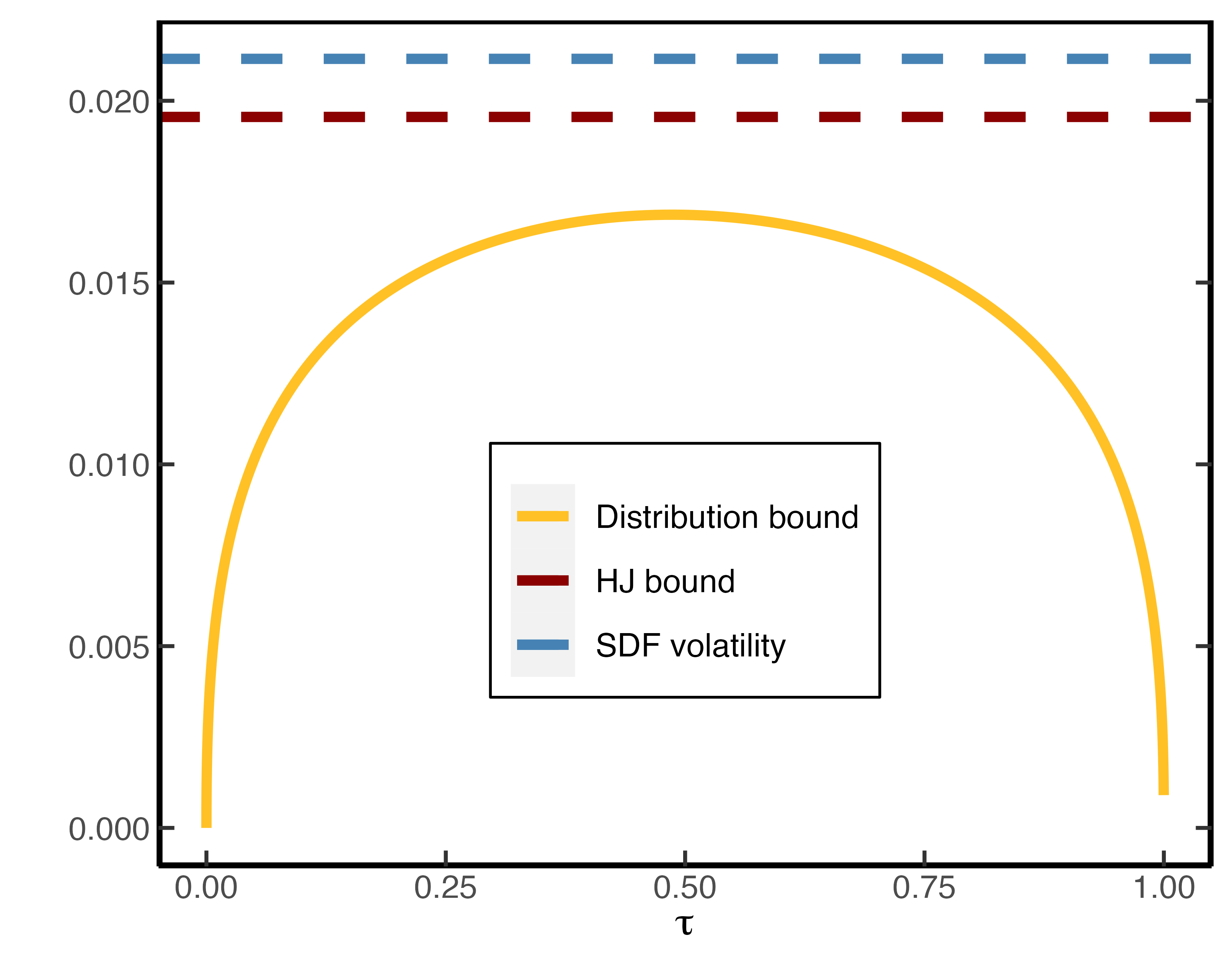}
			\caption{Without jumps}
			\label{fig:lrr_local}
		\end{subfigure}
		\begin{subfigure}[b]{0.49\textwidth}
			\includegraphics[width=\textwidth]{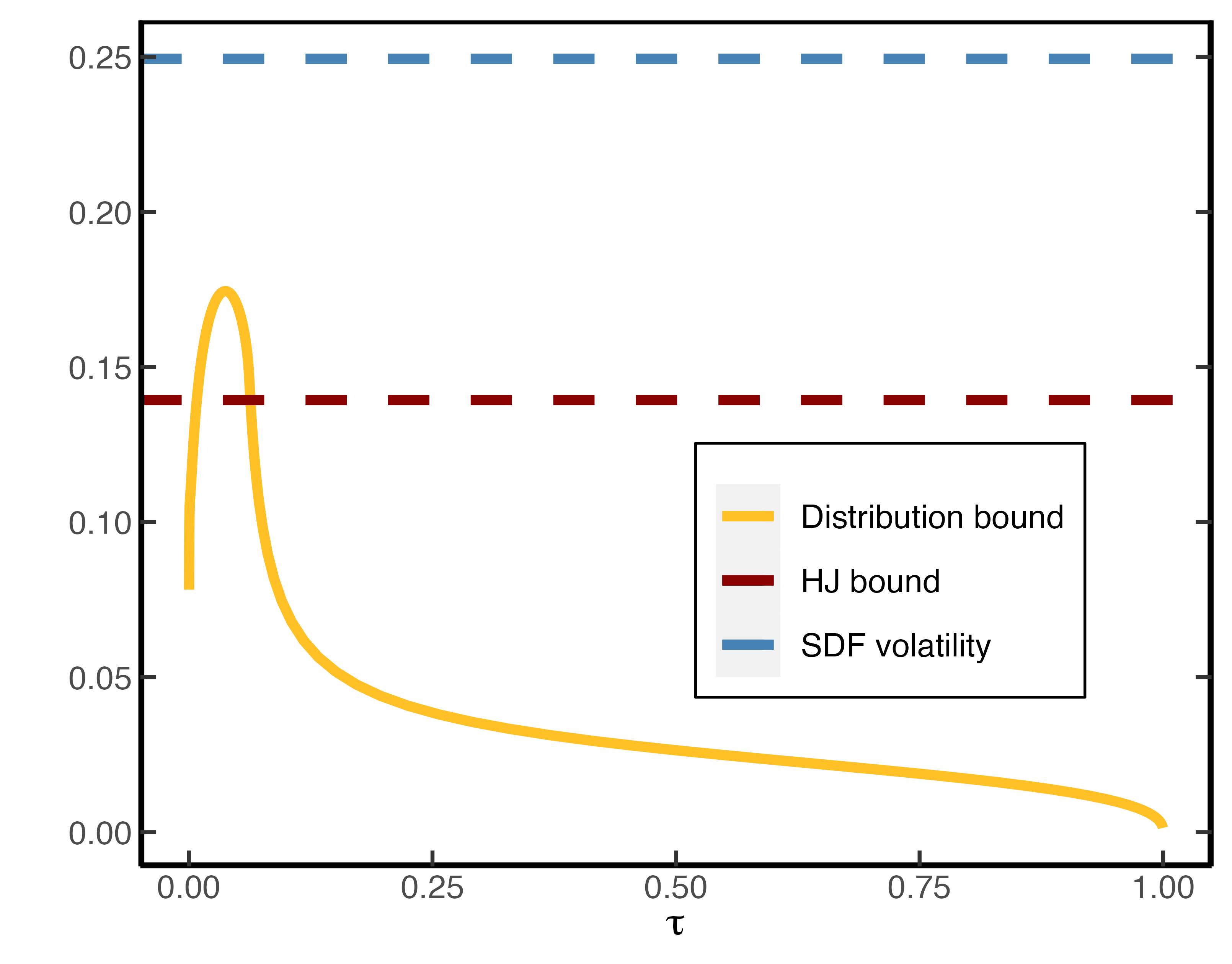}
			\caption{With jumps}
			\label{fig:dis_loc}
		\end{subfigure}
		\caption{\textbf{HJ and distribution bound in disaster risk model without and with jumps.} \footnotesize Panels (a) and (b) show the HJ and distribution bound for the disaster risk model (Example {\protect \ref{ex:disaster_risk_jumps}})   without and with jumps respectively.  The bounds and true SDF volatility are reported in monthly units. Parameters are calibrated according to \citet[Model \rom{2}]{backus2011disasters}.}
		\label{fig:disaster}
	\end{figure}

\end{example}

\subsection{Data and Empirical Estimation of  the Distribution Bound}
To further illustrate the presence of disaster risk in the data, I estimate the distribution bound \eqref{eq:quantile_bound_RiskFree} empirically, using the same 30-day S\&P500 returns as discussed in Section \ref{sec:data}.  However, in this case, I use non-overlapping returns that cover the period 1996--2021.\footnote{I use non-overlapping returns in this section to facilitate testing and to make the results comparable to other nonparametric bounds, which are typically estimated based on non-overlapping returns (see e.g. \citet{liu2020index}).} These returns are sampled at the middle of each month, resulting in a total of 312 observations.  Over this period, the Sharpe ratio is 13\%, and the HJ bound therefore implies that the monthly SDF is quite volatile.\\

The distribution bound consists of three unknowns that need to be estimated: \begin{inparaenum}[(i)]
\item the physical distribution $(F)$; \item the risk-neutral quantile function $(\tilde{Q}_\tau)$, and; \item the risk-free rate ($\uncfree$). \end{inparaenum} 
To estimate the unconditional risk-free rate, denoted by $\uncfreehat$, I rely on the historical average of monthly interest rates. Next, to obtain an estimate of the physical distribution, I employ a kernel (CDF) estimator, given by:
\begin{equation}\label{eq:cdf_unc}
	\func(x) \coloneqq \frac{1}{T} \sum_{t=1}^{T} \Phi\lro{\frac{x-\omrkt}{h}},
\end{equation}
where $\Phi(\cdot)$ is the Epanechnikov kernel and $h$ is the bandwidth determined by cross-validation. This choice of estimator ensures that the distribution bound is a smooth function of $\tau$, which reduces the impact of outliers relative to the discontinuous empirical CDF. \\

Finally, I apply the procedure outlined in Section \ref{sec:data} to estimate $\tilde{F}_t$ (the conditional risk-neutral CDF). Subsequently, I average the conditional distributions to estimate the unconditional CDF:
\begin{equation*}
	\funct(x) \coloneqq \frac{1}{T} \sum_{t=1}^{T} \tilde{F}_t(x).
\end{equation*}
Under appropriate assumptions about the distribution of returns,  $\funct$ converges to $\tilde{F}$ as $T \to \infty$.  An estimate of the unconditional risk-neutral quantile function can then obtained from
\begin{equation}\label{eq:q_unc}
	\qunct(\tau) \coloneqq \inf\left\{x \in \mathbb{R} : \tau \le \funct(x) \right\}.
\end{equation}
Finally, based on the physical CDF \eqref{eq:cdf_unc} and risk-neutral quantile function \eqref{eq:q_unc}, I estimate the distribution bound by
\begin{equation}\label{eq:sup_smooth}
	\quantunc(\tau) \coloneqq \frac{\abs{\tau - \ordunc(\tau)}}{\sqrt{\ordunc(\tau)  (1-\ordunc(\tau))} \uncfreehat}, \qquad   \tau \in [\varepsilon,1-\varepsilon] \subseteq (0,1),
\end{equation}
where $\ordunc(\tau) \coloneqq \func(\qunct(\tau))$ is the estimated ordinal dominance curve  and $\varepsilon$ is a small positive number.

\subsection{Unconditional Evidence of Disaster Risk}\label{sec:empirical}
\begin{figure}[!htb]
	\centering
	\begin{subfigure}[b]{0.49\textwidth}
		\centering
		\includegraphics[width=\textwidth]{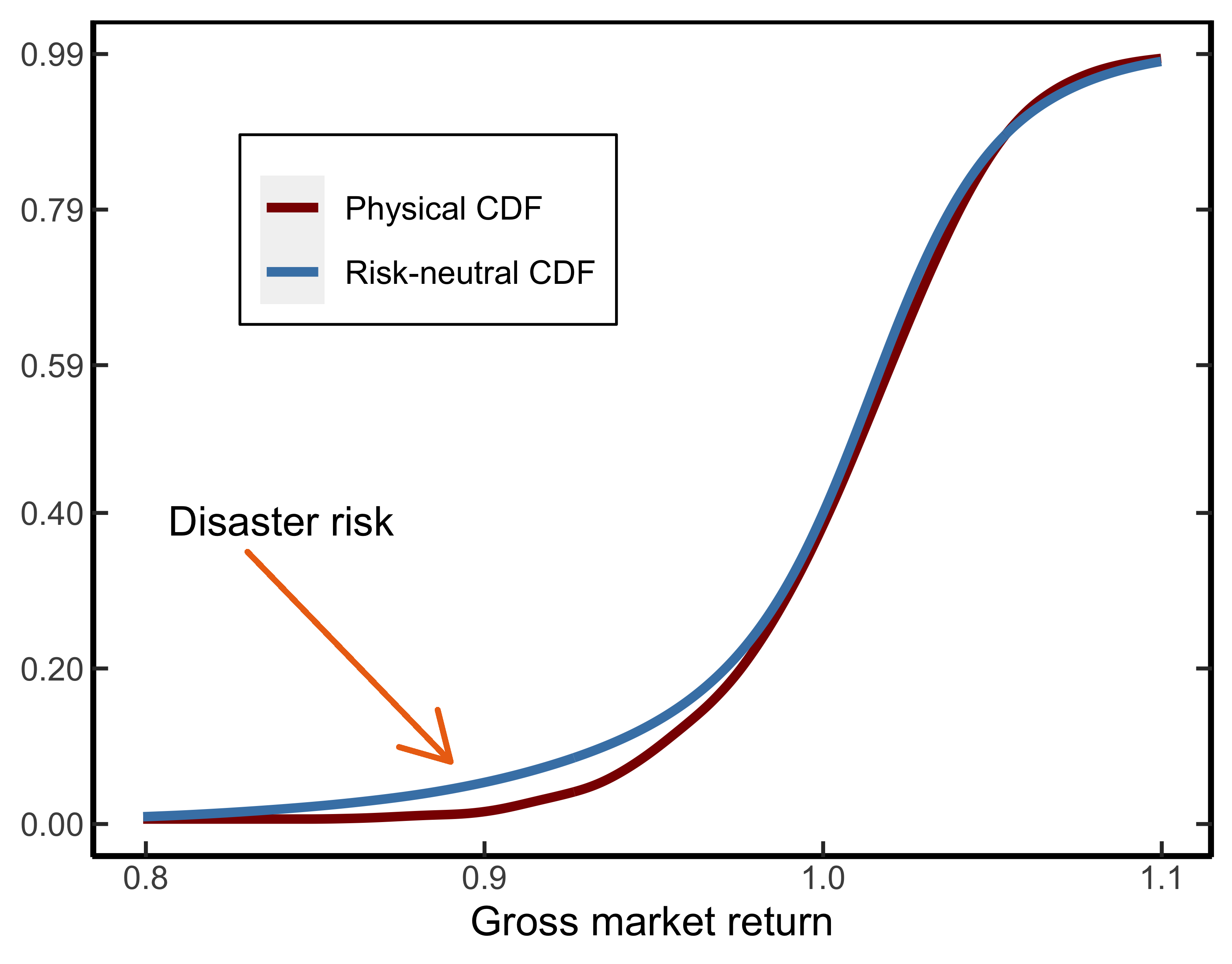}
		\caption{}
		\label{fig:unc_cdf}
	\end{subfigure}
	\begin{subfigure}[b]{0.49\textwidth}
		\centering
		\includegraphics[width=\textwidth]{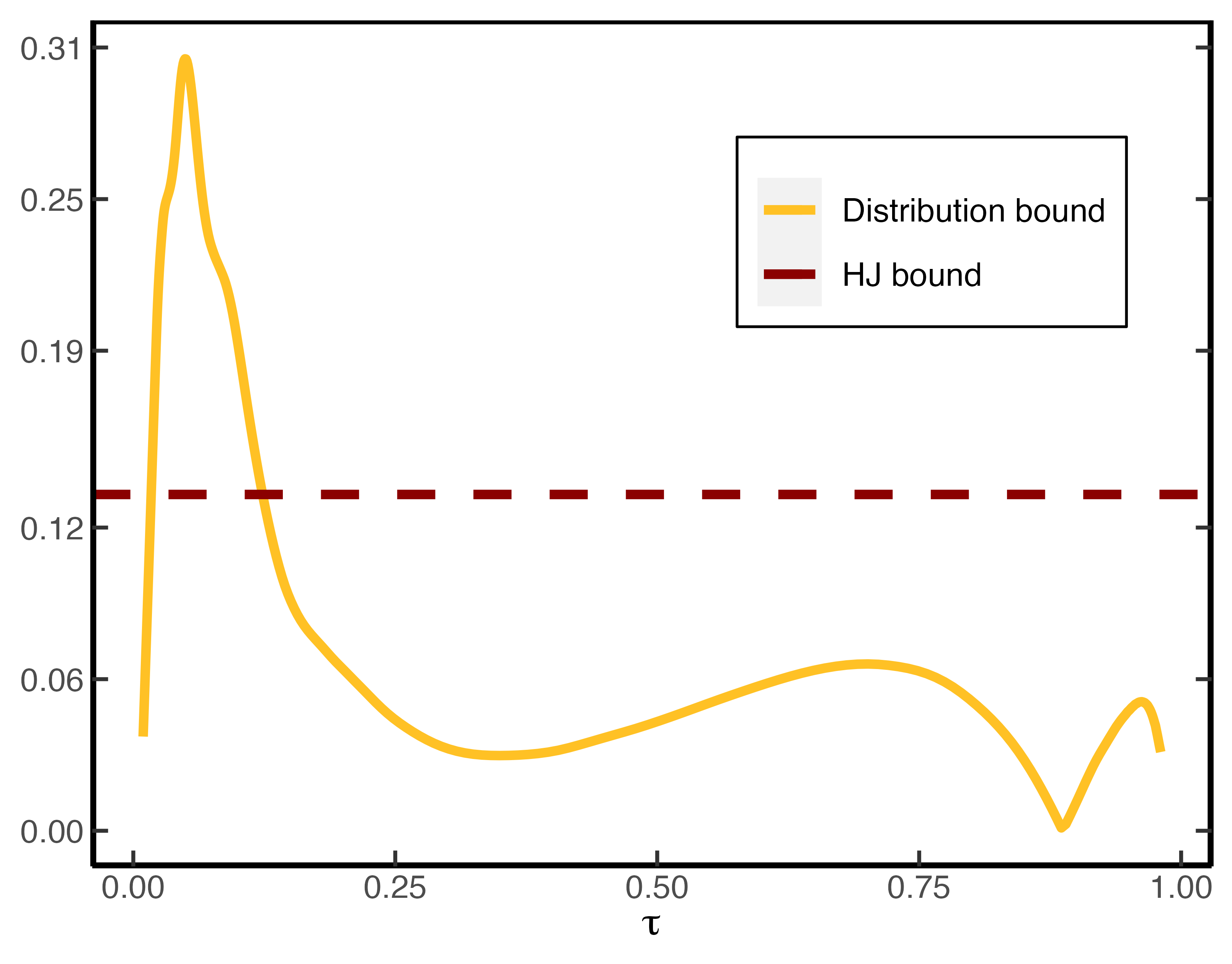}
		\caption{}
		\label{fig:unc_loc}
	\end{subfigure}
	\hfill
	\caption{\textbf{Physical/risk-neutral CDF and distribution bound for monthly S\&P500 returns}. \footnotesize Panel (a) shows the unconditional phyisical and risk-neutral CDF for monthly S\&P500 returns, over the period 1996-2021. Panel (b) shows the distribution bound as function of $\tau$, together with the HJ bound.}
	\label{fig:quant_bound}
\end{figure}

Figure \ref{fig:unc_cdf} illustrates the estimated physical and risk-neutral measures, which differ most in the left-tail. The distribution bound shows that this difference leads to a volatile SDF, which is shown in Figure \ref{fig:unc_loc}. The lower bound on the SDF volatility implied by the distribution bound is much stronger than the HJ bound in the left-tail. This finding aligns with empirical evidence documenting that high Sharpe ratios can be attained by selling out-of-the money put options (see \citet{broadie2009understanding} and the references therein). The supremum of the distribution bound occurs around the 5th percentile, implying that the monthly SDF volatility must exceed 31\%. This value is more than twice the level indicated by the sample HJ bound. Moreover, the shape of the distribution bound is quite similar to the distribution bound implied by the disaster risk model in Figure \ref{fig:dis_loc}.\footnote{The non-monotonicity in the right-tail of the distribution bound occurs because $\tilde{F}(x) > F(x)$, for $x$ large enough. That is, the physical distribution does not first-order stochastically dominates the risk-neutral distribution. This result is consistent with the negative $\widebar{\mathrm{Hit}}$ estimates in Table \ref{tab:only.rn.quantile}. } \\

The graphical evidence suggests that the distribution bound renders a stronger bound on the SDF volatility than the HJ bound. To test this hypothesis more formally, I fix a priori the probability level at 0.037 ($\tau = 0.037$), which renders the sharpest bound on the SDF volatility in the disaster risk model (Example {\protect \ref{ex:disaster}}). At this probability level, the distribution bound is 26\% in the data, which is roughly double the level implied by the HJ bound.\\

To see whether this difference is statistically significant, I consider the following test statistic
\begin{equation}\label{eq:bootstrap_distance}
	\mathcal{T} \coloneqq \quantunc(0.037) - \frac{\abs{\bar{R}_m - \uncfreehat}}{\hat{\sigma} \uncfreehat}.
\end{equation}
The first term on the right denotes the estimated distribution bound \eqref{eq:sup_smooth} evaluated at the 3.7th percentile, using the entire time series of returns $\{\omrkt \}$. The second term denotes the estimated HJ bound, using $\bar{R}_m$ and $\hat{\sigma}$ as the respective sample mean and standard deviation of \{$\omrkt$\}. A value of $\mathcal{T}  > 0$ indicates that the distribution bound is stronger than the HJ bound. To test this restriction,  consider the null and alternative hypothesis:
\begin{align}\label{eq:H0}
	&H_0: \mathcal{T} \le 0 \\
	&H_1: \mathcal{T} > 0. \nonumber
\end{align}
Since the distribution of \eqref{eq:bootstrap_distance} is difficult to characterize, I use stationary bootstrap to approximate the $p$-value under the null hypothesis. The stationary bootstrap is used to generate time indices from which we recreate (with replacement) bootstrapped returns $\{\omrkt^{\star} \}$ \citep{politis1994stationary}. The same bootstrapped time indices are used to re-estimate the physical CDF and risk-neutral quantile function. I repeat the bootstrap exercise 100,000 times and for each bootstrap sample, I calculate the test statistic $\mathcal{T}^\star$. Finally, the empirical $p$-value is obtained as the fraction of times $\mathcal{T}^\star  \le 0$. The last column in Table \ref{tab:boot_pval} shows that the $p$-value is 7.5\%, which provides preliminary evidence that the distribution bound significantly exceeds the HJ bound.\\

\begin{table}[!htb]
	\captionsetup{width=12cm}	
	\centering
	\caption{\textbf{Sample bounds and bootstrap result}}
	\label{tab:boot_pval}
	\begin{adjustbox}{max width=\textwidth}
		\begin{threeparttable}
			\renewcommand{\TPTminimum}{\linewidth}
			\makebox[\linewidth]{%
				\begin{tabular}{@{}cccc@{}}
					\toprule
					\midrule
					Sample size & HJ bound   &distribution bound  & $p$-value \\
					312 & 0.133  & 0.260 & 0.075 \\ \bottomrule
			\end{tabular}}%
			\begin{tablenotes}
				\footnotesize
				\item  \textit{Note}: This table reports the HJ and distribution bound for monthly S\&P500 returns over the period 1996--2021. The distribution bound is evaluated at $\tau = 0.0374$. The final column denotes the $p$-value of the null hypothesis in \eqref{eq:H0}. The $p$-value is obtained from 100,000 bootstrap samples and counts the fraction of times that $\mathcal{T}^\star \le 0$.
			\end{tablenotes}
		\end{threeparttable}
	\end{adjustbox}
\end{table}

\begin{Remark}
When the HJ bound is stronger than the distribution bound, many of the bootstrap samples may not include disaster shocks. Over the entire sample period, there are only two instances where returns were less than -20\%: in September 2008 and February 2020. When considering bootstrap samples that include both of these months, the $p$-value is only 3.6\%. In contrast, the $p$-value increases to 22\%  for bootstrap samples that exclude these months. These findings underscore the sensitivity of the test to the presence of disaster shocks. Overall, the results suggest that, unconditionally, the SDF needs to be highly volatile to be consistent with local differences between the physical and risk-neutral measure in the left-tail. 
\end{Remark}

\section{A Model-Free Lower Bound on  Disaster Risk Premia}\label{sec:dark}
The previous findings indicate that the risk-neutral quantile function is not a good approximation of the physical quantile function in the left-tail. In this section, I derive a lower bound on disaster risk premia observed from option prices.  This lower bound does not require parameter estimation and relaxes the assumption of a time-homogeneous linear relation between the physical and risk-neutral quantiles in \eqref{eq:tail}.\\

\subsection{Approximating the Quantile Difference}
To analyze the difference between $\cpquant$ and $\cquant$, I use some elementary tools from functional analysis. The quantile function can be regarded as a map $\varphi$ between normed spaces, taking as input a distribution function and returning the quantile function: $\varphi(F_t) = F_t^{-1} = Q_{t,\tau}$. Expanding $\varphi$ around the observed risk-neutral CDF yields
\begin{equation}\label{eq:vonMises1}
	\cpquant - \cquant   = \varphi(F_t) -  \varphi(\tilde{F}_t) =  \varphi_{\tilde{F}_t}'(F_t-\tilde{F}_t) + o\lro{\norm{F_t - \tilde{F}_t}},
\end{equation}
where $\norm{\cdot}$ is a norm on a suitable linear space\footnote{Formally, the space can be defined as $\{\Delta: \Delta = c(F - G),F,G \in \mathbb{D}, c \in \mathbb{R}\}$ and $\mathbb{D}$ is the space of distribution functions \citep{serfling2009approximation}. See \citet[Section 20.1]{van2000asymptotic} and \citet[p. 217]{serfling2009approximation} for further details about the approximation.} and $\varphi_{\tilde{F}_t}'(F_t-\tilde{F}_t)$ is the G\^{a}teaux derivative of $\varphi$ at $\tilde{F}_t$ in the direction of $F_t$:
\begin{align}\label{eq:Gateax}
	\varphi_{\tilde{F}_t}'(F_t-\tilde{F}_t) &\coloneqq \lim_{\lambda \downarrow 0} \frac{\varphi \lr{(1-\lambda)\tilde{F}_t + \lambda F_t}}{\lambda} \nonumber\\ 
	&= \frac{\partial}{\partial \lambda} \varphi\lro{(1-\lambda) \tilde{F}_t + \lambda F} \bigg|_{\lambda = 0}.
\end{align}

Heuristically, the G\^{a}teaux derivative can be thought of as measuring the change in the quantile function when  the risk-neutral distribution is moved in the direction of the physical distribution. Appendix \ref{app:mises} shows that the G\^{a}teaux derivative is given by
\begin{equation}\label{eq:vonMises2}
	\varphi_{\tilde{F}_t}'(F_t-\tilde{F}_t)  = \frac{\tau - F_t(\cquant) }{\tilde{f}_t(\cquant)} = \frac{\tau - \phi_t(\tau)}{\tilde{f}_t(\cquant)},
\end{equation}
where $\phi_t(\tau) = F_t(\cquant)$ denotes the conditional ordinal dominance curve (ODC). I proceed under the working hypothesis that the remainder term in \eqref{eq:vonMises1} is ``small'' in the sup-norm, $\norm{g}_{\infty} = \sup_x\abs{g(x)}$.
\begin{assumption}\label{ass:small}
	The remainder term in \eqref{eq:vonMises1} can be neglected.  
\end{assumption}
\begin{Remark}
	The assumption implies that the first order approximation in \eqref{eq:vonMises1} is accurate. The condition that $||F_t - \tilde{F}_t||_{\infty}$ is small can be understood as excluding near-arbitrage opportunities, since the distribution bound in Proposition \ref{prop:dist_bound} shows that substantial pointwise differences between $F_t(\cdot)$ and $\tilde{F}_t(\cdot)$ lead to a very volatile SDF.  Appendix \ref{app:bs_sim_evidence} illustrates the approximation in the \citet{black1973pricing} model.
\end{Remark}

I combine \eqref{eq:vonMises1} and \eqref{eq:vonMises2} in conjunction with Assumption \ref{ass:small} to obtain the approximation
\begin{equation}\label{eq:q_risk_adjust}
	\cpquant -  \cquant \approx  + \underbrace{\frac{\tau - F_t(\cquant) }{\tilde{f}_t(\cquant)}}_{\text{risk-adjustment}}.
\end{equation}
The second term on the right can be thought of as a risk-adjustment term to capture the unobserved wedge between $\cpquant$ and $\cquant$. The approximation in \eqref{eq:q_risk_adjust} contains the terms $\cquant$ and $\tilde{f}_t(\cquant)$, which are directly observed at time $t$ using  the \citet{breeden1978prices} formula in \eqref{eq:breeden}. However, $F_t(\cdot)$ is unknown and hence \eqref{eq:q_risk_adjust} cannot be used directly to approximate $\cpquant$.\\

\subsection{A Lower Bound on Disaster Risk Premia}
To make further progress, I show that the numerator term, $\tau - F_t(\cquant)$, can be lower bounded with option data under economically motivated constraints. This bound, combined with the approximation in  \eqref{eq:q_risk_adjust}, will then imply a lower bound on disaster risk premia.\\

I start from the observation that the SDF in representative agent models  can be expressed as a function of the market return \citep{chabi2020conditional}:
\begin{equation}\label{eq:sdf_chabi}
	\frac{\cexp{\osdf}}{\osdf} = \frac{\frac{u'(W_t x_0)}{u'(W_t x)}}{\texpneut\lr{\frac{u'(W_t x_0)}{u'(W_t x)}}}  \qquad \text{with } x = \omrkt \ \text{and }   x_0  = \ofree,
\end{equation}
where $W_t$ is the agent's wealth at time $t$ and $u(x)$ represents the agent's utility function. Define
\begin{equation}\label{eq:notation}
	\zeta(x) \coloneqq \frac{u'(W_t \ofree)}{u'(W_t x)} \quad  \text{and } \quad  \theta_k = \frac{1}{k!} \left(\frac{\partial^k \zeta(x)}{\partial x^k}\right)_{x=\ofree}.
\end{equation} 
Notice that $\zeta(\cdot)$ is simply the inverse of the intertemporal marginal rate of substitution (IMRS) and $\theta_k$ are the coefficients of the its Taylor expansion  around $\ofree$. I make the following assumptions about the market return and the IMRS of the representative agent. 
\begin{assumption}\label{ass:f}
	In the representative agent model, it holds that
	\begin{inparaenum}[(i)]
		\item $\qcexp{\omrkt^3} < \infty$; and  \label{item:mean3}	
		\item 	$\zeta^{(4)}(x) \le 0$. \label{item:2}  
	\end{inparaenum}
\end{assumption}
Assumption \ref{ass:f}(\ref{item:mean3}) allows for fat tails in the risk-neutral distribution as long as the third moment exists. This assumption relaxes the implicit assumption made by \citet{chabi2020conditional} that infinitely many moments exist. Figure  \ref{fig:lee_bound} in the Appendix illustrates that the risk-neutral distribution frequently exhibits a finite number of moments, some of which may not exceed 4, particularly in turbulent market conditions. \citet{chabi2020conditional} present sufficient conditions for \ref{ass:f}(\ref{item:2}) to hold, which relate to the sign of the fifth derivative of the utility function of the representative agent. Specifically, for common utility functions such as CRRA or HARA utility, parameter restrictions are needed to ensure that  \ref{ass:f}(\ref{item:2}) holds.\footnote{For example, for CRRA utility, the risk-aversion coefficient cannot be too large. See Appendix \ref{app:verify} for a detailed discussion.}\\

I need one additional assumption to bound disaster risk premia. To state this assumption and the resulting lower bound, I use the following notation for high-order risk-neutral moments and truncated high-order risk-neutral moments, respectively.
\begin{align}\label{eq:trunc_mrkt}
\chabi{n} & \coloneqq \texpneut\lr{ \left(\omrkt - \ofree\right)^n} \nonumber\\
\trunc{n}{k_0} &\coloneqq \texpneut\lr{\ind{\omrkt \le k_0}  (\omrkt - \ofree)^n  }.
\end{align}

\begin{assumption}\label{ass:skewness}
In the representative agent model, the following holds:
\begin{enumerate}[(i)]
	\item $(-1)^{k-1} \theta_k \ge \frac{1}{\ofree^k}$ for $k=1,2,3$  \label{item:chabi1} 
	\item $\chabi{3} \le 0$. \label{item:chabi2}
\end{enumerate}
\end{assumption}

\citet[Table 6]{chabi2020conditional} provide empirical evidence that  \ref{ass:skewness}(\ref{item:chabi1}) holds with equality when estimating the conditional equity premium. Assumption  \ref{ass:skewness}(\ref{item:chabi2}) is a very mild restriction on risk-neutral skewness, which is almost always negative at every date and time horizon. This empirical fact is well known.\footnote{\citet{chabi2020conditional} argue that all odd risk-neutral moments should be negative, since they expose the investor to unfavorable market conditions.} \\

The following two propositions show how option data can be employed to establish bounds on the difference between the physical and risk-neutral measures in the left-tail.

\begin{proposition}[Lower Bound on CDF]\label{prop:lower_bound}
	Suppose Assumptions \ref{ass:f} and  \ref{ass:skewness} hold, and assume that the risk-neutral density exists. Then, 
	\begin{equation}\label{eq:lower_bound}
		\tau - F_t\lro{\cquant} \ge \frac{\sum_{k=1}^3  \frac{(-1)^{k-1}}{\ofree^k} \left( \tau \chabi{k} -\trunc{k}{\cquant} \right)    }{1+\sum_{k=1}^{3} \frac{(-1)^{k-1}}{\ofree^k} \chabi{k}}     \eqqcolon \lrb,
	\end{equation}
	for all $\tau \le \tau'$, where $\tau'$ is defined implicitly by
	\begin{equation*}
		\tilde{Q}_{t,\tau'} = \min\left(\ofree - \sqrt{\vartilde(\omrkt)},\cquantstar\right),
	\end{equation*}
	and $\cquantstar$ is defined in Theorem \ref{thm:lower_bound}.
\end{proposition}
\begin{proof}
	See Appendix \ref{app:proof_q_lrb}.
\end{proof}

\begin{proposition}[Lower Bound on Disaster Risk Premia]\label{prop:quantile}
	Consider the same assumptions in Proposition \ref{prop:lower_bound} and assume additionally that Assumption \ref{ass:small} holds. Then, for all $\tau \le \tau'$
	\begin{equation}\label{eq:q_first_order}
		\cpquant - \cquant \ge \overbrace{\frac{\lrb}{\tilde{f}_t(\cquant)}}^{\text{\emph{risk-adjustment}}} \eqqcolon \lrbdp.
	\end{equation}
\end{proposition}

\begin{proof}
By Assumption  \ref{ass:small}, the approximation in \eqref{eq:q_risk_adjust} holds, which in combination with Proposition \ref{prop:lower_bound} renders
\begin{align*}
\cpquant - \cquant &\overset{\eqref{eq:q_risk_adjust}}{\approx} \frac{\tau - F_t(\cquant)}{\tilde{f}_t(\cquant)} \\
&\overset{\eqref{eq:lower_bound}}{\ge } \frac{1}{\tilde{f}_t(\cquant)} \left(   \frac{\sum_{k=1}^3  \frac{(-1)^{k+1}}{\ofree^k} \left( \tau \chabi{k} -\trunc{k}{\cquant} \right)    }{1+\sum_{k=1}^{3} \frac{(-1)^{k+1}}{\ofree^k} \chabi{k}}     \right). \qedhere
\end{align*}
\end{proof}

Proposition \ref{prop:lower_bound} provides a bound on the physical CDF that requires no parameter estimation and relies solely on time $t$ information. This result complements recent work on belief recovery. \citet{ross2015recovery} demonstrated CDF recovery under the assumption of transition independence, but subsequent research has questioned this assumption \citep{borovivcka2016misspecified, qin2018long, jackwerth2020does}. In contrast, Proposition \ref{prop:lower_bound} establishes a lower bound on the left-tail of the physical distribution using a different set of mild economic constraints. Additionally, Section \ref{sec:data} showed that the right-tail of $F_t$ can be approximately recovered from the risk-neutral distribution due to the minimal need for risk-adjustment. These findings suggest the potential for approximate recovery of $F_t$ using option prices.\\


I will test this hypothesis using the lower bound on disaster risk premia in Proposition \ref{prop:quantile}. Specifically, a tight lower bound in \eqref{eq:q_first_order}  would enable direct inference on both the physical distribution ($\cpquant$) and disaster risk premia ($\cpquant - \cquant$). While Section \ref{sec:rn_qr_reg} proposed the quantile model \eqref{eq:tail} to estimate $\cpquant$, it can be criticized for having time-homogeneous coefficients. Proposition \ref{prop:quantile} relaxes that assumption. Furthermore, the lower bound in \eqref{eq:q_first_order} is not prone to the historical sample bias critique of \citet{welch2008comprehensive}.   Alternatively, one can estimate a disaster risk model to infer $\cpquant$, but this approach is also susceptible to misspecification concerns and faces challenges in estimation due to the scarcity of disaster events in the data \citep{Julliard2012,martin2013consumption}.\\

\subsection{Calculating the Lower Bound}
Before assessing how tight the lower bound is in Proposition  \ref{prop:quantile}, I  outline the procedure to calculate it, which depends on $\lrb$ and $\tilde{f}_t(\cquant)$. Both functions can be derived from $\cquant$, which is estimated using the same data and procedure of Section \ref{sec:data}. To see that $\tilde{f}_t(\cquant)$ can be derived from $\cquant$, notice that $\frac{d}{d \tau} \tilde{Q}_t(\tau) = 1/\tilde{f}_t(\cquant)$. The latter term can thus be approximated by\footnote{I slightly abuse notation to emphasize that the derivative is taken w.r.t.\ $\tau$, so that $\tilde{Q}_t(\tau+h)$ denotes $\tilde{Q}_{t,\tau+h}$.}
\begin{equation*}
	\frac{1}{\tilde{f}_t(\cquant)} \approx \frac{\tilde{Q}_t(\tau +h) - \tilde{Q}_t(\tau-h)}{2h},
\end{equation*}
where $h$ is the bandwidth of the $\tau$-grid. Second, to calculate $\lrb$ in \eqref{eq:lower_bound}, I use $\cquant$, as well as the formula for high-order risk-neutral moments in Appendix \ref{app:chabi}.\\

Given the evidence in Table \ref{tab:only.rn.quantile} that $\cpquant > \cquant$ in the left-tail, Proposition \ref{prop:quantile} has nontrivial content in the data if $\lrbdp \ge 0$. Appendix Table \ref{tab:summary_risk_adjustment} contains summary statistics of $\lrbdp$, which show that the lower bound is always positive, right-skewed, more pronounced in the right-tail and economically meaningful in magnitude, with outliers that can spike up to 29\%.\\

\subsection{Tightness of the Lower Bound: In-sample Evidence}\label{sec:q_test}
To test whether the lower bound in Proposition \ref{prop:quantile} is tight, I form \emph{excess quantile returns}: $\omrkt - \cquant$. Since $\cquant$ is observed at time $t$, it follows that  $Q_{t,\tau}(\omrkt - \cquant) = Q_{t,\tau}(\omrkt) - \cquant$.  Subsequently, I use QR to estimate the model
\begin{align}\label{eq:exc_quant}
	Q_{t,\tau}(\omrkt) - \cquant(\omrkt)  = \beta_0(\tau) + \beta_1(\tau) \lrbdp, \nonumber \\
	[\hat{\beta}_0(\tau),\hat{\beta}_1(\tau)] = \argmin_{(\beta_0,\beta_1)  \in \mathbb{R}^2} \sum_{t=1}^{T} \rho_\tau(\omrkt -\cquant - \beta_0 - \beta_1 \lrbdp). 
\end{align}
Regression \eqref{eq:exc_quant} is a quantile analogue of the mean excess return regressions of \citet{welch2008comprehensive}. Under the null hypothesis that the lower bound is tight, it holds that
\begin{equation}\label{eq:H0_tight_lrb}
	H_0: \quad [\beta_0(\tau), \beta_1(\tau)] = [0,1].
\end{equation}
Less restrictive, one can test whether $\beta_0(\tau) = 0$ and $\beta_1(\tau) > 0$, which implies that the statistical ``factor'' $\lrbdp$ explains the conditional quantile wedge.\footnote{For example, if we start with a quantile factor model $\cpquant = \cquant + \beta(\tau) \lrbdp$, the model has one testable implication for the data: the intercept in a quantile regression of $\omrkt - \cquant$ on $\lrbdp$ should be zero. Quantile factor models have recently been proposed by \citet{chen2021quantile}.} \\

Table \ref{tab:robustness_lrb} presents the results of regression \eqref{eq:exc_quant}. The null hypothesis of a tight lower bound in  \eqref{eq:H0_tight_lrb}  is not rejected for $\tau = 0.2$, but it is rejected for $\tau \in \{0.05,0.1\}$ across all horizons. When the null hypothesis is rejected, the $\beta_1(\tau)$-coefficient exceeds 1, consistent with the theory that $\lrbdp$ represents a lower bound on disaster risk premia.  In all cases, the lower bound is economically meaningful, since $\beta_1(\tau)$ is significantly different from 0, while $\beta_0(\tau) = 0$ can never be rejected. However, the explanatory power of the regression in Table  \ref{tab:robustness_lrb} is modest, as shown by the $R^1(\tau)$ measure-of-fit:
\begin{equation}\label{eq:R1_RA}
	R^1(\tau) = 1 - \frac{\min_{b_0,b_1} \sum \rho_\tau(\omrkt - b_0 - b_1 \lrbdp)  }{\min_{b_0} \sum \rho_\tau(\omrkt - b_0)}.
\end{equation}
But, following the reasoning of Section \ref{sec:q_predict}, the predictive power in the left-tail cannot be too big, for otherwise near-arbitrage opportunities exist.\\

\begin{table}[!htb]
\captionsetup{width=12cm}
\centering
\caption{\textbf{Quantile regression with lower bound}}
\label{tab:robustness_lrb}
\begin{adjustbox}{max width=\textwidth}
\begin{threeparttable}
\begin{tabular}{@{}lcccccc@{}}
\toprule
\midrule
Horizon &  $\tau $ & $\hat{\beta}_0(\tau)$ & $\hat{\beta}_1(\tau)$ & $\underset{(p\text{-value})}{\text{Wald test}}$ & $R^1(\tau)[\%]$ & Obs \\ 
\cmidrule(lr){1-7}
&  &  &  &  &  &  \\
\underline{30 days} & 0.05 & $\underset{( 0.005 )}{\text{ -0.01 }}$ & $\underset{( 0.349 )}{ \text{ 4.43 }}$ & 0.00 & 6.03 & 4333 \\
& 0.1 & $\underset{( 0.006 )}{\text{ -0.01 }}$ & $\underset{( 0.450 )}{ \text{ 2.17 }}$ & 0.03 & 3.18 &  \\
& 0.2 & $\underset{( 0.006 )}{ \text{ -0.01 }}$ & $\underset{( 0.400 )}{\text{ 1.33 }}$ & 0.02 & 0.41 &  \\
&  &  &  &  &  &  \\
\underline{60 days} & 0.05 & $\underset{( 0.013 )}{\text{ -0.01 }}$ & $\underset{( 0.571 )}{ \text{ 5.53 }}$ & 0.00 & 3.60 & 4312 \\
& 0.1 & $\underset{( 0.011 )}{ \text{ -0.02 }}$ & $\underset{( 0.540 )}{ \text{    3.25 }}$ & 0.00 & 2.23 &  \\
& 0.2 & $\underset{( 0.009 )}{\text{ -0.02 }}$ & $\underset{( 0.398 )}{\text{  1.50   }}$ & 0.27 & 0.48 &  \\
&  &  &  &  &  &  \\
\underline{90 days} & 0.05 & $\underset{( 0.032 )}{\text{ -0.02 }}$ & $\underset{( 1.113 )}{ \text{ 6.37 }}$ & 0.00 & 4.91 & 4291 \\
& 0.1 & $\underset{( 0.018 )}{\text{ -0.02 }}$ & $\underset{( 0.528 )}{ \text{ 3.05 }}$ & 0.00 & 4.43 &  \\
& 0.2 & $\underset{( 0.019 )}{\text{ -0.02 }}$ & $\underset{( 0.626 )}{\text{  1.36   }}$ & 0.69 & 1.46 &  \\ \bottomrule
\end{tabular}%
\begin{tablenotes}
\footnotesize
\item \textit{Note}: This table reports the QR estimates of \eqref{eq:exc_quant} over the sample period 2003-2021 at different horizons, using overlapping returns. Standard errors are shown in parentheses and calculated using SETBB with a block length equal to the prediction horizon. \emph{Wald test} denotes the $p$-value of the joint restriction $[\beta_0(\tau),\beta_1(\tau)] =[0,1]$. $R^1(\tau)$ denotes the goodness-of-fit measure \eqref{eq:R1_RA}.
\end{tablenotes}
\end{threeparttable}
\end{adjustbox}
\end{table}

I also directly test the predictive power of the lower bound in estimating the physical quantile function. To this end, I use the following model-free quantile forecast:
\begin{equation}\label{eq:predict_quantile}
	\widehat{Q}_{t,\tau} \coloneqq  \cquant + \lrbdp.
\end{equation}
To evaluate the accuracy of this forecast, I use QR to estimate the model
\begin{equation}\label{eq:qr_qhat}
	Q_{t,\tau}(\omrkt) = \beta_0(\tau) + \beta_1(\tau) \widehat{Q}_{t,\tau}.
\end{equation}
An accurate forecast would imply the joint restriction 
\begin{equation}\label{eq:H0Quantile}
	H_0 : \  \beta_0(\tau) = 0, \quad \beta_1(\tau) = 1.
\end{equation}

Table \ref{tab:quantile__regression} summarizes the estimates of \eqref{eq:qr_qhat} for several percentiles. The results compare favorably  to the risk-neutral estimates in Table \ref{tab:only.rn.quantile}. First, the point estimates are closer to the $[0,1]$ benchmark. Second, the Wald test on the joint restriction in \eqref{eq:H0Quantile} is never rejected except for $\tau = 0.05$ at the 60-day horizon. Third, the in-sample explanatory power is higher. The same conclusion applies when comparing the predictive results to the expanding quantile regression from Table \ref{tab:rolling_window}. Collectively, these findings suggest that  $\widehat{Q}_{t,\tau}$ can be considered as a good lower bound on  the physical quantile function in the left-tail.\\


\begin{table}[!htb]
\captionsetup{width=12cm}	
\centering
\caption{\textbf{Quantile regression with model-free quantile forecast}}
\label{tab:quantile__regression}
\begin{adjustbox}{max width=\textwidth}
\begin{threeparttable}
\begin{tabular}{lccccccl}
\toprule
\midrule
Horizon & $\tau$ & $\hat{\beta}_0(\tau)$ & $\hat{\beta}_1(\tau)$ & $\underset{(p\text{-value})}{\text{Wald test}}$ & $R^1(\tau)[\%]$ & $R_{oos}^1(\tau)[\%]$ & Obs \\ 
\midrule
30 days & 0.05 & $\underset{( 0.249 )}{\text{ 0.29 }}$ & $\underset{( 0.265 )}{\text{ 0.70 }}$ & 0.06 & 6.28 & 9.94 & 4333 \\
& 0.1 & $\underset{( 0.250 )}{\text{ 0.28 }}$ & $\underset{( 0.260 )}{\text{ 0.72 }}$ & 0.18 & 3.57 & 4.02 &  \\
& 0.2 & $\underset{( 0.381 )}{\text{ 0.57 }}$ & $\underset{( 0.388 )}{\text{ 0.43 }}$ & 0.29 & 0.58 & 2.53 &  \\
&  &  &  &  &  &  &  \\
60 days & 0.05 & $\underset{( 0.382 )}{\text{ 0.30 }}$ & $\underset{( 0.426 )}{\text{ 0.71 }}$ & 0.02 & 3.40 & 17.81 & 4312 \\
& 0.1 & $\underset{( 0.352 )}{\text{ 0.38 }}$ & $\underset{( 0.373 )}{\text{ 0.61 }}$ & 0.13 & 2.35 & 9.22 &  \\
& 0.2 & $\underset{( 0.487 )}{\text{ 0.44 }}$ & $\underset{( 0.498 )}{\text{ 0.56 }}$ & 0.21 & 0.57 & 4.28 &  \\
&  &  &  &  &  &  &  \\
90 days & 0.05 & $\underset{( 0.520 )}{\text{ 0.36 }}$ & $\underset{( 0.602 )}{\text{ 0.64 }}$ & 0.05 & 4.26 & 21.98 & 4291 \\
& 0.1 & $\underset{( 0.482 )}{\text{ 0.31 }}$ & $\underset{( 0.521 )}{\text{ 0.70 }}$ & 0.06 & 4.19 & 13.22 &  \\
& 0.2 & $\underset{( 0.696 )}{\text{ 0.23 }}$ & $\underset{( 0.710 )}{\text{ 0.78 }}$ & 0.48 & 0.70 & 5.99 &  \\ \bottomrule
\end{tabular}%
\begin{tablenotes}
\footnotesize
\item \textit{Note}:  This table reports the QR estimates of \eqref{eq:qr_qhat} over the sample period 2003-2021. Standard errors are shown in parentheses and calculated using the SETBB, with block length equal to the prediction horizon. \emph{Wald test} gives the $p$-value of the Wald test on the joint restriction: $\hat{\beta}_0(\tau) = 0, \hat{\beta}_1(\tau) = 1$. $R^1(\tau)$ denotes the in-sample goodness-of fit criterion \eqref{eq:R1tau}. $R_{oos}^1(\tau)$ is the out-of-sample goodness-of fit, using a rolling window size equal to 10 times the return horizon.
\end{tablenotes}
\end{threeparttable}
\end{adjustbox}
\end{table}

\subsection{Tightness of the Lower Bound: Out-of-sample Evidence}
Given that the in-sample results from Table \ref{tab:quantile__regression} suggest that $\widehat{Q}_{t,\tau}$ is a good lower bound for $\cpquant$, it is natural to assess its out-of-sample performance by using $\widehat{Q}_{t,\tau}$ to directly predict $\cpquant$, which does not require any parameter estimation.\\

To assess the out-of-sample performance, I use the $R_{oos}^1(\tau)$ measure of fit defined in \eqref{eq:Roos_rn} with $\widehat{Q}_{t,\tau}$ instead of $\cquant$.  Table \ref{tab:quantile__regression} shows that $\widehat{Q}_{t,\tau}$ improves upon the historical rolling quantile out-of-sample in all cases. In particular, this outperformance is most pronounced at the 5th percentile, which is expected since option data are known to provide useful information about extreme downfalls in the stock market \citep{bates2008market,bollerslev2011tails}. In Appendix \ref{app:robust_lb}, I run a battery of robustness tests which show that, out-of-sample, $\lrbdp$ better predicts the conditional quantile function than other benchmarks such as the risk-neutral quantile or the VIX index. The latter result is particularly encouraging since the VIX predictor uses in-sample information.


\subsection{Robustness of the QR Estimates}
The in- and out-of-sample results support $\lrbdp$ as a robust lower bound for disaster risk premia. It is instructive to compare this lower bound to the disaster risk premia reported in Figure \ref{fig:dp}, which are inferred from the quantile regression in \eqref{eq:qr_est}. Consistent with the theory, the estimated disaster risk premium at $\tau = 0.05$ exceeds the lower bound in 99\% of cases for 30-day returns and 99.9\% for 60-day returns. When violations of the lower bound occur, the differences are typically small.\\

\begin{figure}[!htb]
	\centering
	\begin{subfigure}[b]{0.49\textwidth}
		\centering
		\includegraphics[width=\textwidth]{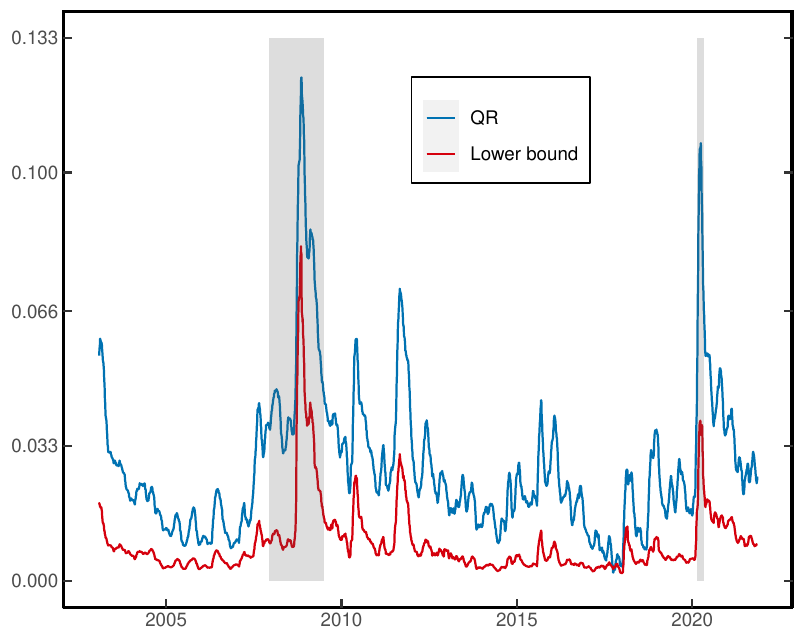}
		\caption{30-day returns}
		\label{fig:q_diff30}
	\end{subfigure}
	\begin{subfigure}[b]{0.49\textwidth}
		\centering
		\includegraphics[width=\textwidth]{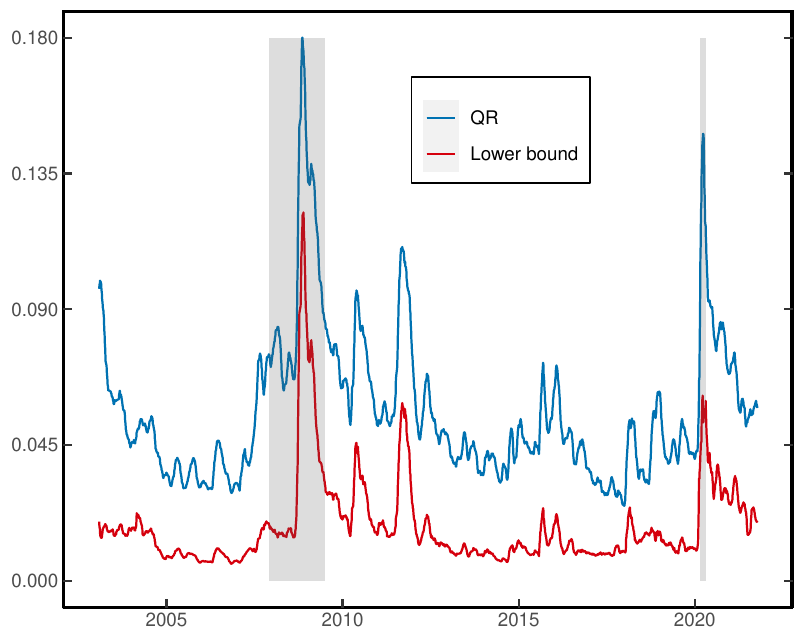}
		\caption{60-day returns}
		\label{fig:q_diff60}
	\end{subfigure}
	\caption{\textbf{Lower bound on disaster risk premium at 5th percentile}. \footnotesize Panel (a) shows the lower bound on the disaster risk premium for 30-day returns, at $\tau = 0.05$. QR denotes the estimated disaster risk premium from the quantile regression \eqref{eq:qr_est}. The right panel shows a similar graph for 60-day returns. Both figures are smoothed using a 30-day rolling window. The two shaded bars signify the Great Recession period (Dec 2007 -- June 2009) and Covid-19 crisis (Feb 2020 -- April 2020).}
\end{figure}

Figures \ref{fig:q_diff30} and \ref{fig:q_diff60} show the lower bound for 30- and 60-day returns, respectively, alongside the disaster risk premium estimated from the quantile regression \eqref{eq:qr_est}. In both cases, there is a substantial correlation between the lower bound and the disaster risk premium obtained from the QR estimates. Especially during the global financial crisis and the Covid-19 crisis, both methods predict significant increases in the disaster risk premium. Outside these crisis periods, the lower bound is more conservative. Overall, the model-free lower bound corroborates the robustness of the estimated disaster risk premium in Figure \ref{fig:dp}.\\

\section{Conclusion}\label{sec:conclusion}
I use return and option data on the S\&P500 in combination with quantile regression to estimate local differences between the conditional risk-neutral and physical quantile functions. Empirically, these differences are substantial in the left-tail, whereas in the right-tail, they are barely discernible. Therefore,  the lion's share of the equity premium is driven by downside returns, which is model-free evidence for disaster risk.\\

By tracking these quantile differences over time, the results also demonstrate that disaster risk is time-varying, pervasive, and a driving force behind much of the equity premium, even outside crisis periods. Additionally, my findings show that disaster risk is more nuanced than previous literature suggests. Much of the disagreement can be attributed to the incorporation of conditioning information.  While prior research primarily focused on unconditional estimation, my approach accounts for  conditioning information embedded in the risk-neutral quantile function, which is crucial to obtain accurate estimates of disaster risk.\\

To build on this finding, I show that disaster risk makes the SDF highly volatile. In particular, option strategies involving a short position in an asset that pays one dollar in case of a disaster exhibit substantially higher Sharpe ratios compared to a direct investment in the market portfolio. The data reveal that such investment strategies yield a monthly Sharpe ratio of 30\%, more than doubling the Sharpe ratio of the market return.\\

Finally, I suggest a model-free lower bound on disaster risk premia observed from option prices.  This lower bound serves as a good predictor of the quantile wedge, exhibiting spikes during crises and significant fluctuations over time. Furthermore, the lower bound closely aligns with estimates of disaster risk premia based on quantile regression, thereby reinforcing the robustness of my findings.

\bibliographystyle{plainnat}
\markboth{}{}
\bibliography{refer}

\newpage
\appendix

\setcounter{table}{0}
\renewcommand{\thetable}{\Alph{section}\arabic{table}}

\setcounter{figure}{0}
\renewcommand{\thefigure}{\Alph{section}\arabic{figure}}

\section{Proofs}
This section contains proofs and detailed calculations of results used in the main paper.

\subsection{Decomposing the Equity Premium}\label{subsec:tail_risk}
For any atomless integrable random variable $X$ with CDF $F(\cdot)$ and quantile function $Q = F^{-1}$, we have
\begin{equation*}
\mathbb{E}(X) = \int_{\mathbb{R}} x \diff F(x)  = \int_0^1  Q(\tau) \diff \tau.
\end{equation*}
The second identity holds by the change of variables formula for the Lebesgue-Stieltjes integral. In case $F$ has a density, the formula follows from a simple substitution $x \to Q(\tau)$. Hence,
\begin{equation*}
\cexp{\omrkt} - \ofree =  \cexp{\omrkt} - \texptilde{\omrkt} = \int_{0}^{1} \lro{ \cpquant - \cquant } \diff \tau.
\end{equation*}

\subsection{Stochastic Dominance and Pricing Kernel Monotonicity}\label{app:stoch_dom}
In this section I provide more details on the relation between stochastic dominance and pricing kernel monotonicity. To begin with, recall that the physical distribution is first-order stochastic dominant (FOSD) over the risk-neutral distribution if and only if $F_t(x) \le \tilde{F}_t(x)$. The latter definition is equivalent to $F_t(\cquant) \le \tau$ for all $ \tau \in (0,1)$, which follows from the substitution $x \to \cquant$. \\

\begin{figure}[!htb]
	\centering
	\includegraphics[width=0.7\textwidth]{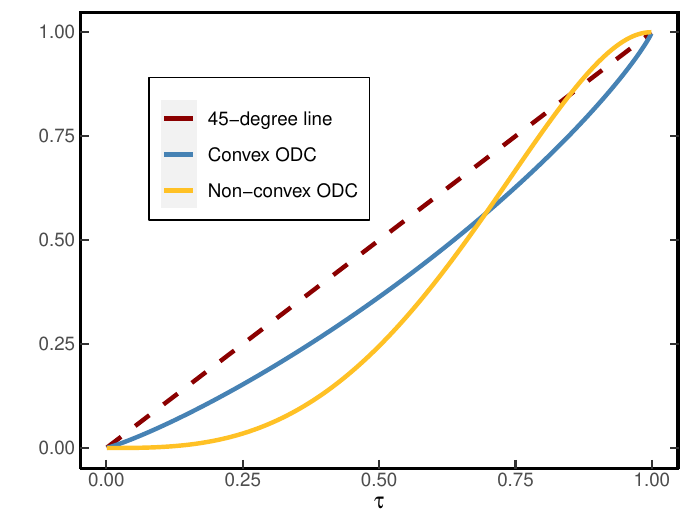}
	\caption{\textbf{Ordinal dominance curve with and without first-order stochastic dominance.} \footnotesize This figure shows two different ordinal dominance curves. The blue ODC corresponds to a situation where the physical measure FOSD the risk-neutral measure, whereas the yellow line shows a situation where FOSD fails. }
	\label{fig:convex_odc}
\end{figure}

To see the connection with pricing kernel monotonicity, we recall from \citet{beare2016empirical} that pricing kernel monotonicity is equivalent to $\phi_t(\tau) \coloneqq F_t(\cquant)$ being a convex function for all $\tau$.\footnote{ \citet{beare2016empirical} actually consider  the reverse function $\phi_t(\tau) = \tilde{F}_t(\cpquant)$, so that pricing kernel monotonicity is equivalent to $\phi_t(\cdot)$ being concave.} Figure \ref{fig:convex_odc} shows two different ODCs; the blue line corresponds to a situation where FOSD holds and the pricing kernel is monotonic (hence convex), whereas the yellow line shows a scenario where FOSD does not hold and  convexity automatically fails. The geometric argument for why non-monotonicity is implied by a failure of FOSD is conveyed by the figure: if FOSD fails, the yellow line must cross the 45-degree line for some $\tau \in (0,1)$, which automatically implies that the ODC is non-convex since the ODC has to satisfy $\phi_t(1) = 1$, because the physical and risk-neutral measures are equivalent. The proposition below thus follows. 

\begin{proposition}\label{prop:monFOSD}
If the pricing kernel is a monotonically decreasing function of the market return,  the physical measure first-order stochastically dominates the risk-neutral measure. Conversely, a violation of FOSD implies a violation of pricing kernel monotonicity. 
\end{proposition}

A violation of FOSD is puzzling from the viewpoint of expected utility maximization. In this framework, the SDF is given by $u'(\omrkt)/\mathbb{E}_t(u'(\omrkt))$, where $u(\cdot)$ is a utility function and the initial endowment is normalized to one for simplicity. The following proposition shows that a sufficient (but not necessary) condition for FOSD to hold is that $u'(\cdot)$ is non-increasing; a rather ubiquitous assumption in asset pricing models.

\begin{proposition}\label{prop:sufFOSD}
In the expected utility framework, a sufficient condition for the physical measure to first-order stochastically dominate the risk-neutral measure is that $u'(\cdot)$ is non-increasing.
\end{proposition}

\begin{proof}
Using the SDF to change from physical to risk-neutral measure, it follows that FOSD is equivalent to
\begin{align*}
	F_t(x)  &\le \tilde{F}_t(x)  \\
	\iff  \cexp{\ind{\omrkt \le x}} &\le \cexp{\frac{u'(\omrkt)}{\cexp{u'(\omrkt)}} \ind{\omrkt \le x} } \\
	\iff 0 &\le \cov(\ind{\omrkt \le x}, u'(\omrkt) ).
\end{align*}
By Lemma \ref{lemma:cheby},  the covariance above is nonnegative if $u'(\cdot)$ is non-increasing.
\end{proof}

\subsection{Proof of Proposition \ref{lemma:bs}}\label{app:proof_logn}
I separately show (\ref{item:logn1})  and (\ref{item:logn2}) of Proposition \ref{lemma:bs}. To prove these results, I use the following lemma.

\begin{lemma}\label{lemma:logn_quant}
	In the lognormal model, the physical and risk-neutral quantile functions conditional on $\mu_t,\sigma_t$ are given by, respectively
	\begin{align}
		\cpquant &= \exp\lr{(\mu_t - \frac{1}{2}\sigma_t^2)N + \sigma_t \sqrt{N} \Phi^{-1} (\tau)}  \label{eq:logn_p}\\
		\cquant &= \exp\lr{(r_f - \frac{1}{2}\sigma_t^2)N + \sigma_t \sqrt{N} \Phi^{-1} (\tau)} \label{eq:logn_r},
	\end{align}
	where $\Phi^{-1}(\cdot)$ denotes the quantile function of the standard normal distribution. If $\mu_t \sim \mathcal{N}(\mu,\sigma_\mu^2)$ and independent from $\sigma_t$,  the physical quantile function conditional on $\sigma_t$, but not $\mu_t$, equals
	\begin{equation}\label{eq:logn_p_s}
		\cpquant(\sigma_t,\sigma_\mu) = \exp\lr{(\mu - \frac{1}{2}\sigma_t^2)N + \lro{\sqrt{\sigma_{\mu}^2 N^2 + \sigma_t^2 N}} \Phi^{-1} (\tau)}.
	\end{equation}
\end{lemma}
\begin{proof}
	The quantile function of a random variable $X$ such that $\log X \sim \mathcal{N}(a,b^2)$, is given by $\exp(a + b \Phi^{-1}(\tau))$. Therefore, the quantile functions conditional on $\mu_t,\sigma_t$ in \eqref{eq:logn_p} and \eqref{eq:logn_r} follow immediately from the conditional lognormal assumption. In \eqref{eq:logn_p_s}, the function is conditioned on $\sigma_t$, but not $\mu_t$. Since $\mu_t$ is assumed to be normally distributed and independent from $\sigma_t$, it follows that 
	\begin{equation*}
		(\mu_t - \frac{1}{2} \sigma_t^2)N + \sigma_t \sqrt{N} Z_{t+N} | \sigma_t \sim \mathcal{N}\lro{(\mu - \frac{1}{2}\sigma_t^2)N, \sigma_{\mu}^2 N^2 + \sigma_t^2 N}.
	\end{equation*}
	The expression in \eqref{eq:logn_p_s} can now be obtained again using the general formula of the lognormal quantile function. 
\end{proof}

\begin{proof}[Proof of Proposition \ref{lemma:bs}(\ref{item:logn1})]
	Recall that $\sqrt{a^2 + b^2} \le \sqrt{a^2 + b^2 + 2ab} = a+b$, provided $a,b \ge 0$. This inequality shows that
	\begin{align*}
		&\exp\lr{\lro{\sqrt{\sigma_\mu^2 N^2 + \sigma_t^2 N} - \sigma_t \sqrt{N}} \Phi^{-1}(\tau)   } \\
		&\le  \exp\lr{\lro{\sqrt{\sigma_\mu^2 N^2 + \sigma_t^2 N} - \sigma_t \sqrt{N}} \abs{\Phi^{-1}(\tau)}   }\\
		&\le  \exp\lro{ \sigma_\mu N  \abs{\Phi^{-1}(\tau) }  }\\
		&= 1+\bigoh{\sigma_\mu N},
	\end{align*}
	uniformly in $\tau \in \mathcal{I}$ and the support of $\sigma_t$. In combination with Lemma \ref{lemma:logn_quant}, it follows that
	\begin{align*}
		\cpquant(\sigma_t,\sigma_\mu) &= \cquant e^{(\mu - r_f)N}  \exp\lr{\lro{\sqrt{\sigma_\mu^2 N^2 + \sigma_t^2 N} - \sigma_t \sqrt{N}} \Phi^{-1}(\tau)   } \\
		&= \cquant e^{(\mu - r_f)N}  \lro{1 + \bigoh{\sigma_\mu N}}. \qedhere
	\end{align*}
\end{proof}

In order to prove Proposition \ref{lemma:bs}(\ref{item:logn2}), I need additional regularity conditions stated in Assumption \ref{ass:sig_small} below. The following notation for the quantile empirical process will be used:
\begin{align*}
\LTtau(\beta,\sigma_\mu) &\coloneqq \frac{1}{T} \sum_{t=1}^{T} \rho_\tau(\omrkt - \beta_0 - \beta_1 \cquant)\\
\Ltau(\beta,\sigma_\mu) &\coloneqq \lim_{T \to \infty} \frac{1}{T} \sum_{t=1}^{T}  \uexp{\rho_\tau(\omrkt - \beta_0 - \beta_1 \cquant)}.
\end{align*}

\begin{assumption}\label{ass:sig_small}
	In the lognormal model, assume additionally that
	 \begin{enumerate}[(i)]
		\item $\uexp{\omrkt}$ and $\uexp{\cquant}$ are finite, \label{item:Flogn}
		\item \label{item:IDlogn} $\Ltau(\beta,0)$ has an identifiably unique minimum $\beta^*$ at $\sigma_\mu = 0$, i.e., for all $\varepsilon>0$
		\begin{equation*}
	    \inf_{\norm{\beta - \beta^*} > \varepsilon} \Ltau(\beta,0) - \Ltau(\beta^*,0) > 0.
		\end{equation*}
		 \item as $T \to \infty$, for any compact set $\mathcal{B}$ and sequence $b_T \searrow 0$, 
		 \begin{subequations}
		\begin{align}
			&\sup_{\beta \in \mathcal{B}} \norm{\Ltau(\beta,\sigma_\mu^T) - \Ltau(\beta,0)} = o(1) \  \text{(Uniform continuity)}. \label{eq:Clogn} \\
		\sup_{\sigma_\mu \le b_T} &\sup_{\beta \in \mathcal{B}} \norm{\LTtau(\beta,\sigma_\mu) - \Ltau(\beta,\sigma_{\mu})} = o_p(1) \ \text{(Uniform LLN)}.  \label{eq:Ulogn} \
		\end{align}
	\end{subequations}
	\end{enumerate} 
\end{assumption}
\begin{proof}[Proof of Proposition \ref{lemma:bs}(\ref{item:logn2})]
	Consider the population minimization problem of quantile regression at $\sigma_\mu = 0$
	\begin{equation}\label{eq:unique}
		[\beta_0^*(0 ;\tau),\beta_1^*(0 ; \tau)]  \coloneqq \argmin_{(\beta_0,\beta_1) \in \mathbb{R}^2} \Ltau(\beta,0).
	\end{equation}
	Assumptions  \ref{ass:sig_small}(\ref{item:Flogn},\ref{item:IDlogn})  ensure that the objective function is well defined and  the solution in \eqref{eq:unique} is unique for all $\tau \in \mathcal{I}$. At $\sigma_\mu = 0$, $\cpquant = e^{(\mu-r)N}$, so  that $[\beta_0^*(0 ;\tau),\beta_1^*(0 ; \tau)] = [0,e^{(\mu-r)N}]$. To ease notation in the following derivation, I write $\hat{\beta}(\sigma_{\mu}^T) \coloneqq \argmin_{\beta} \LTtau(\beta,\sigma_\mu^T)$ and $\beta^*(0) = \argmin_\beta \Ltau(\beta,0)$. It then follows that for every $\varepsilon >0$ there exists a $\delta > 0$ such that
	\begin{align*}
		&\Pr\lro{\norm{\hat{\beta}(\sigma_{\mu}^T) - \beta^*(0)} > \varepsilon }\\
		&\le \Pr\lro{ \Ltau(\hat{\beta}(\sigma_{\mu}^T),0) - \Ltau(\beta^*(0),0)  > \delta } \\
		&= \Pr\lro{ \Ltau(\hat{\beta}(\sigma_{\mu}^T),0) -  \LTtau(\hat{\beta}(\sigma_{\mu}^T),\sigma_\mu^T) +\LTtau(\hat{\beta}(\sigma_{\mu}^T),\sigma_\mu^T)    - \Ltau(\beta^*(0),0)  > \delta } \\
		&\le \Pr\lro{  \Ltau(\hat{\beta}(\sigma_\mu^T),0) -  \LTtau(\hat{\beta}(\sigma_\mu^T),\sigma_\mu^T) +\LTtau(\beta^*(0),\sigma_\mu^T)    - \Ltau(\beta^*(0),0)  > \delta }\\
		&\le \Pr\lro{ 2  \sup_{\beta \in \mathcal{B}}  \norm{ \Ltau(\beta,0) -  \LTtau(\beta,\sigma_\mu^T)} > \delta }.
	\end{align*}
The second line follows from identification and the second to last line from the minimization property of $\hat{\beta}(\sigma_\mu^T)$. Therefore, it suffices to show that 
\begin{equation*}
 \sup_{\beta \in \mathcal{B}}  \norm{ \Ltau(\beta,0) -  \LTtau(\beta,\sigma_\mu^T)} = o_p(1).
\end{equation*}
This claim follows from
\begin{align*}
 &\sup_{\beta \in \mathcal{B}} \norm{ \Ltau(\beta,0) -  \LTtau(\beta,\sigma_\mu^T)} \\
 \le  &\sup_{\beta \in \mathcal{B}}  \norm{ \Ltau(\beta,0)  -  \Ltau(\beta,\sigma_\mu^T)} + \norm{ \Ltau(\beta,\sigma_{\mu}^T) -  \LTtau(\beta,\sigma_\mu^T)}\\
 \le  &\sup_{\beta \in \mathcal{B}}  \norm{ \Ltau(\beta,0)  -  \Ltau(\beta,\sigma_\mu^T)} + \sup_{\sigma_\mu \le b_T}  \sup_{\beta \in \mathcal{B}}  \norm{ \Ltau(\beta,\sigma_{\mu}) -  \LTtau(\beta,\sigma_\mu)}.
\end{align*}
The first term is $o(1)$ by \eqref{eq:Clogn} and the second term is $o_p(1)$ by \eqref{eq:Ulogn}, which completes the proof. The claim in \eqref{eq:q_forecast} easily follows from \eqref{eq:claim}.

\end{proof}

\subsection{Proof of Proposition \ref{prop:dist_bound}}\label{app:proof_qbound}
\begin{proof}
 Starting from the definition of a risk-neutral quantile, it follows that
	\begin{align}\label{eq:quantile}
	\tau &= \tilde{\mathbb{P}}_t\lr{\omrkt \le \cquant} \nonumber\\
	 &= \texpneut \lr{\ind{\omrkt \le  \cquant}} \nonumber \\
	 &= \frac{1}{\cexp{\osdf}} \cexp{\osdf \ind{\omrkt \le \cquant}} \nonumber\\
	&= \frac{1}{\cexp{\osdf}} \lro{\cov\left(\osdf,\ind{\omrkt \le \cquant}\right) +  \cexp{\osdf} \cexp{\ind{\omrkt \le \cquant}} } \nonumber\\
	&=\frac{1}{\cexp{\osdf}} \cov\left(\osdf,\ind{\omrkt \le \cquant}\right) + \underbrace{\cexp{\ind{\omrkt \le \cquant}}}_{= \phi_t(\tau)}.
	\end{align}
	Rearranging then yields
	\begin{equation*}
 \frac{1}{\cexp{\osdf}} \cov\left(\osdf, \ind{\omrkt \le \cquant}   \right) = 	\tau - \phi_t(\tau) .
	\end{equation*}
	Using Cauchy-Schwarz renders the inequality
	\begin{align}\label{eq:q_bnd}
	\frac{1}{\cexp{\osdf}} \sigma_t(\osdf) \sigma_t\left( \ind{\omrkt \le \cquant}\right) & \ge \abs{\tau - \phi_t(\tau)}  \nonumber\\
	 \frac{\sigma_t(\osdf)}{\cexp{\osdf}} &\ge \frac{\abs{\tau - \phi_t(\tau)}}{\sigma_t\left( \ind{\omrkt \le \cquant}\right) } .
	\end{align}
	Finally, since $\ind{\omrkt \le \cquant}$ is a Bernoulli random variable, it follows that 
	\begin{equation}\label{eq:b2}
	\sigma_t\left( \ind{\omrkt \le \cquant }\right)  = \sqrt{\phi_t(\tau) (1- \phi_t(\tau))}.
	\end{equation}
	Proposition \ref{prop:dist_bound} now follows after substituting \eqref{eq:b2} into \eqref{eq:q_bnd}. The bound formulated in terms of the CDFs in \eqref{eq:CDF_rep} follows from the substitution $\cquant \to x$. 
\end{proof}

\subsection{Distribution Bound when SDF and Return are Jointly Normal}\label{app:joint_norm}
In this Section I derive \eqref{eq:stein} and  \eqref{eq:rel_eff} , when $M$ and $\uncmrkt$ are jointly normal. First consider \eqref{eq:rel_eff}. The proof of the distribution bound in Proposition \ref{prop:dist_bound} gives the following identity
\begin{equation*}
	\frac{\abs{\tau - \phi(\tau)}}{\uncfree} = \abs{\unccov\left(\ind{\uncmrkt \le \tilde{Q}_\tau},\uncsdf \right)}.
\end{equation*}
Standard SDF properties also yield the well known result
\begin{equation*}
	\frac{\abs{\mathbb{E}(\uncmrkt) - \uncfree}}{\uncfree} = \abs{\unccov\lro{\uncmrkt,\uncsdf}}.
\end{equation*}
These results, combined with \eqref{eq:stein} prove \eqref{eq:rel_eff}, since
\begin{align*}
	\frac{\text{HJ bound}}{\text{distribution bound}}  &=  \frac{\frac{\abs{\uexp{\uncmrkt}-\uncfree}}{\sigma_\uncrisky \uncfree}}{\frac{\abs{\tau - \phi(\tau)}}{\sqrt{\phi(\tau)(1-\phi(\tau))}\uncfree}   } \nonumber\\
	&\overset{\eqref{eq:stein}}{=} \frac{\sqrt{\phi(\tau)(1-\phi(\tau))}}{\sigma_\uncrisky f_\uncrisky(\tilde{Q}_\tau)},
\end{align*}
where $f_\uncrisky(\tilde{Q}_\tau)$ is the marginal density of $\uncmrkt$.\\

Finally, I make use of the following covariance identities to prove \eqref{eq:stein}.
\begin{lemma}[Hoeffding]\label{lemma:hoeff}
	For any square integrable random variable $X$ and $Z$ with marginal CDFs $F_X, F_Z$ and joint CDF $F_{X,Z}$,	it holds that 
	\begin{align}
		\unccov\lr{\ind{Z \le z},X} &= - \int_{-\infty}^\infty \lr{F_{X,Z}(x,z) - F_X(x) F_Z(z)}\diff x \label{eq:hof}\\
		\unccov\lr{Z,X} &= - \int_{-\infty}^\infty \unccov\lr{\ind{Z \le z},X}\diff z. \label{eq:avg}
	\end{align}
\end{lemma}
\begin{proof}
	See \citet{lehmann1966some}.
\end{proof}
I also need a relation for the bivariate normal distribution. Suppose that $X,Z$ are jointly normal with correlation $\rho$, mean $\mu_X, \mu_Z$ and variance $\sigma_X^2, \sigma_Z^2$, then 
\begin{equation}\label{eq:sungur}
	\pde{\Phi_2(x,z;\rho,\mu_X,\mu_Z,\sigma_X^2,\sigma_Z^2)}{\rho} = \sigma_X \sigma_Z \phi_2(x,z;\rho,\mu_X,\mu_Z,\sigma_X^2,\sigma_Z^2),
\end{equation} 
where $\Phi_2(\cdot)$ denotes the bivariate normal CDF and $\phi_2(\cdot)$ denotes the bivariate normal PDF \citep{sungur1990dependence}. We can now prove a covariance identity for jointly normal random variables.
\begin{proposition}\label{prop:hoeff}
	Suppose $\uncmrkt$ and $M$ are jointly normal with correlation $\rho$, then
	\begin{equation}\label{eq:loc_dep}
		-\unccov\lr{\ind{\uncmrkt \le x},M} =  \unccov\lr{\uncmrkt,M}    \phi_R(x) ,
	\end{equation}
	where $\phi_R(\cdot)$ is the marginal density of $\uncmrkt$.
\end{proposition}
\begin{proof}
	To lighten notation, I suppress the dependence on $\mu_R,\mu_M,\sigma_R^2,\sigma_M^2$ in the joint CDF and PDF. We then have
	\begin{align*}
		-\unccov\lr{\ind{\uncmrkt \le x},M} &= \int_{-\infty}^\infty \Phi_2(x,m;\rho) - \Phi_2(x,m;0) \diff m\\
		&= \int_{-\infty}^\infty \int_0^\rho \sigma_R \sigma_M \phi_2(x,m;y) \diff y \diff m\\
		&= \sigma_R \sigma_M \rho \phi_R(x)\\
		&= \unccov\lr{\uncmrkt,M} \phi_R(x),
	\end{align*}
	where, in the first line, I use \eqref{eq:hof} together with $F_R(r) F_M(m) = \Phi_2(r,m;0)$, the second line follows from \eqref{eq:sungur} and the third line follows from Fubini's theorem to swap the order of integration and $\int_{-\infty}^\infty \phi_2(x,m;y)\diff m = \phi_R(x)$.
\end{proof}
\begin{Remark}
	The second covariance identity in \eqref{eq:avg} shows that $\unccov\lr{\ind{\uncmrkt\le x},M}$ is a measure of local dependence. In case of joint normality \eqref{eq:loc_dep}, the weight is given by the marginal PDF. For other distributions, the weighting factor is more complicated, but sometimes can be given an explicit form using a local Gaussian representation (see \citet{chernozhukov2018distribution}).
\end{Remark}

\subsection{Minimum Percentile of Distribution Bound with Normal SDF }\label{app:min_qbound_normal}
This section shows that the relative efficiency between the HJ bound and distribution bound is minimized when $\tilde{Q}_\tau = \mu_R$. To see this, write  $x = \tilde{Q}_\tau$, and use  $F(\cdot)$ to denote the physical CDF of $\uncmrkt$. I also drop the $R$ subscript for $f$ to avoid notational clutter. Consider
	\begin{equation*}
		\Gamma(x) = \frac{F(x)(1 - F(x))}{f(x)^2}.
	\end{equation*}
	Minimizing $\Gamma(x)$ is equivalent to minimizing \eqref{eq:rel_eff} and first order conditions imply that the optimal $x^*$ satisfies
	\begin{equation}\label{eq:aux}
		[f(x^*)-2F(x^*)f(x^*)]f(x^*)^2 - 2f(x^*)f'(x^*)[F(x^*)(1-F(x^*))]=0.
	\end{equation}
	Since $f,F$ are the respective PDF and CDF of the normal random variable $\uncmrkt$, it follows that $f'(\mu_R) = 0$ and $F(\mu_R) = 1/2$. As a result, \eqref{eq:aux} holds when $\tilde{Q}_{\tau^*} = x^* = \mu_R$.

\subsection{Distribution Bound when SDF and Return are Log-normal}\label{app:lognormal}
This section provides a closed form approximation for the relative efficiency between the HJ and distribution bound under joint lognormality. The result depends on Stein's Lemma \citep[Lemma 3.6.5]{casella2002statistical}:\footnote{I use the form of Stein's Lemma reported in \citet[p. 163]{cochrane2009asset}, which follows from Stein's lemma as reported in \citet{casella2002statistical}.}
\begin{lemma}[Stein's Lemma]
	If $X_1,X_2$ are bivariate normal, $g: \mathbb{R} \to \mathbb{R}$ is differentiable and $\mathbb{E}\abs{g'(X_1)} < \infty$, then
	\begin{equation*}
		\unccov\lro{g(X_1),X_2} = \mathbb{E}\lr{g'(X_1)} \unccov(X_1,X_2).
	\end{equation*}
\end{lemma}

To prove the approximation, we approximate $M$ by a first order Taylor expansion, which gives
\begin{equation*}
	\hat{M} = e^{-(r_f + \frac{\sigma_M^2}{2})\lambda} + Z_M \sigma_M \sqrt{\lambda} e^{-(r_f + \frac{\sigma_M^2}{2}) \lambda}.
\end{equation*}
Notice that $\hat{M} = M + o_p(\sqrt{\lambda})$.  Consequently, by Stein's Lemma
\begin{align*}
	&\unccov(\uncmrkt,M) \approx \unccov(\uncmrkt, \hat{M} ) =  \sigma_M \sqrt{\lambda} e^{-(r_f + \frac{\sigma_M^2}{2})\lambda}\unccov(\uncmrkt,Z_M)\\
	&= \sigma_M \sqrt{\lambda} e^{-(r_f + \frac{\sigma_M^2}{2}) \lambda} \uexp{\sigma_R \sqrt{\lambda} \exp\lro{\lr{\mu_R - \frac{\sigma_R^2}{2}}\lambda + \sigma_R \sqrt{\lambda} Z_R}} \unccov(Z_R,Z_M)\\
	&= \sigma_M \sigma_R \lambda e^{-(r_f + \frac{\sigma_M^2}{2}) \lambda} e^{\mu_R \lambda } \unccov(Z_R,Z_M).
\end{align*}
By Proposition \ref{prop:hoeff},
\begin{align*}
	&\unccov(\ind{\log \uncmrkt \le x}, M) \approx \unccov\left(\ind{\log \uncmrkt \le x}, \hat{M} \right)\\
	&= \sigma_M \sqrt{\lambda} e^{-(r_f + \frac{\sigma_M^2}{2})\lambda} \unccov\lro{\ind{\log \uncmrkt \le x}, Z_M}\\
	&= \sigma_M \sqrt{\lambda} e^{-(r_f + \frac{\sigma_M^2}{2})\lambda} \unccov\lro{\ind{(\mu_R - \sigma_R^2/2)\lambda + \sigma_R \sqrt{\lambda} Z_R \le x}, Z_M}\\
	&= -\sigma_M \sqrt{\lambda} e^{-(r_f + \frac{\sigma_M^2}{2})\lambda} f\lro{x} \unccov\lro{Z_R,Z_M}.
\end{align*}
Here, $f$ is the density of a normal random variable with mean $(\mu_R - \sigma_R^2/2)\lambda$ and variance $\lambda \sigma_R^2$. As a result,
\begin{equation}\label{eq:ratio_approx}
	\abs{\frac{\uexp{\uncmrkt} - e^{\lambda r_f}}{\tau - \phi(\tau)}} \approx \frac{\sigma_R \sqrt{\lambda} e^{\mu_R \lambda}}{f(x)}.
\end{equation} 
The same reasoning in Example \ref{ex:joint_norm} implies that the relative efficiency between the HJ and distribution bound can be approximated by 
\begin{align}\label{eq:eff_logn}
	\frac{\text{HJ bound}}{\text{distribution bound}}  &=  \frac{\frac{\abs{\uexp{\uncmrkt}-R_f}}{\sigma(\uncmrkt) R_f}}{\frac{\abs{\tau - \phi(\tau)}}{\sqrt{\phi(\tau)(1-\phi(\tau))}R_f}   } \\
	&\overset{\eqref{eq:ratio_approx}}{\approx} \frac{\sqrt{\mathbb{P}(r \le x)\cdot(1-\mathbb{P}(r \le x))}}{\sigma(\uncmrkt)} \times  \frac{\sigma_R \sqrt{\lambda} e^{\mu_R \lambda}}{f(x)},
\end{align}
where $r = \log R$ and $x = \log \tilde{Q}_\tau$. Using the same reasoning as in Example \ref{ex:joint_norm}, the expression on the right hand side of \eqref{eq:eff_logn} is minimized by choosing $x = \log \tilde{Q}_\tau^*$ s.t. $\mathbb{P}(\uncmrkt \le \tilde{Q}_\tau^*) = 1/2$. In that case the relative efficiency equals
\begin{equation*}
	\frac{\sqrt{2 \pi \sigma_R^2}  \sqrt{\lambda} e^{\mu_R \lambda}}{2 \sqrt{[\exp(\sigma_R^2 \lambda)-1] \exp(2 \mu_R \lambda)}} = \frac{1}{2} \sqrt{\frac{2 \pi \sigma_R^2 \lambda}{\exp(\sigma_R^2 \lambda)-1}}.
\end{equation*}

\subsection{Distribution Bound with Pareto Distribution}\label{app:pareto}
This section derives an explicit expression of the distribution bound when the return and SDF follow the Pareto distribution. 

\begin{example}[Pareto distribution]
Let $U \sim \unif{0}{1}$ (Uniform distribution on [0,1]) and consider the following specification:
	\begin{equation}\label{eq:paretoSDF}
		M = A U^\alpha, \ \uncmrkt =  B U^{-\beta} \qquad \text{with} \qquad \alpha, \beta,A,B > 0 .
	\end{equation}
	A random variable $X \sim \pareto{C}{\zeta}$ follows a Pareto distribution with scale parameter $C > 0$ and shape parameter $\zeta > 0$ if the CDF is given by 
	\begin{equation*}
		\mathbb{P}(X \le x) = \begin{cases}
			1 - \lro{x/C}^{-\zeta} & x \ge C\\
			0 &  x <C.
		\end{cases}
	\end{equation*}
	The assumption \eqref{eq:paretoSDF} implies that returns follow a Pareto distribution, both under the physical and risk-neutral measure. This fact allows me to obtain an explicit expression for the distribution bound. I summarize these properties in the Proposition below. 
	
	\begin{proposition}\label{prop:par} Let the SDF and return be given by \eqref{eq:paretoSDF}. Then,
		\begin{enumerate}[(i)]
			\item Under $\mathbb{P}$, the distribution of returns is Pareto: $\uncmrkt \sim \pareto{B}{\frac{1}{\beta}}$. 
			\item Under $\tilde{\mathbb{P}}$, the distribution of returns is Pareto: $\uncmrkt \sim \pareto{B}{\frac{\alpha +1}{\beta}}$.
			\item The Sharpe ratio on the asset return is given by 
			\begin{equation}\label{eq:sharpe_ratio}
				\frac{\uexp{\uncmrkt} - R_f}{\sigma(\uncmrkt)} = \frac{\frac{B}{1-\beta} - \frac{\alpha+1}{A}}{\sqrt{\frac{B^2}{1-2\beta} - \lro{\frac{B}{1-\beta}}^2}}.
			\end{equation}
			\item \label{item:2Par} The distribution bound is given by
			\begin{equation*}
				\frac{1}{R_f}\frac{\abs{\tau -\phi(\tau)}}{ \sqrt{\phi(\tau) (1- \phi(\tau))}} = \frac{A}{1+\alpha} \frac{\abs{\tau - 1 + (1-\tau)^{\frac{1}{\alpha+1}}}}{\sqrt{(1 -(1-\tau)^{\frac{1}{\alpha+1}})(1-\tau)^{\frac{1}{\alpha+1}}}}.
			\end{equation*}
			\item \label{item:3Par} If $\beta \nearrow \frac{1}{2}$,  the HJ bound converges to $0$.
		\end{enumerate}
	\end{proposition}
	\begin{proof}
		See the end of this section.
	\end{proof}
	
	Proposition \ref{prop:par}(\ref{item:2Par}) shows that the distribution bound is independent of the Pareto tail index $\beta$. Properties (\ref{item:2Par}) and (\ref{item:3Par}) provide some intuition when the distribution bound is stronger than the HJ bound. Namely, heavier tails of the distribution of $\uncmrkt$ (as measured by $\beta$) lead to a lower Sharpe ratio. However, the distribution bound is unaffected by $\beta$ since it only depends on the tail index $\alpha$. Therefore, when $\beta$ gets close to $1/2$, the HJ bound is rather uninformative whereas the distribution bound may fare better. Moreover, no additional restrictions on the parameter space are necessary to calculate the distribution bound, while the HJ bound requires $\beta < 1/2$.\footnote{The latter restriction is not unreasonable for asset returns, since typical tail index estimates suggest $\beta \in [1/4,1/3]$ \citep{danielsson2000value}.} \\
	
	Figure \ref{fig:pareto_graph} shows two instances of the distribution and HJ bound using different parameter calibrations. Both calibrations are targeted to match an equity premium of 8\% and risk-free rate of 0\%, but in Panel \ref{fig:pareto_heavyTail}, the distribution of returns has a fatter tail compared to Panel \ref{fig:pareto_ligthTail}. In both calibrations, the distribution bound has a range of values for which it is stronger than the HJ bound. In line with Proposition \ref{prop:par}, we see that the range is larger in Panel (b), since the HJ bound is less informative owing to the heavier tails of $\uncmrkt$. However, the distribution bound attains its maximum in the right-tail since that is the region where the physical and risk-neutral measure differ most. This result is inconsistent with the empirical results from Table \ref{tab:only.rn.quantile}, which indicate that the physical and risk-neutral measure are nearly identical in the right-tail.

	\begin{figure}[htbp] 
		\centering
		\begin{subfigure}[b]{0.49\textwidth}
			\centering
			\includegraphics[width=\textwidth]{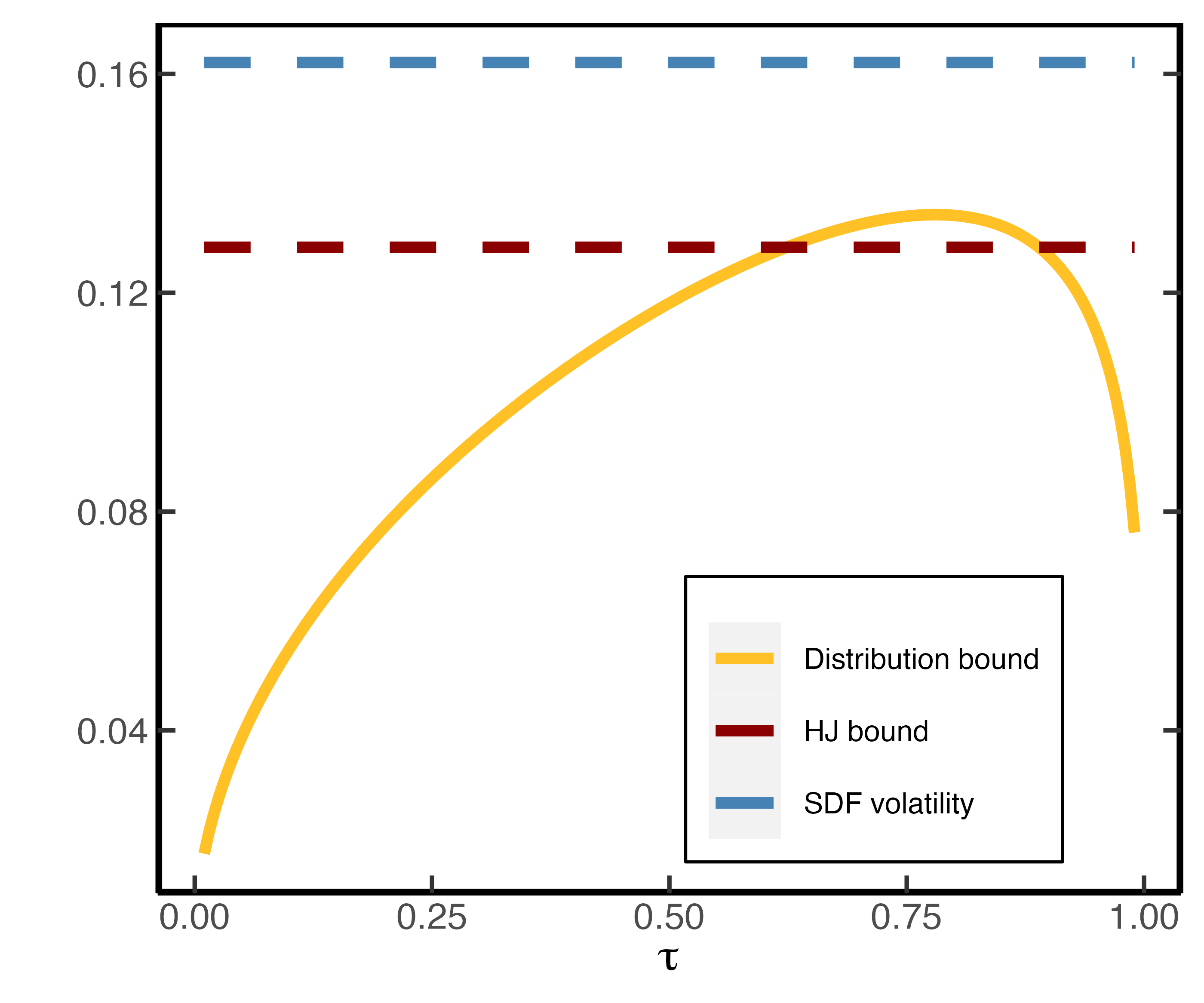}
			\caption{$\beta = 0.33$}
			\label{fig:pareto_ligthTail}
		\end{subfigure}
		\hfill
		\begin{subfigure}[b]{0.49\textwidth}
			\centering
			\includegraphics[width=\textwidth]{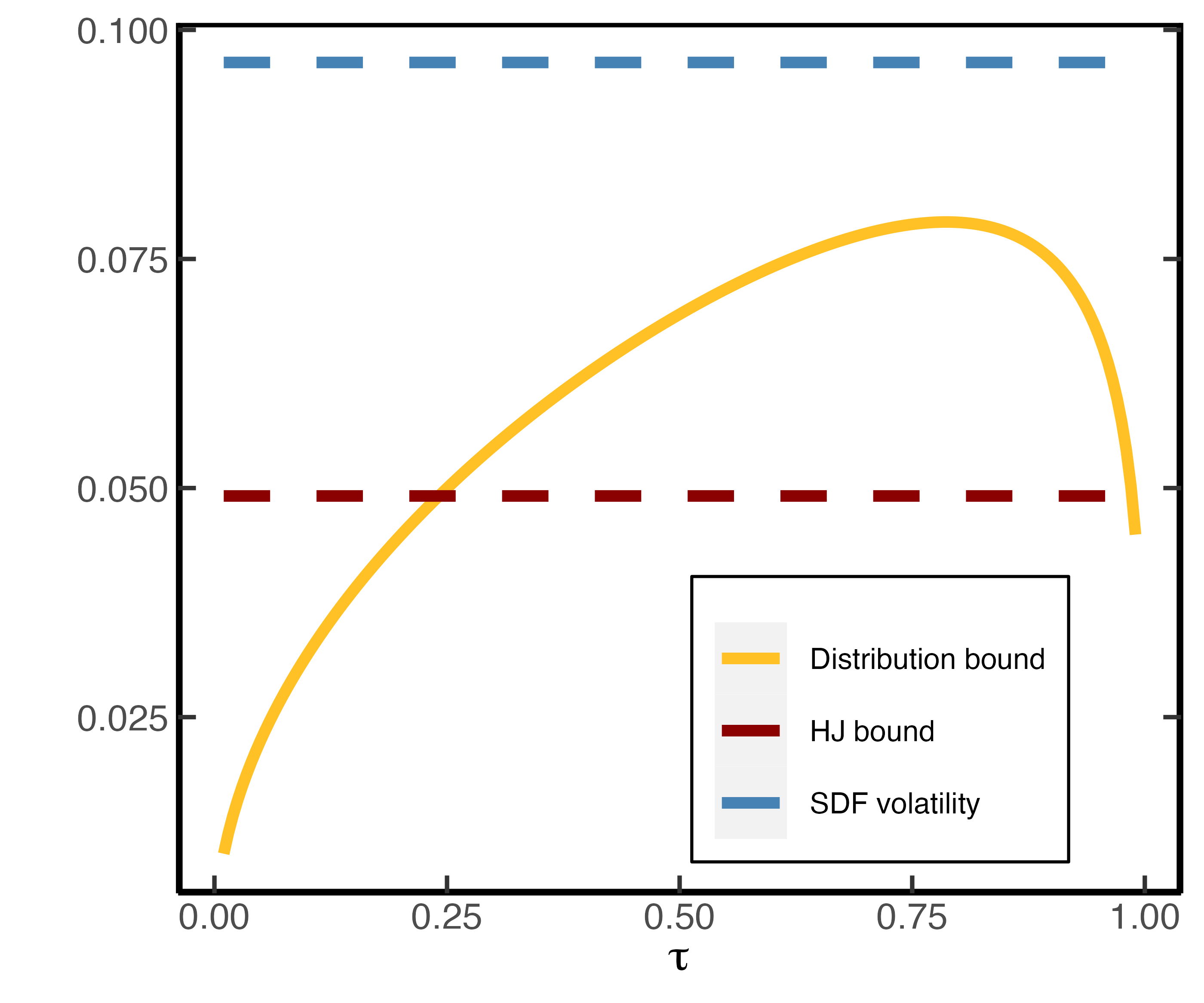}
			\caption{$\beta = 0.45$}
			\label{fig:pareto_heavyTail}
		\end{subfigure}
		\caption{\textbf{HJ and distribution  bound for heavy tailed returns}. \footnotesize Both panels plot the distribution bound, HJ bound  and true SDF volatility for the Pareto model \eqref{eq:paretoSDF}. In Panel (b), the distribution of returns has a fatter tail compared to Panel (a). Panel (a) uses the parameters $[A,\alpha,B,\beta] = [1.19,0.19,0.72,0.33]$. Panel (b) uses the parameters $[A,\alpha,B,\beta] = [1.11,0.11,0.59,0.45]$. Both calibrations imply an equity premium of 8\% and (net) risk-free rate of 0\%.}
		\label{fig:pareto_graph}
	\end{figure} 
	
\end{example}

\begin{proof}[Proof of Proposition \ref{prop:par}]
	\begin{enumerate}[(i)]
		\item 
		The distribution of returns is Pareto, since
		\begin{align*}
		\mathbb{P} (\uncmrkt \le x) &= \mathbb{P}\left(U^{-\beta} \le x/B \right) \\
		& = \mathbb{P}\left(U \ge \lro{x/B}^{-\frac{1}{\beta}}\right) = 1- \lro{\frac{x}{B}}^{-\frac{1}{\beta}}, \qquad x \ge B.
		\end{align*}
		
			\item  Since $R_f M$ is the Radon-Nikodym derivative that induces a change of measure from $\mathbb{P}$ to $\tilde{\mathbb{P}}$, it follows that
		\begin{align*}
		\tilde{\mathbb{P}}(\uncmrkt \le x) &= R_f \uexp{M \ind{\uncmrkt \le x}} \\
		&= R_f \int_0^1 A u^\alpha \ind{B u^{-\beta} \le x} \diff u\\
		&= R_f A \int_0^1 u^\alpha \ind{u \ge \lro{\frac{x}{B}}^{-\frac{1}{\beta}}} \diff u\\
		&= \frac{R_f A}{\alpha + 1} \lro{1 - \lro{\frac{x}{B}}^{-\frac{\alpha+1}{\beta}} } \\
		&= {1 - \lro{\frac{x}{B}}^{-\frac{\alpha+1}{\beta}} }.
		\end{align*}
		The last line follows from \eqref{eq:temp_constraint} below.
		
		\item 
		Routine calculations show that the mean and variance of $\uncmrkt$ are given by (provided $\beta < 1/2$)
		\begin{equation}\label{eq:rdist}
		\uexp{\uncmrkt} = \frac{B}{1-\beta} \qquad \sigma^2(\uncmrkt) = \frac{B^2}{1-2\beta} - \lro{\frac{B}{1-\beta}}^2.
		\end{equation}
		Likewise, the distribution of the SDF follows from
		\begin{equation*}
		\mathbb{P}\lro{M \le x} = \mathbb{P}\lro{A U^\alpha \le x} = \lro{\frac{x}{A}}^{\frac{1}{\alpha}}, \quad 0 \le x \le A.
		\end{equation*} 
		In this case, $M$ is said to have a Pareto lower tail. The expectation is given by 
		\begin{equation*}
		\uexp{M} = \frac{A}{\alpha+1}.
		\end{equation*} 
		The constraint $\uexp{M \uncmrkt} = 1$ forces 
		\begin{equation}\label{eq:sdf2}
		\frac{AB}{\alpha - \beta +1} = 1.
		\end{equation}
		In addition  from $\uexp{M} = \frac{1}{R_f}$ it follows
		\begin{equation}\label{eq:temp_constraint}
		\frac{A}{\alpha+1} = \frac{1}{R_f}.
		\end{equation}
		The Sharpe ratio can now be computed from \eqref{eq:rdist} and \eqref{eq:temp_constraint}.

	\item  It is straightforward to show that the quantiles of a $\pareto{C}{\zeta}$ distribution are given by 
	\begin{equation*}
	Q_\tau = C  (1-\tau)^{-1/\zeta}.
	\end{equation*}
	It therefore follows that the risk-neutral quantile function is equal to 
	\begin{equation*}
	\tilde{Q}_\tau = B  (1-\tau)^{-\frac{\beta}{\alpha+1}}.
	\end{equation*}
	As a result
	\begin{align*}
	\mathbb{P}(\uncmrkt \le \tilde{Q}_\tau) &= \mathbb{P}\lro{\uncmrkt \le B (1-\tau)^{-\frac{\beta}{\alpha+1}}}\\
	&= 1- \lro{\frac{B}{B (1-\tau)^{\frac{-\beta}{\alpha+1}}}}^{\frac{1}{\beta}}\\
	&= 1 -(1-\tau)^{\frac{1}{\alpha+1}}.
	\end{align*}
	Hence, the distribution bound evaluates to
	\begin{equation*}
	\frac{1}{R_f} \frac{\abs{\tau - \phi(\tau)}}{\sqrt{\phi(\tau)(1-\phi(\tau))}} = \frac{A}{1+\alpha} \frac{\abs{\tau - 1 + (1-\tau)^{\frac{1}{\alpha+1}}}}{\sqrt{(1 -(1-\tau)^{\frac{1}{\alpha+1}})(1-\tau)^{\frac{1}{\alpha+1}}}}.
	\end{equation*}

	\item  The HJ bound, as given by the Sharpe ratio in \eqref{eq:sharpe_ratio}, goes to $0$ as $\beta \nearrow 1/2$ since $\sigma(\uncmrkt)  \nearrow \infty $. \qedhere
	\end{enumerate} 
\end{proof}

\subsection{Derivation of G\^{a}teaux Derivative}\label{app:mises}
In this Section I derive \eqref{eq:vonMises2}. For ease of exposition, I drop the time subscripts. For $\lambda \in [0,1]$, define $\tilde{F}_\lambda \coloneqq (1-\lambda)\tilde{F} + \lambda F$. The following (trivial) identity will prove helpful\footnote{This ``equality'' may actually only be an inequality for some $\tau$, but this is immaterial to the argument.}
\begin{equation}\label{eq:tauF}
	\tau = \tilde{F}_\lambda \tilde{F}_\lambda^{-1}.
\end{equation}
To further simplify notation, write $q(\lambda) \coloneqq \tilde{F}_\lambda^{-1}$. Then \eqref{eq:tauF} becomes
\begin{equation*}
	\tau = (1-\lambda)\tilde{F}(q(\lambda)) + \lambda F(q(\lambda)).
\end{equation*}
Applying the implicit function theorem, we obtain
\begin{equation*}
	q'(\lambda) = - \frac{-\tilde{F}(q(\lambda)) + F(q(\lambda))}{(1-\lambda) \tilde{f}(q(\lambda)) + \lambda f(q(\lambda))}.
\end{equation*}	
Plug in $\lambda = 0$ to get
\begin{equation}\label{eq:1}
	q'(0)  = - \frac{-\tilde{F}(q(0)) + F(q(0))}{\tilde{f}(q(0))}.
\end{equation}
Notice that 
\begin{equation}\label{eq:2}
	\tilde{F}_\lambda \big|_{\lambda = 0} = \tilde{F} \implies q(\lambda) \big|_{\lambda = 0} = q(0) = \tilde{F}^{-1}.
\end{equation}	
Substitute \eqref{eq:2} into \eqref{eq:1} to obtain
\begin{equation}\label{eq:3}
	q'(0) = - \frac{-\tilde{F}(\tilde{F}^{-1})  + F(\tilde{F}^{-1})  }{\tilde{f}(\tilde{F}^{-1})} = \frac{\tau - F(\tilde{F}^{-1})  }{\tilde{f}(\tilde{F}^{-1})}.
\end{equation}
Notice that $q'(0)$ is exactly equal to the G\^{a}teaux derivative from the definition in \eqref{eq:Gateax}, since
\begin{equation*}
	\frac{\partial}{\partial \lambda} \varphi\lr{(1-\lambda)\tilde{F} + \lambda F} \bigg|_{\lambda =0} = \frac{\partial}{\partial \lambda} q(\lambda) \bigg|_{\lambda = 0} = q'(0).
\end{equation*}

\subsection{Proof of Proposition \ref{prop:lower_bound}}\label{app:proof_q_lrb}
 In the proofs that follow, I repeatedly use Taylor's theorem with integral remainder, which is stated for completeness. 
\begin{lemma}[Taylor's theorem]\label{thm:mean}
	Let $\zeta^{(3)}(\cdot)$ be absolutely continuous on the closed interval between $a$ and $x$, then 
	\begin{equation*}
		\zeta(x) = \sum_{k=0}^{3} \frac{\zeta^{(k)}(a)}{k!} (x-a)^k + \int_{a}^{x}\frac{\zeta^{(4)}(t)}{3!}(x-t)^3 \diff t.
	\end{equation*}
\end{lemma}

The proof of Proposition \ref{prop:lower_bound} proceeds in several stages, by first proving an infeasible lower bound on $\tau - F_t(\cquant)$, which is later refined into a feasible lower bound under additional assumptions.  Before doing so, I collect several results about the SDF in representative agent models.
\begin{lemma}\label{lemma:chabi_F}
	Assume a representative agent model with SDF given by \eqref{eq:sdf_chabi}, then
	\begin{equation}\label{eq:chabi_F}
		\tau - F_t(\cquant) = -\frac{\covhat{\ind{\omrkt \le \cquant}}{\zeta(\omrkt)}}{\qcexp{\zeta(\omrkt)}},
	\end{equation}
	where $\zeta(\cdot)$ is defined in \eqref{eq:notation}. 
\end{lemma}
\begin{proof}
	Use the reciprocal of the SDF to pass from physical to risk-neutral measure
	\begin{align}
		F_t(\cquant) &= \cexp{\ind{\omrkt \le \cquant}} = \texpneut\lr{ \ind{ \omrkt \le \cquant} \frac{\cexp{\osdf}}{\osdf }   }   \nonumber\\
		&=\covhat{\ind{\omrkt \le \cquant}}{\frac{\cexp{\osdf}}{\osdf  }} + \tau. \label{eq:cov}
	\end{align}
	Rearranging the above and using the definition of $\zeta(\cdot)$ in \eqref{eq:notation}, as well as \eqref{eq:sdf_chabi}, we obtain \eqref{eq:chabi_F}.
\end{proof}

\begin{lemma}\label{lemma:zeta_all}
	Under Assumption \ref{ass:f},
	\begin{equation*}
		\qcexp{\zeta(\omrkt)} \le \sum_{k=0}^{3} \theta_k \chabi{k} = 1 + \sum_{k=1}^{3} \theta_k \chabi{k},
	\end{equation*} 
where $\zeta(x)$ is the IMRS defined in \eqref{eq:notation}.
\end{lemma}
\begin{proof}
	In the integral of Lemma  \ref{thm:mean}, substitute $s = (t-a)/(x-a)$ to get
	\begin{align*}
		\zeta(x) &= \sum_{k=0}^{3} \frac{\zeta^{(k)}(a)}{k!} (x-a)^k + (x-a)^4 \int_{0}^{1}\frac{\zeta^{(4)}(a + s(x-a))}{3!}\lro{1-s}^3  \diff s\\
		& \le \sum_{k=0}^{3} \frac{\zeta^{(k)}(a)}{k!} (x-a)^k,
	\end{align*}
	since $\zeta^{(4)}(x) < 0$ by Assumption \ref{ass:f}(\ref{item:2}). Using this result with $a = \ofree$ and taking expectations, we obtain
	\begin{equation*}
		\qcexp{\zeta(\omrkt)} \le \sum_{k=0}^{3} \theta_k \chabi{k}. \qedhere
	\end{equation*}
\end{proof} 

Under Assumption \ref{ass:f},  the difference between the physical and risk-neutral distribution in the left-tail can be bounded as follows. 
\begin{theorem}[Infeasible Lower Bound]\label{thm:lower_bound}
	Let Assumption \ref{ass:f} hold and assume that the risk-neutral CDF is absolutely continuous with respect to Lebesgue measure. Define $\tau^*$ so that $ G(\cquantstar) = \texptilde{G(\omrkt)}$, where
	\begin{equation*}
		G(\omrkt) \coloneqq \int_{\ofree}^{\omrkt}  \zeta^{(4)}(t) (\omrkt - t)^3 \diff t. 
	\end{equation*}
	Then for all $\tau \le \tau^*$,
	\begin{equation}\label{eq:inf_lower}
		\tau - F_t\lro{\cquant} \ge \frac{\sum_{k=1}^3  \theta_k \left(  \tau \chabi{k} -\trunc{k}{\cquant} \right)    }{1+\sum_{k=1}^{3} \theta_k \chabi{k}},
	\end{equation}
	where $\chabi{k}, \trunc{k}{\cquant}$ are defined in \eqref{eq:trunc_mrkt}.
\end{theorem}

\begin{proof}[Proof of Theorem \ref{thm:lower_bound}]
	By Taylor's theorem,
	\begin{equation}\label{eq:uneq}
		\begin{split}
			 &-\covhat{\ind{\omrkt \le \cquant}}{\zeta(\omrkt)} =  \sum_{k=1}^3 \theta_k \lro{ \tau \chabi{k} - \trunc{k}{\cquant} } \\
			& - \widetilde{\cov} \bigg[\ind{\omrkt \le \cquant},\frac{1}{3!}\int_{\ofree}^{\omrkt}  \zeta^{(4)}(t) (\omrkt - t)^3 \diff t  \bigg]\\
			&\ge \sum_{k=1}^3 \theta_k \lro{ \tau \chabi{k} - \trunc{k}{\cquant} }. 
		\end{split}
	\end{equation}
	The last line follows from Lemma \ref{lemma:neg_cov} below. Hence, 
	\begin{align*}
		\tau - F_t(\cquant) &= -\frac{\covhat{\ind{\omrkt \le \cquant}}{\zeta(\omrkt)}}{\qcexp{\zeta(\omrkt)}}\\
		&\ge  \frac{\sum_{k=1}^3  \theta_k \left(  \tau \chabi{k} -\trunc{k}{\cquant} \right)    }{1+\sum_{k=1}^{3} \theta_k \chabi{k}},
	\end{align*}
	where the first identity follows from Lemma \ref{lemma:chabi_F} and the inequality follows from \eqref{eq:uneq} and Lemma \ref{lemma:zeta_all}. 
\end{proof}

\begin{Remark}
	The condition that $\tau \le \tau^*$ is sufficient but not necessary, as the proof of Theorem \ref{thm:lower_bound} shows. Furthermore, the proof also shows that $\tau^* > 0$ exists regardless of the utility function. In practice, however, $\tau^*$ is unknown since $G(\cdot)$ depends on the unknown utility function of the representative agent. Appendix \ref{app:lrb_CRRA} shows that $\tau^* \approx 0.5$ in the data for CRRA utility and different levels of risk-aversion. In light of this result, it seems that $\tau \in \{0.05,0.1,0.2\}$ is sufficiently conservative for the lower bound to hold, and I use these values in the empirical application in Section \ref{sec:q_test}. 
\end{Remark}

\begin{lemma}\label{lemma:neg_cov}
	Suppose that Assumption \ref{ass:f} holds. In addition, define $\tau^*$ so that $ G(\cquantstar) = \texptilde{G(\omrkt)}$, where
	
	\begin{equation*}
		G(\omrkt) \coloneqq \int_{\ofree}^{\omrkt}  \zeta^{(4)}(t) (\omrkt - t)^3 \diff t. 
	\end{equation*}
	 Then for all $\tau \le \tau^*$,
	\begin{equation}\label{eq:help_cov}
		\covhat{\ind{\omrkt \le \cquant}}{\int_{\ofree}^{\omrkt}  \zeta^{(4)}(t) (\omrkt - t)^3 \diff t } \le 0.
	\end{equation}
\end{lemma}
\begin{proof}
If $\zeta^{(4)} \equiv 0$, then \eqref{eq:help_cov} trivially holds. Hence, assume that $\zeta^{(4)}$ is not identically equal to zero. First we show that $G(\omrkt)$ is increasing on $(0,\ofree)$, since by Leibniz' rule
\begin{equation*}
	G'(\omrkt) = - 3\int_{\omrkt}^{\ofree}  \zeta^{(4)}(t)(\omrkt - t)^2 \diff t \ge 0.
\end{equation*}
The inequality follows since $\zeta^{(4)}(t) < 0$ by Assumption \ref{ass:f}(\ref{item:2}). Temporarily write $K = \cquant$ to ease notation and consider
\begin{equation*}
\Gamma(K) = 	\covhat{\ind{\omrkt \le K}}{\int_{\ofree}^{\omrkt}  \zeta^{(4)}(t) (\omrkt - t)^3 \diff t }.
\end{equation*} 
By Leibniz' rule again, we get
\begin{equation*}
\Gamma'(K) = \tilde{f}_t(K) \lro{G(K) - \texptilde{G(\omrkt)}}.
\end{equation*}
Since $G(\ofree) = 0$, $G(\omrkt) \le 0$  and $G(\omrkt)$ is increasing on $(0,\ofree)$, we know that $\Gamma'(K)  \le 0 $  for all $K \le K^* < \ofree$, where $K^*$  is defined such that $G(K^*) =  \texptilde{G(\omrkt)}$.  To complete the proof, define $\tau^*$ so that it satisfies $\tilde{Q}_{\tau^*} = K^*$. 
\end{proof}

\begin{Remark}
The bound in \eqref{eq:inf_lower} is infeasible since $\{\theta_k\}_{k=1}^3$ is unknown.\footnote{In Appendix \ref{app:crash}, I use comparative statics for common utility functions to analyze the tail difference between the physical and risk-neutral distribution.} However, \citet{chabi2020conditional} show that these unknowns relate to the coefficient of {relative risk-aversion}, {relative prudence} and {relative temperance} of the representative agent. Based on this observation and using results from the expected utility literature \citep{Eeckhoudt2006}, the authors propose an additional restriction on $\theta_k$ that allows me to prove the feasible lower bound in Proposition \ref{prop:lower_bound}. 
\end{Remark}

\begin{proof}[Proof of Proposition \ref{prop:lower_bound}]
Using Assumption \ref{ass:skewness}(\ref{item:chabi1}) and \ref{ass:skewness}(\ref{item:chabi2}), we get $\theta_2 \chabi{2} \le -1/\ofree^2 \chabi{2}$ and $\theta_3 \chabi{3} \le 1/\ofree^3 \chabi{3}$, from which it follows that 
\begin{equation}\label{eq:pos_moment}
1 + \sum_{k=1}^{3} \theta_k \chabi{k} \le 1 - \frac{1}{\ofree^2} \chabi{2} +  \frac{1}{\ofree^3} \chabi{3}.
\end{equation}
Second, recall that for $K>0$
\begin{equation*}
\tilde{F}_t(K) \chabi{k} - \trunc{k}{K} = -\covhat{\ind{\omrkt \le K}}{(\omrkt - \ofree)^k}.
\end{equation*}
If $k = 1,3$, then Chebyshev's sum inequality \ref{lemma:cheby} implies that
\begin{equation*}
\Gamma(K) \coloneqq \covhat{\ind{\omrkt \le K}}{(\omrkt - \ofree)^k} \le 0.
\end{equation*}
Hence under Assumption (\ref{item:chabi1}),
\begin{equation}\label{eq:neg_cor}
	\theta_k\lro{\tilde{F}_t(K) \chabi{k} - \trunc{k}{K}} \ge 	\frac{1}{\ofree
	^k} \lro{\tilde{F}_t(K) \chabi{k} - \trunc{k}{K}} \quad \text{for } k = 1,3.
\end{equation}
If $k=2$, we obtain from Leibniz' rule
\begin{equation}\label{eq:Gamma}
\Gamma'(K) = \tilde{f}_t(K)\lr{(K-\ofree)^2 - \vartilde(\omrkt)}.
\end{equation}
It follows that \eqref{eq:Gamma} is positive if $K \le \ofree - \sqrt{\vartilde(\omrkt)} \eqqcolon K^{**}$. Combining \eqref{eq:neg_cor} and \eqref{eq:Gamma}, we get for $K \le K^{**}$
\begin{equation}\label{eq:cor_always_pos}
\theta_k \lro{\tilde{F}_t(K) \chabi{k} - \trunc{k}{K} } \ge \frac{(-1)^{k+1}}{\ofree^k} \lro{\tilde{F}_t(K) \chabi{k} - \trunc{k}{K} }.
\end{equation}
Collecting the results from \eqref{eq:pos_moment} and \eqref{eq:cor_always_pos} and using the general upper bound \eqref{eq:inf_lower} from Theorem \ref{thm:lower_bound}, it follows that 
\begin{align*}
	\tau - F_t\lro{\cquant} &\overset{\eqref{eq:inf_lower}}{\ge} \frac{\sum_{k=1}^3  \theta_k \left(  \tau \chabi{k} -\trunc{k}{\cquant} \right)    }{1+\sum_{k=1}^{3} \theta_k \chabi{k}}\\
	&\ge \frac{\sum_{k=1}^3  \frac{(-1)^{k-1}}{\ofree^k} \left( \tau \chabi{k} -\trunc{k}{\cquant}  \right)    }{1+\sum_{k=1}^{3} \frac{(-1)^{k-1}}{\ofree^k} \chabi{k}} ,
\end{align*}
for all $\tau$ such that $\tilde{Q}_{t,\tau} \le \min(K^*,K^{**})$, where $K^{*}$ is defined in Theorem \ref{thm:lower_bound}. 	
\end{proof}

\begin{Remark}
The bound only holds for quantiles far enough in the left-tail. Compared to Theorem \ref{thm:lower_bound}, the additional condition needed for the bound to hold is that  $\cquant \le \ofree - \sqrt{\vartilde(\omrkt)}$, which covers a wide range of quantiles in the left-tail, since in the data $\sqrt{\vartilde(\omrkt)}$ is in the order of $10^{-3}$ for 90-day returns, whereas the risk-free rate is typically around 1.\footnote{At the 30- and 60-day horizon, the risk-neutral standard deviation is even smaller.}
\end{Remark}

\subsection{Formulas for market moments}\label{app:chabi}
This Section presents formulas for the (un)truncated risk-neutral moments of the excess market return.  I use a slight abuse of notation and write $\tilde{Q}(\tau) \coloneqq \tilde{Q}_\tau(\omrkt)$, to emphasize that the integrals below are taken with respect to $\tau$.
\begin{proposition}
	Any risk-neutral moment can be computed from the risk-neutral quantile function, since
	\begin{equation}\label{eq:b3}
	\texpneut \lr{(\omrkt - \ofree)^n} = \int_0^1 [\tilde{Q}_{\tau}(\omrkt-\ofree)]^n \diff \tau = \int_0^1 [\tilde{Q}(\tau) - \ofree]^n \diff \tau.
	\end{equation}
	Moreover, any truncated risk-neutral moment can be calculated by
	\begin{equation*}
	\texpneut\lr{(\omrkt - \ofree)^n \ind{\omrkt \le k_0}} = \int_0^{\tilde{F}_t(k_0)} [\tilde{Q}(\tau)-\ofree]^n \diff \tau.
	\end{equation*}
\end{proposition}
\begin{proof}
	For any random variable $X$ and integer $n$ such that the $n$-th moment exists, we have
	\begin{equation*}
	\uexp{X^n} = \int_0^1 [Q_X(\tau)]^n \diff \tau.
	\end{equation*}
	This follows straightforward from the substitution $x = Q(\tau)$. Now use that for any constant $a \in \mathbb{R}$, $Q_{X-a}(\tau) = Q_X(\tau) -a$ to derive \eqref{eq:b3}. The truncated formula follows similarly.
\end{proof}
\begin{Remark}
	Frequently I use $k_0 = \tilde{Q}_\tau$, in which case the truncated moment formula reduces to 
	\begin{equation*}
		\texpneut\lr{(\omrkt - \ofree)^n \ind{\omrkt \le \tilde{Q}_\tau}} = \int_0^{\tau} [\tilde{Q}(p)-\ofree]^n \diff p.
	\end{equation*}
\end{Remark}

\section{Estimating the Risk-neutral Quantile Function}\label{app:martingale}
\subsection{Data Description}\label{app:data_description}
To estimate the risk-neutral quantile curve for each point in time, I use daily option prices from OptionMetrics covering the period 01-01-1996 until 12-31-2021. The data consist of European Put and Call options on the S\&P 500 index. The option contract further contains information on the highest closing bid and lowest closing ask price and price of the forward contract on the underlying security. I use the midpoint of the bid and ask price to proxy for the unobserved option price.  In addition, I obtain data on the daily risk-free rate from Kenneth French' website.\footnote{See \url{http://mba.tuck.dartmouth.edu/pages/faculty/ken.french/data_library.html\#Research}} Finally, I obtain stock price data on the closing price of the S\&P 500 from WRDS.\\

I use an additional cleaning procedure for the option data, prior to estimating the martingale measure.  All observations are dropped for which the highest closing bid price equals zero, as well as all option prices that violate no-arbitrage bounds. Subsequently, I drop all option prices with maturity less than 7 days or greater than 500 days. After the cleaning procedure, I'm left with 23,264,113 option-day observations. \\

I discard all observations prior to 2003 for the quantile regression application, since there are many days in the period 1996-2003 that have insufficient option data to estimate $\cquant$ at the 30, 60 and 90-day horizon. Occasionally it happens that I cannot estimate the risk-neutral quantile on a specific day in the post 2003 period and I discard these days as well.\footnote{The number of days I cannot estimate the risk-neutral PDF is very small, about 2\% in total. Most of these days occur at the beginning of the sample period.}

\subsection{Estimation Procedure}\label{app:risk_neutral_quantile_function}
There is a substantial literature on how to extract the martingale measure from option prices. I use the \texttt{RND Fitting Tool} application on MATLAB, which is developed by \citet{barletta2018analyzing}.\footnote{The application can be downloaded from the author's GITHUB page: \url{https://github.com/abarletta/rndfittool}}. The tool is based on the orthogonal polynomial expansion of \citet{filipovic2013density}. In short, the idea is to approximate the conditional risk-neutral density function by an expansion of the form
\begin{equation*}
\tilde{f}_t(x) \approx \phi(x) \lr{1 + \sum_{k=1}^K \sum_{i=0}^{k} c_k w_{i,k} x^k},
\end{equation*}
where $\phi(x)$ is an arbitrary density and the polynomial term serves to tilt the density function towards the risk-neutral distribution. Further details about the estimation of the coefficients $w_{i,k}$ and $c_k$ can be found in \citet{filipovic2013density}.\\

For my purpose, I need to choose the kernel function $\phi(\cdot)$, the estimation method for $c_k$ and the degree of the expansion $K$. I follow the recommendation of \citet{barletta2018analyzing} and use the {double beta} distribution for the kernel and principal component analysis to estimate $c_k$. This is the most robust method for S\&P500 options. To avoid overfitting, I use $K = 3$ if the number of option data is less than 70, $K=6$ if the number is less than 100 and $K=8$ otherwise. This choice renders a good approximation for most time periods.\\

I interpolate the estimated risk-neutral densities for a given time horizon. Occasionally, there are no two interpolation points. In such cases, I drop the observations to avoid negative density estimates due to extrapolation. Since the RND Fitting Tool is designed for an equal number of put and call options, I use Put-Call parity to convert in-the-money call prices to put prices and vice versa. Subsequently, I use Black-Scholes implied volatilities to interpolate the Call-Put option price curve near the forward price. This transformation ensures that the risk-neutral density does not have a discontinuity for strike prices that are close to being at-the-money \citep{figlewski2008estimating}. Finally, I integrate the density function and take the inverse to obtain the risk-neutral quantile curve 
\begin{equation*}
	\cquant \coloneqq \inf\left\{x \in \mathbb{R} : \tau \le \tilde{F}_t(x) \right\}, \quad \text{where } \tilde{F}_t(x) = \int_{0}^{x} \tilde{f}_t(y) \diff y.
\end{equation*}

\section{Verifying Assumption  \ref{ass:f}(\ref{item:2}) in Representative Agent Models}\label{app:verify}
The proof of Theorem \ref{thm:lower_bound} relies on Assumption \ref{ass:f}(\ref{item:2}). This section derives parameter restrictions for common utility functions that are needed so that Assumption \ref{ass:f}(\ref{item:2}) is satisfied. Most of these restrictions closely resemble those of \citet{chabi2020conditional}. I also illustrate the lower bound with actual data assuming CRRA utility.

\subsection{Log utility}
In this case $u(x) = \log x$. It follows that $\zeta(x) = x/\ofree$. Clearly $\zeta^{(4)}(x) = 0$ and Assumption \ref{ass:f}  holds.

\subsection{CRRA utility}
More generally, consider $u(x) = \frac{x^{1-\gamma}}{1-\gamma}$ for $\gamma \ge 0$. It follows that  $\zeta(x) = (\frac{x}{\ofree})^\gamma$ and hence
\begin{equation*}
	\zeta^{(4)}(x) = \frac{1}{\ofree^\gamma} \gamma(\gamma-1)(\gamma-2)(\gamma-3) x^{\gamma-4}.
\end{equation*} 
Part (\ref{item:2})  of Assumption \ref{ass:f} holds if $\gamma \in [0,1]$, but also if $\gamma \in [2,3]$. Notice that the additional restrictions in the feasible lower bound in Proposition \ref{prop:lower_bound} {cannot} be accommodated by this model. To see this, observe that $\theta_2 \le -1/\ofree^2$ implies that $\gamma(\gamma-1)/2 \le  -1/\ofree^2$, which cannot hold for any reasonable interest rate. This failure illustrates that a representative agent model with CRRA utility is misspecified in that it cannot produce a sizable risk-premium on skewness.\footnote{See in particular \citet[Equation (A.5)]{chabi2020conditional}, which shows that $\theta_2$ is related to the risk-premium on market skewness.}

\subsection{CARA utility}
In this case, $u(x) = 1-e^{-\gamma x}$ and $\zeta(x) = e^{\gamma^*(x-\ofree)}$, where $\gamma^* = W_t \gamma$. Since $\zeta^{(4)} > 0$, Assumption \ref{ass:f} does not hold. 

\subsection{HARA utility}
The utility function is given by $u(x) = \frac{1-\gamma}{\gamma}\lro{\frac{ax}{1-\gamma} + b}^\gamma$, where $a>0$ and $\frac{ax}{1-\gamma} + b > 0$.  Successive differentiation renders
\begin{equation*}
	\zeta^{(4)}(x) = 	\frac{- \gamma  ( \gamma + 1)  (\gamma  + 2)   (aW_t)^4 \left(\frac{a W_t x}{1-\gamma }+b\right)^{-\gamma -3} \left(\frac{a W_t \ofree}{1-\gamma }+b\right)^{\gamma -1}}{(1-\gamma )^3}.
\end{equation*}
We see that $\gamma \in [0,1)$ is a sufficient condition for $\zeta^{(4)}(x) \le 0$.

\subsection{Lower Bound in the Data for CRRA utility}\label{app:lrb_CRRA}
\begin{figure}[!htb]
	\centering
	\begin{subfigure}[b]{0.49\textwidth}
		\centering
		\includegraphics[width=\textwidth]{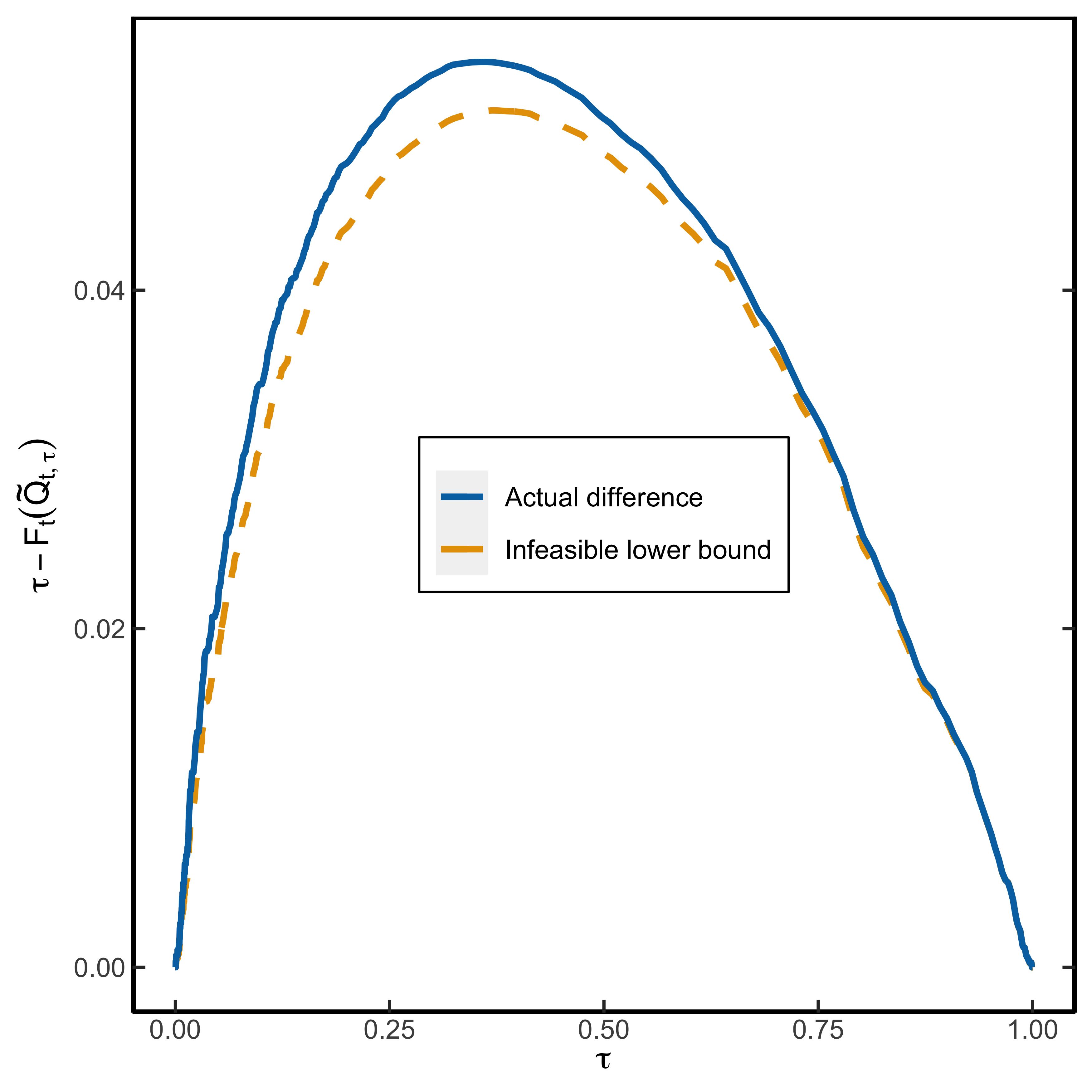}
		\caption{Infeasible lower bound, $\gamma = 2.2$}
		\label{fig:lrb_infeas}
	\end{subfigure}
	\hfill
	\begin{subfigure}[b]{0.49\textwidth}
		\centering
		\includegraphics[width=\textwidth]{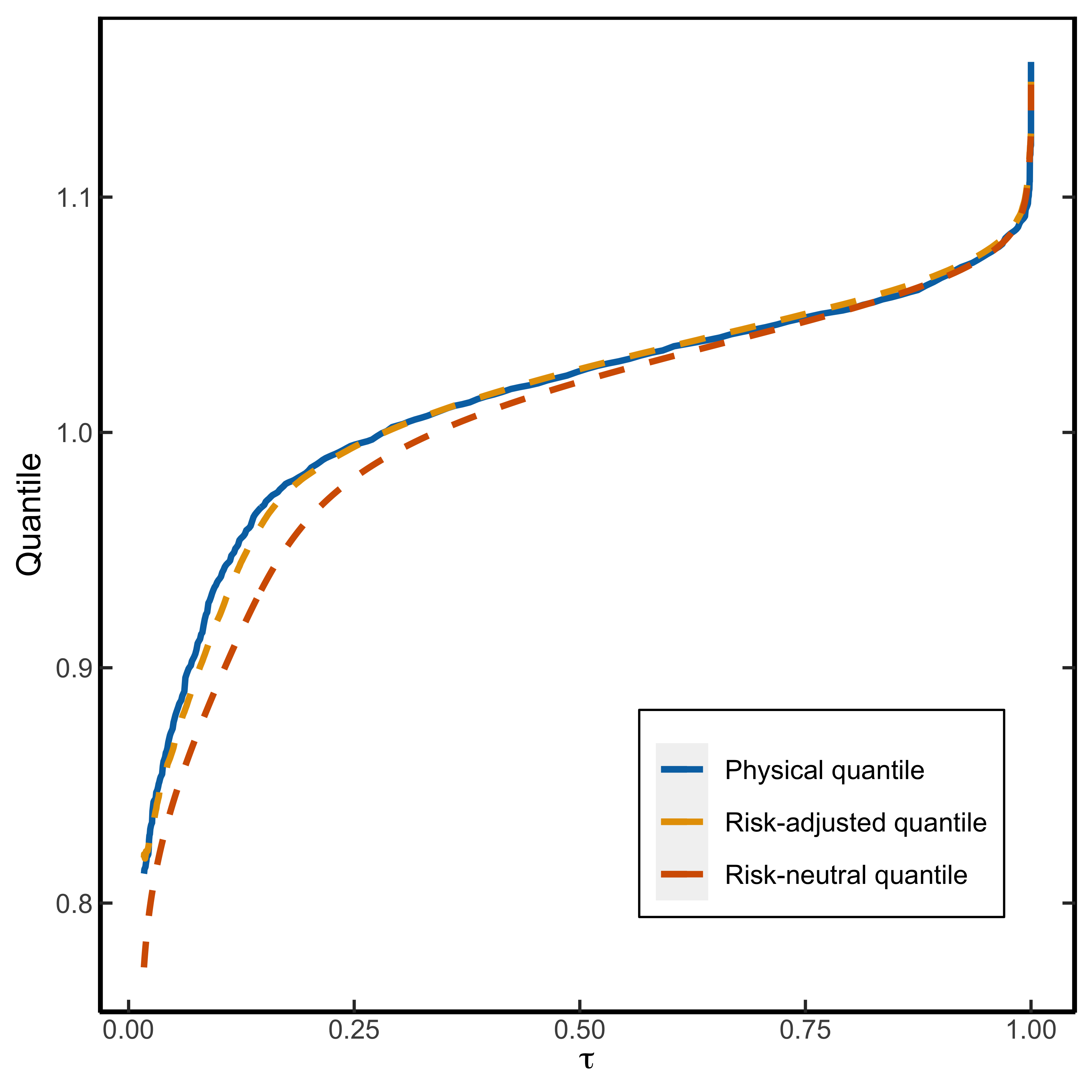}
		\caption{Quantile function,  $\gamma = 2.2$}
		\label{fig:quant_infeas}
	\end{subfigure}
\begin{subfigure}[b]{0.49\textwidth}
	\centering
	\includegraphics[width=\textwidth]{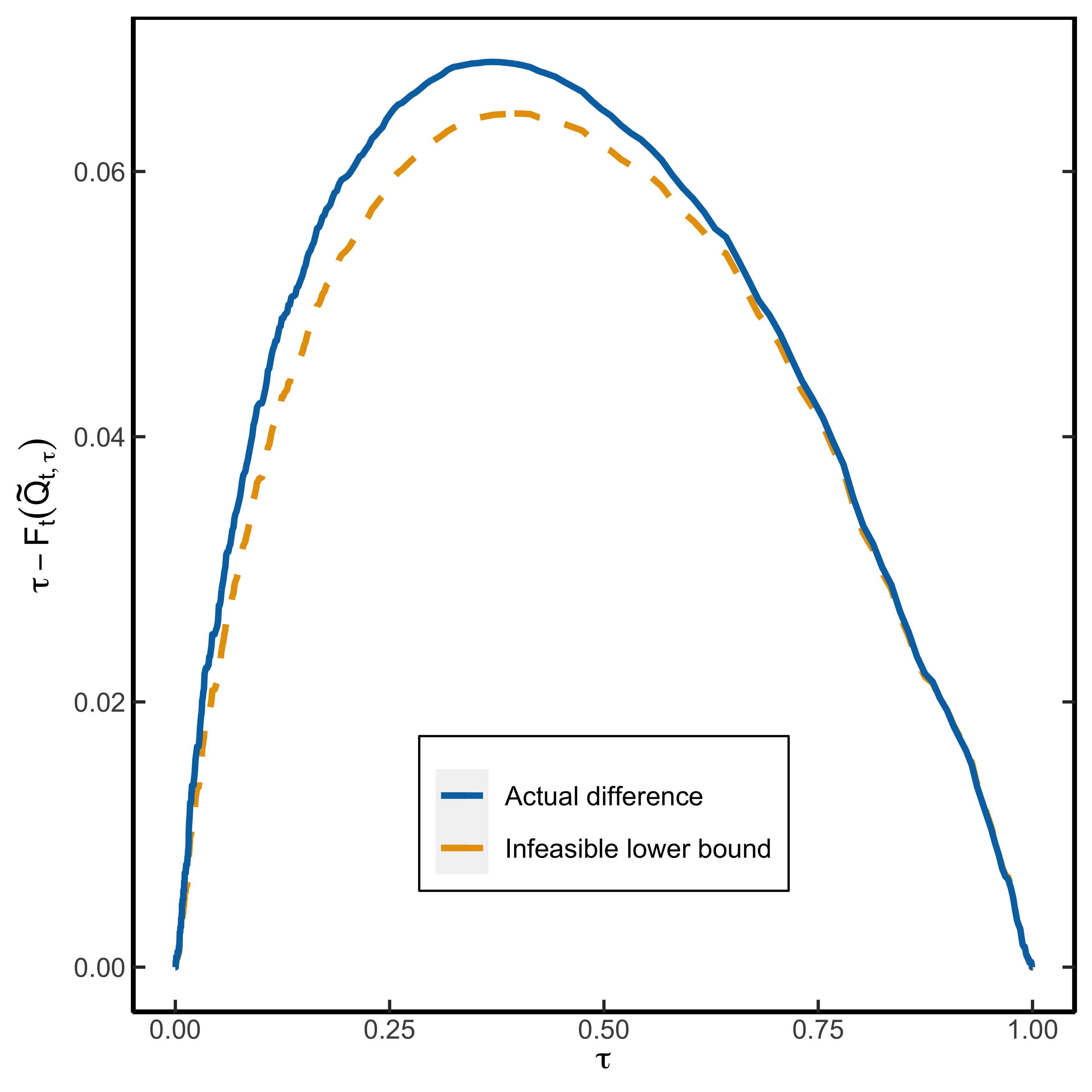}
	\caption{Infeasible lower bound, $\gamma = 2.9$}
	\label{fig:lrb_infeas_high}
\end{subfigure}
\hfill
\begin{subfigure}[b]{0.49\textwidth}
	\centering
	\includegraphics[width=\textwidth]{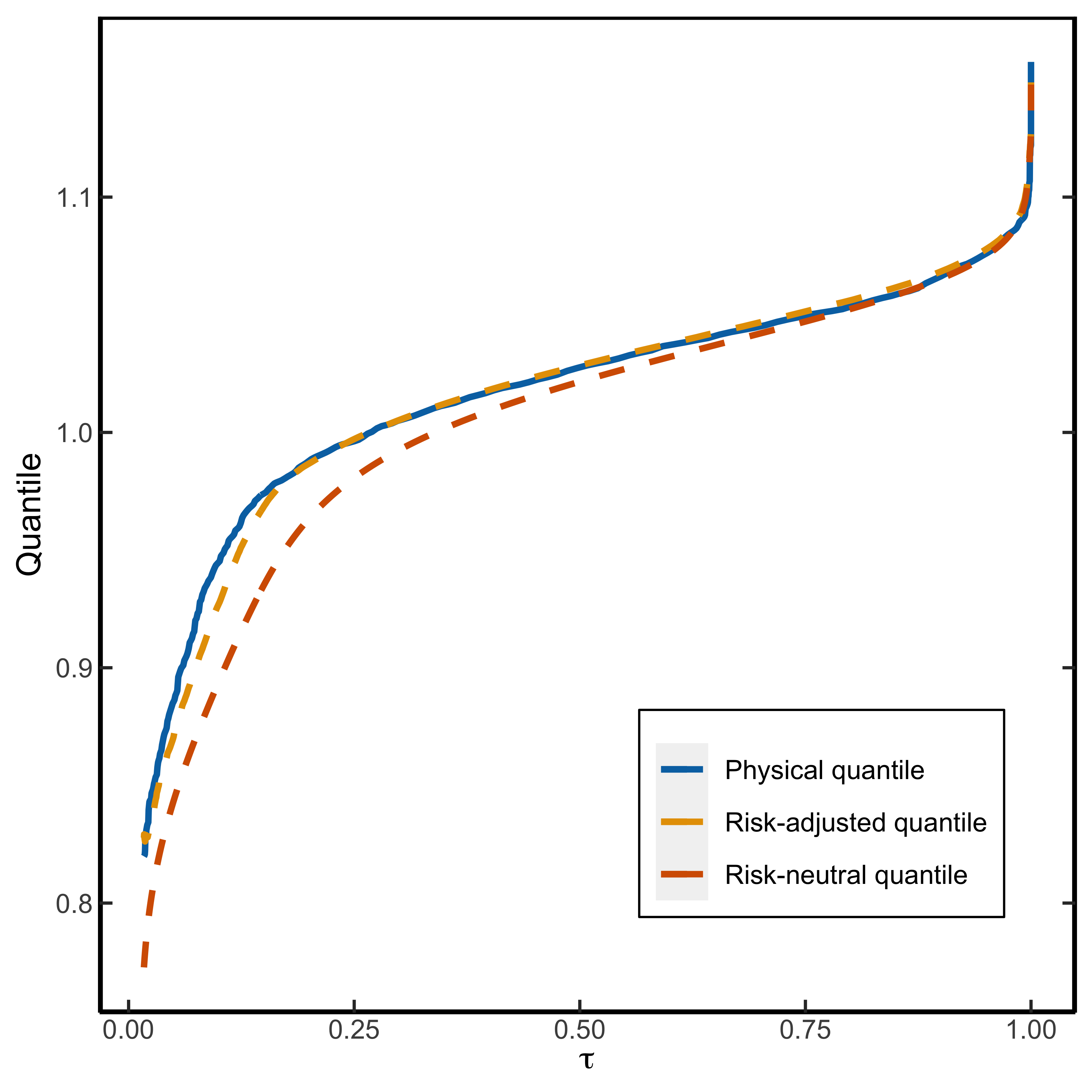}
	\caption{Quantile function,  $\gamma = 2.9$}
\end{subfigure}
	\caption{\textbf{Lower bound with CRRA utility for 90-day returns}. \footnotesize This figure shows the lower bound on $\tau - F_t(\cquant)$ as well as the quantile approximation $\cpquant \approx \cquant + \lrbdp$ in a representative agent model with CRRA utility function, $u(x) = x^{1-\gamma}/(1-\gamma)$, for $\gamma \in \{2.2,2.9\}$. The left panels show the infeasible lower bound $\lrb$, and the true risk-adjustment, $\tau - F_t(\cquant)$. The right panels show the physical, risk-neutral and risk-adjusted quantile functions. The risk-adjusted quantile function uses the infeasible lower bound. The risk-neutral distribution is coming from option data on the S\&P500 on October 28, 2015 with a maturity of 90 days.}
	\label{fig:ra}
\end{figure}

Figure \ref{fig:ra} illustrates the infeasible lower bound as well as the quantile approximation for CRRA utility with different levels of risk aversion.  The risk-neutral distribution is obtained from option data over a 90-day horizon on  October 28, 2015. Panels \ref{fig:lrb_infeas} and \ref{fig:lrb_infeas_high}  show the infeasible lower bound from Theorem \ref{thm:lower_bound} when risk aversion is 2.2 and 2.9 respectively. Consistent with the theorem, the infeasible lower bound is below $\tau - F_t(\cquant)$ in the left-tail, and seems to hold for a large range of $\tau$'s, in particular for all $\tau \le 0.5$. The right panels show the quantile approximation \eqref{eq:q_first_order} based on the infeasible lower bound. We see that the risk-adjusted quantile approximation comes much closer to the physical quantile relative to the risk-neutral quantile function. \\

\section{Crash Probability in Representative Agent Models}\label{app:crash}
In this Section, I derive several results about conditional tail probabilities in representative agent models. Specifically, I show how these probabilities can be calculated using common utility functions and how they are affected by a change in the underlying parameter (comparative statics).  The results do not assume a specific distribution of the market return and generalize some known results in the literature which assume log-normality.

\subsection{Crash Probability with Log Utility}\label{sec:martin}
\citet[Remark 1]{chabi2020conditional} show that their bounds on the equity premium equal the bounds of \citet{martin2017expected} when the representative agent has log preferences. Here, I derive the analogous result for the subjective crash probability of a log investor reported by \citet[Result 2]{martin2017expected}. In our notation, \citet{martin2017expected} shows that 
\begin{equation}\label{eq:temp_mart}
\mathbb{P}_t\lro{\omrkt < \alpha} = \alpha \lr{\mathrm{Put}_{t}'(\alpha S_t) - \frac{\mathrm{Put}_{t}(\alpha S_t)}{\alpha S_t}},
\end{equation}
where $\mathrm{Put}_t'$ is the derivative of the put option price curve seen as a function of the strike. Under log preferences and using \eqref{eq:cov}, it follows that
\begin{align}
&\cprob(\omrkt  < \cquant) = \tau + \frac{1}{\ofree} \covhat{\ind{\omrkt \le \cquant}}{\omrkt} \nonumber \\
&=\tau + \frac{1}{\ofree} \lro{\texpneut\lr{\ind{\omrkt \le \cquant} \omrkt } - \texpneut(\omrkt)\texpneut\left(\ind{\omrkt \le \cquant}\right)} \nonumber \\
&= \frac{1}{\ofree} \texpneut\lr{\ind{\omrkt \le \cquant} \omrkt}. \label{eq:7.16}
\end{align}
The result now follows upon substituting $\tilde{Q}_\tau = \alpha$, since \citet{martin2017expected} shows that \eqref{eq:7.16} equals the right hand side of \eqref{eq:temp_mart}.

\subsection{Crash Probability with CRRA utility}\label{sec:CRRA}
I now consider the case in which the representative agent has constant relative risk aversion (CRRA) utility, $u(x) = x^{1-\gamma}/(1-\gamma)$, where $\gamma$ is the relative risk aversion parameter. First, I show that the excess market return is non-decreasing in $\gamma$ \emph{regardless} of the distribution of the market return.\footnote{\citet{cochrane2009asset} derives this result when the distribution is lognormal.} The proof uses the following lemma, which is a special case of the FKG inequality \citep[Theorem 1.3]{hsu1999probability}.

\begin{lemma}[Chebyshev sum inequality]\label{lemma:cheby}
Let $X$ be a random variable and let $g,h$ both be non-increasing or non-decreasing. Then,
\begin{equation*}
\mathbb{E}\lro{g(X) h(X)} \ge \mathbb{E}\lro{g(X)} \mathbb{E}\lro{h(X)}.
\end{equation*}
The inequality is reversed if one is non-increasing and the other is non-decreasing.
\end{lemma}
\begin{proof}
Let $X_1,X_2$ be \iid copies of $X$ and assume that $g,h$ are non-decreasing. It follows that
\begin{equation}\label{eq:cheby}
\lro{g(X_1) - g(X_2)}\lro{h(X_1) - h(X_2)} \ge 0.
\end{equation}
Taking expectations on both sides completes the proof. The same proof goes through if $g,h$ are non-increasing. If one is non-increasing and the other is non-decreasing, the inequality in \eqref{eq:cheby} is reversed. 
\end{proof}

\begin{proposition}\label{lemma:crra}
	Assume that a representative investor has CRRA utility, with $\gamma \ge 0$ and $\cexp{\omrkt^{\gamma+1} \log \omrkt} < \infty$. Then, $\cexp{\omrkt} - \ofree$, is non-decreasing in $\gamma$.
\end{proposition}
\begin{Remark}
I suppress the dependence of the physical expectation on $\gamma$ in the notation for readability. 
\end{Remark}
\begin{proof}
	According to \citet[Equation (53)]{chabi2020conditional}, we have
	\begin{equation*}
	\cexp{\omrkt} - \ofree = \frac{\qcexp{\omrkt^{\gamma+1}}}{\qcexp{\omrkt^{\gamma}}}-\ofree \eqqcolon g(\gamma).
	\end{equation*}
	It is enough to show that $g'(\gamma) \ge 0$ for $\gamma \ge 0$. Taking first order conditions, we need to show that
	\begin{equation}\label{eq:crra}
	\qcexp{\omrkt^{\gamma+1} \log \omrkt }\qcexp{\omrkt^\gamma} \ge \qcexp{\omrkt^{\gamma+1}} \qcexp{\omrkt^\gamma \log \omrkt}.
	\end{equation}
	Introduce another probability measure $\mathbb{P}^*$, defined by
	\begin{equation}\label{eq:rn}
	\scexp{Z} \coloneqq \frac{\qcexp{Z \omrkt^{\gamma}}}{\qcexp{\omrkt^\gamma}}.
	\end{equation}
	We can rewrite \eqref{eq:crra} into
	\begin{equation}\label{eq:percolation}
	\scexp{\omrkt^\gamma \log \omrkt} \ge \scexp{\omrkt^\gamma} \scexp{\log \omrkt}.
	\end{equation}
	Inequality \eqref{eq:percolation} now follows from Lemma \ref{lemma:cheby}.
\end{proof}
I mimic the steps above to show that the physical distribution differs more from the risk-neutral distribution at every point in the support, whenever risk aversion is increasing. As before, the dependence of the physical measure on $\gamma$ is omitted. 

\begin{proposition}
Assume that a representative investor has CRRA utility, with $\gamma \ge 0$ and $\cexp{\omrkt^{\gamma} \log \omrkt} < \infty$, then $F_t(x)$ is non-increasing in $\gamma$. In particular, $\tau - F_t(\cquant)$ is non-decreasing in $\gamma$.
\end{proposition} 
\begin{proof}
I start from the relation
\begin{equation*}
F_t(x) = \qcexp{\frac{\omrkt^\gamma}{\qcexp{\omrkt^\gamma}} \ind{\omrkt \le x} }.
\end{equation*}	
From first order conditions, we need to show that	
\begin{multline*}
\qcexp{\log(\omrkt)\ind{\omrkt\le x} \omrkt^\gamma} \qcexp{\omrkt^\gamma} \le\\ \qcexp{\omrkt^\gamma \ind{\omrkt \le x}} \qcexp{\log(\omrkt)\omrkt^\gamma}.
\end{multline*}
Using the same change of measure as in \eqref{eq:rn}, we obtain the equivalent statement
\begin{equation*}
\scexp{\log(\omrkt) \ind{\omrkt \le x}} \le \scexp{\ind{\omrkt \le x}} \scexp{\log \omrkt}.
\end{equation*}
This inequality holds, since $\log(y)$ and $\ind{y \le x}$ are respectively increasing and non-increasing in $y$, hence the result follows from Lemma \ref{lemma:cheby}. Using the substitution $x \to \cquant$, it follows that $\tau - F_t(\cquant)$, is non-decreasing in $\gamma$. 
\end{proof}

\subsection{Exponential utility}\label{sec:Exp}
Here, I assume that the representative agent has exponential utility, $u(x) = 1-e^{-\gamma^* x}$, where $\gamma^*$ is the absolute risk aversion. According to \citet[Equation (55)]{chabi2020conditional}, the following expression for the equity premium obtains
\begin{equation*}
\cexp{\omrkt} - \ofree = \frac{\qcexp{\omrkt e^{\gamma \omrkt}}}{\qcexp{e^{\gamma \omrkt}}} - \ofree,
\end{equation*}
where $\gamma = \gamma^* W_t$ is relative risk aversion and $W_t$ represents the agent's wealth at time $t$. Since there is a one-to-one relation between $\gamma$ and $\gamma^*$, it follows from the results in Section \ref{sec:CRRA} that the equity premium is increasing in $\gamma^*$ and so is the distance between the physical and risk-neutral distribution, as measured by: $\tau - F_t(\cquant)$.

\section{Lower Bound in the Data and Robustness}

\subsection{Lower Bound in the Data}
In the empirical application, I compute the lower bound, $\lrbdp = \lrb/\tilde{f}_t(\cquant)$, for 30-, 60-, and 90-day returns. Table \ref{tab:summary_risk_adjustment} contains summary statistics of $\lrbdp$. The lower bound is right-skewed and is most significant for the 5th and 10th percentile. Moreover, over the 30 day horizon, it can spike up to 25\% and averages to about 1\% in the far left-tail.

\begin{table}[!htb]
\captionsetup{width=12cm}
\centering
\caption{\textbf{Summary statistics of lower bound}}
\label{tab:summary_risk_adjustment}
\begin{adjustbox}{max width=\textwidth}
\begin{threeparttable}
\begin{tabular}{lcccccc}
\toprule
\midrule
Horizon & $\tau$ & Mean & Median & Std. dev. & Min & Max \\ \midrule
30 days & 0.05 & 0.92 & 0.63 & 1.07 & 0.08 & 24.38 \\
& 0.1 & 0.70 & 0.45 & 0.87 & 0.06 & 12.22 \\
& 0.2 & 0.47 & 0.25 & 0.74 & 0.04 & 10.93 \\
&  &  &  &  &  &  \\
60 days & 0.05 & 1.81 & 1.31 & 1.67 & 0.10 & 19.23 \\
& 0.1 & 1.71 & 1.19 & 1.66 & 0.25 & 19.89 \\
& 0.2 & 1.14 & 0.69 & 1.50 & 0.12 & 23.57 \\
&  &  &  &  &  &  \\
90 days & 0.05 & 2.65 & 2.02 & 2.02 & 0.02 & 18.63 \\
& 0.1 & 2.86 & 2.12 & 2.32 & 0.04 & 24.47 \\
& 0.2 & 1.97 & 1.22 & 2.33 & 0.26 & 28.92 \\ \bottomrule
\end{tabular}%
\begin{tablenotes}
\footnotesize
\item \textit{Note}: This table reports summary statistics of the lower bound, $\lrbdp = \lrb/\tilde{f}_t(\cquant)$, in \eqref{eq:predict_quantile} at different time horizons and different quantile levels over the sample period 2003-2021. All statistics are in percentage point.
\end{tablenotes}
\end{threeparttable}
\end{adjustbox}
\end{table}

\subsection{Robustness of the Lower Bound and Risk-neutral Quantile}\label{app:robust_lb}
The lower bound, $\lrbdp$, tries to capture the difference between the physical and risk-neutral quantile in the left-tail. What are some other measures that are available at a daily frequency and contain information about the quantile wedge? One candidate is the VIX index, which is defined as
\begin{equation*}
\mathrm{VIX}_t^2 = \frac{2\ofree}{N} \lr{\int_0^{F_t} \frac{1}{K^2} \mathrm{Put}_t(K) \diff K + \int_{F_t}^\infty \frac{1}{K^2} \mathrm{Call}_t(K) \diff K},
\end{equation*}
where $N$ is the time to expiration, $F_t$ is the forward price on the S\&P500, and $\mathrm{Put}_t(K)$ (resp. $\mathrm{Call}_t(K)$) is the put (resp. call) option price on the S\&P 500 with strike $K$. \citet{martin2017expected} shows that VIX measures risk-neutral entropy
\begin{equation*}
\mathrm{VIX}_t^2 = \frac{2}{N} \tilde{L}_t\lro{\frac{\omrkt}{\ofree}},
\end{equation*}
where entropy is defined as $\tilde{L}_t(X) \coloneqq \log \qcexp{X} - \qcexp{\log X}$. Entropy, just like variance, is a measure of spread in the distribution. However, entropy places more weight on left-tail events than variance, since entropy places more weight on out-of-the money puts. As such, VIX is a good candidate to explain potential differences between $Q_{t,\tau}$ and $\cquant$. Second, the Chicago Board Options Exchange provides daily data on VIX for the 30 day horizon. \\

Table \ref{tab:robustness_lrb_vix} shows the result of the quantile regression
\begin{equation}\label{eq:VIX_inc}
Q_{t,\tau}(\omrkt) - \cquant(\omrkt) = \beta_0(\tau) + \beta_1(\tau)\lrbdp + \beta_{\mathrm{VIX}}(\tau)\mathrm{VIX}_t.
\end{equation}
We see that $\beta_{\mathrm{VIX}}$ is marginally significant in the left-tail. In contrast, $\beta_1(\tau)$ is even more significant compared to Table \ref{tab:robustness_lrb}. Furthermore, the explanatory power of the model that only includes VIX is lower compared to the model that only includes $\lrbdp$ (Table \ref{tab:robustness_lrb}). \\

\begin{table}[!htb]
\captionsetup{width=12cm}
\centering
\caption{\textbf{Quantile regression using Lower Bound and VIX}}
\label{tab:robustness_lrb_vix}
\begin{adjustbox}{max width=\textwidth}
\begin{threeparttable}
\begin{tabular}{lcccc|c}
\toprule
\midrule
& $\hat{\beta}_0(\tau)$ & $\hat{\beta}_1(\tau)$ & $\hat{\beta}_{\text{VIX}}(\tau)$  & $R^1(\tau)[\%]$ & $\underset{(\text{VIX only})}{R^1(\tau)[\%]}$  \\ 
& & & & & \\
$\tau = 0.05$ & $\underset{( 1.889 )}{\text{ -0.20 }}$ & $\underset{( 0.319 )}{ 10.09 }$ & $\underset{( 0.130 )}{\text{  -0.30 }}$ & 6.34 & 5.51 \\
$\tau = 0.1$ & $\underset{( 1.313 )}{\text{  -0.35} }$ & $\underset{( 0.302 )}{  5.06 }$ & $\underset{( 0.089 )}{\text{  -0.22 }}$ & 3.41 & 2.84 \\
$\tau = 0.2$ & $\underset{( 0.955 )}{\text{  -0.28} }$ & $\underset{( 0.256 )}{  3.62 }$ & $\underset{( 0.068 )}{\text{  -0.25 }}$ & 0.61 & 0.18 \\
\bottomrule
\end{tabular}%
\begin{tablenotes}
\footnotesize
\item \textit{Note}: This table reports the QR estimates of   \eqref{eq:VIX_inc} over the 30-day horizon. The sample period is 2003-2021, standard errors are shown in parentheses and calculated using SETBB with a block length equal to the forecast horizon. $R^1(\tau)$ denotes the goodness-of-fit measure \eqref{eq:R1tau}. The last column denotes the goodness-of-fit in the model that only uses VIX as covariate. The standard error and point estimate of $\beta_0$ is multiplied by 100 for readability.
\end{tablenotes}
\end{threeparttable}
\end{adjustbox}
\end{table}

\begin{figure}[!htb]
	\centering
	\begin{subfigure}[b]{0.3\textheight}
		\centering
		\includegraphics[width=\textwidth]{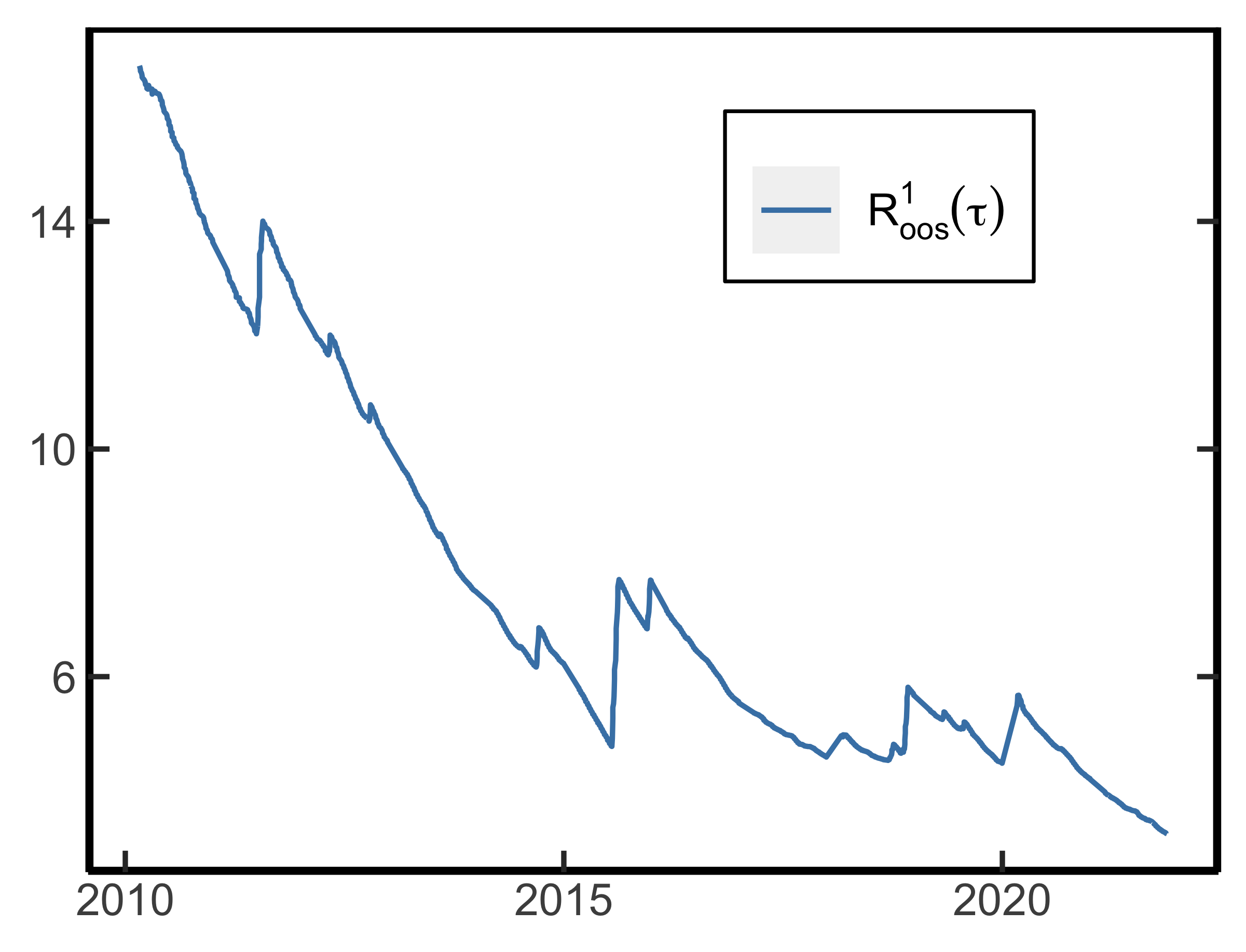}
		\caption{$\tau = 0.05$}
	\end{subfigure}
	\begin{subfigure}[b]{0.49\textwidth}
		\centering
		\includegraphics[width=\textwidth]{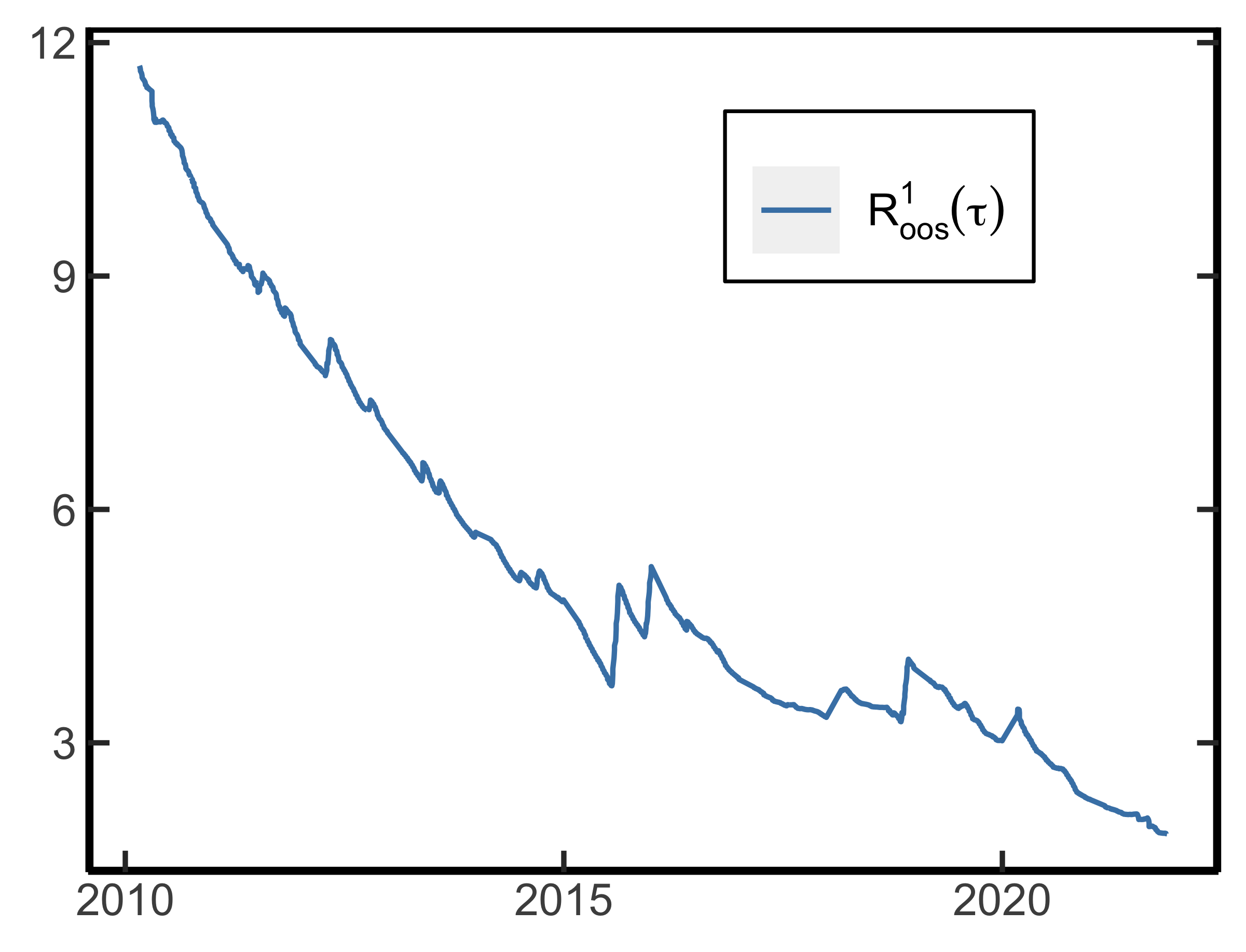}
		\caption{$\tau = 0.1$}
	\end{subfigure}
	\hfill
	\begin{subfigure}[b]{0.49\textwidth}
		\centering
		\includegraphics[width=\textwidth]{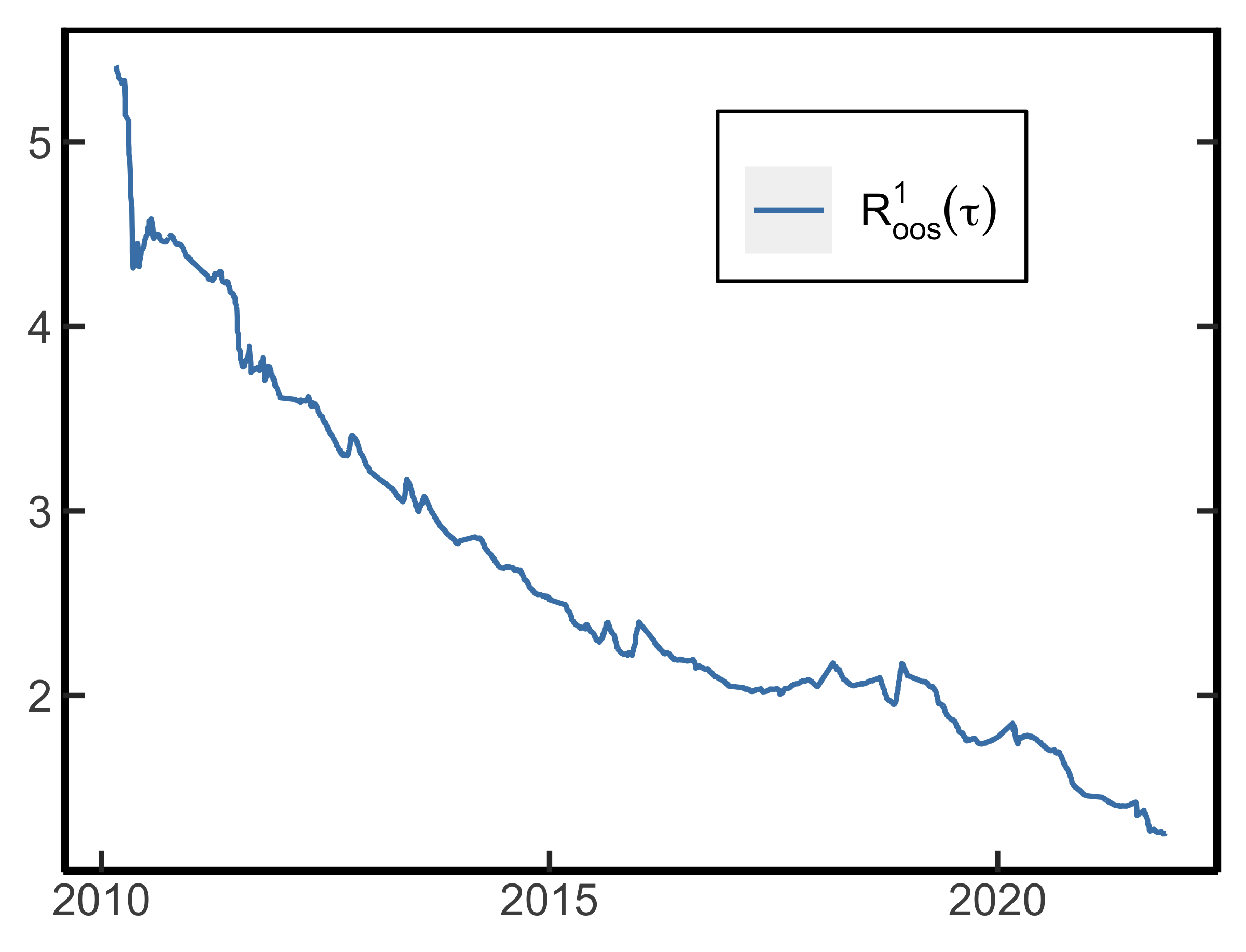}
		\caption{$\tau = 0.2$}
	\end{subfigure}
	\caption{\textbf{Out-of-sample forecast using risk-adjusted quantile with VIX benchmark}. \footnotesize This figure shows the cumulative out-of sample $R^1(\tau)$, defined as $R_{oos}^1(\tau) = 1-\sum_{t=500}^T \rho_\tau(\omrkt - \hat{Q}_{t,\tau})/\sum_{t=500}^T \rho_\tau(\omrkt - \hat{Q}_{t,\tau}^\text{VIX})$, where $\hat{Q}_{t,\tau} = \cquant + \lrbdp$, $\hat{Q}_{t,\tau}^\text{VIX} = \hat{\beta}_0(\tau) + \hat{\beta}_1(\tau) \cdot \text{VIX}_t$, and $\hat{\beta}_0(\tau),\hat{\beta}_1(\tau)$ are the regression estimates from a quantile regression of $\omrkt$ on $\text{VIX}_t$, using data only up to time $t$. The horizon is 30 days and the QR estimates are dynamically updated using an expanding window over the period 2003--2021. The initial sample uses 500 observations.}
	\label{fig:R_oos_left_tail}
\end{figure}

As a second robustness check, I consider how well the direct quantile forecast, $\hat{Q}_{t,\tau} = \cquant + \lrbdp$, compares to the VIX forecast. Since $\hat{Q}_{t,\tau}$ does not require any parameter estimation, this exercise is a measure of out-of-sample performance. However, VIX does not directly measure $Q_{t,\tau}$ and hence I use an expanding window to obtain the VIX benchmark: $\hat{Q}_{t,\tau}^\text{VIX} \coloneqq \hat{\beta}_0(\tau) + \hat{\beta}_1(\tau)  \text{VIX}_t$. Finally, I use the following out-of-sample metric to compare both forecasts
\begin{equation*}
R_{oos}^1(\tau) = 1-\sum_{t=500}^T \rho_\tau(\omrkt - \hat{Q}_{t,\tau})/\sum_{t=500}^T \rho_\tau(\omrkt - \hat{Q}_{t,\tau}^\text{VIX}).
\end{equation*}
Notice that $R_{oos}^1(\tau) > 0$, if $\hat{Q}_{t,\tau}$ attains a lower error than $\hat{Q}_{t,\tau}^\text{VIX}$. This exercise is more ambitious, since $\hat{Q}_{t,\tau}^\text{VIX}$ makes use of in-sample information. Nonetheless, Figure \ref{fig:R_oos_left_tail} shows that $\hat{Q}_{t,\tau}$ outperforms the VIX predictor at all percentiles. \\

\begin{figure}[!htb]
	\centering
	\begin{subfigure}[b]{0.49\textwidth}
		\centering
		\includegraphics[width=\textwidth]{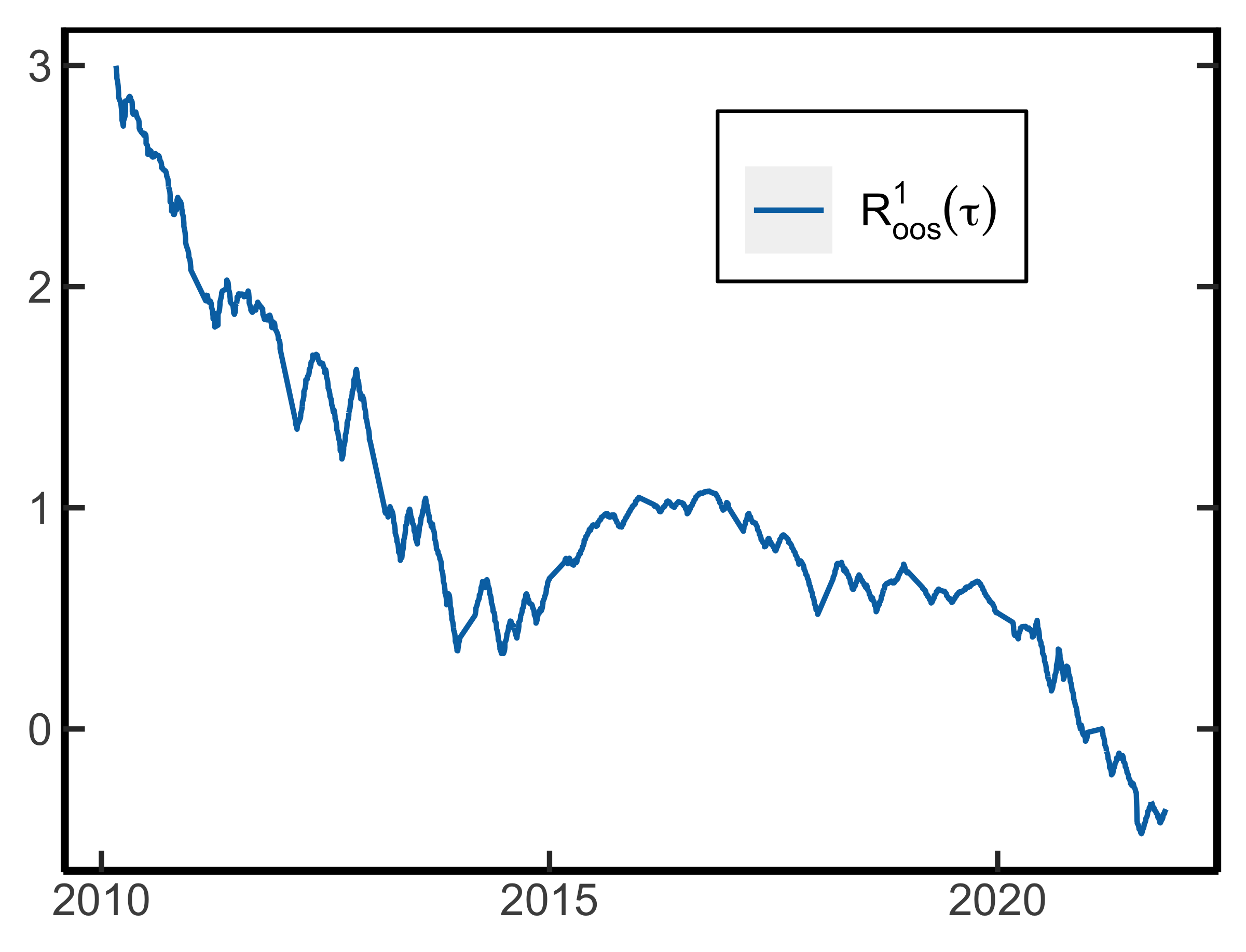}
		\caption{$\tau = 0.5$}
	\end{subfigure}
	\hfill
	\begin{subfigure}[b]{0.49\textwidth}
		\centering
		\includegraphics[width=\textwidth]{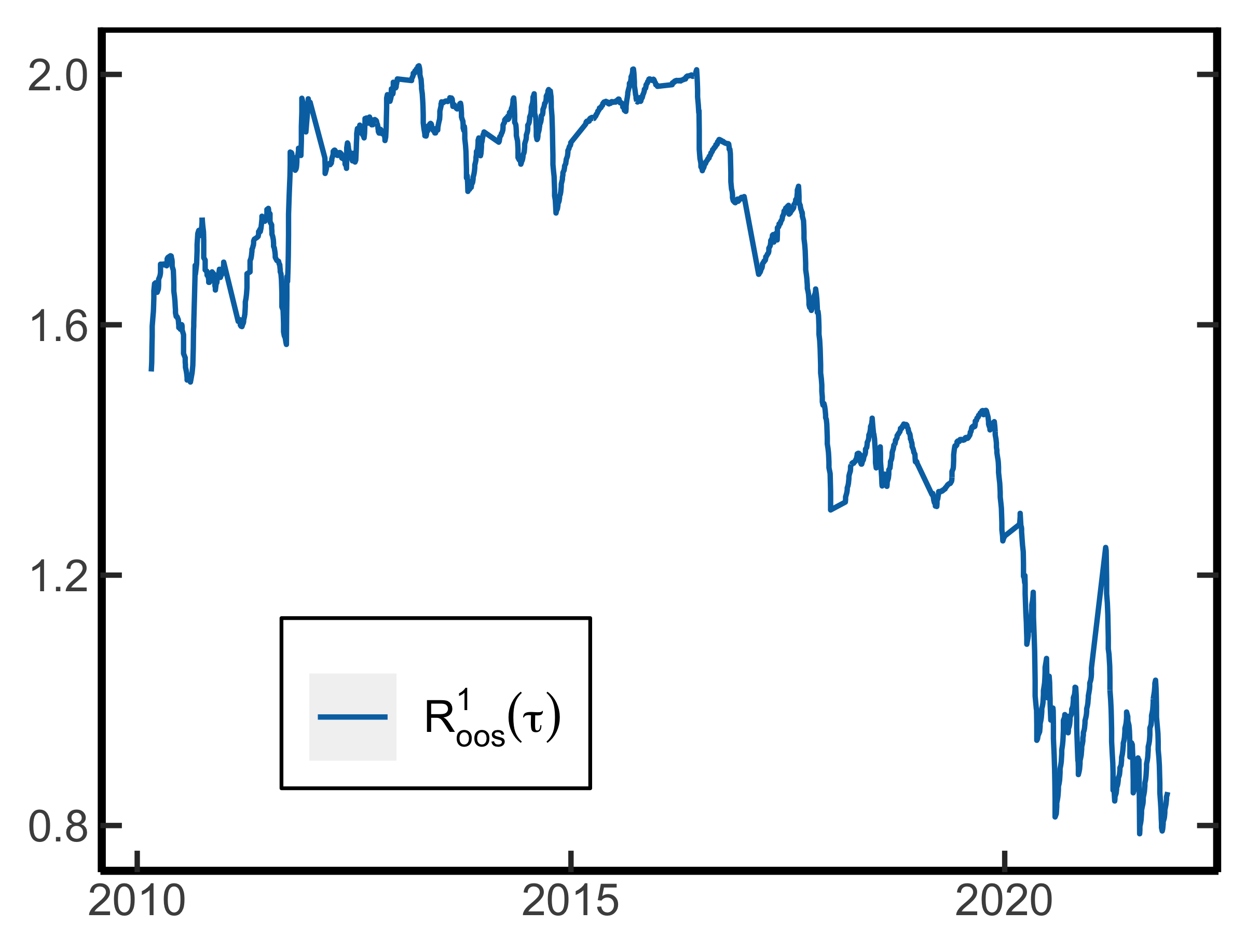}
		\caption{$\tau = 0.8$}
	\end{subfigure}
	\begin{subfigure}[b]{0.49\textwidth}
		\centering
		\includegraphics[width=\textwidth]{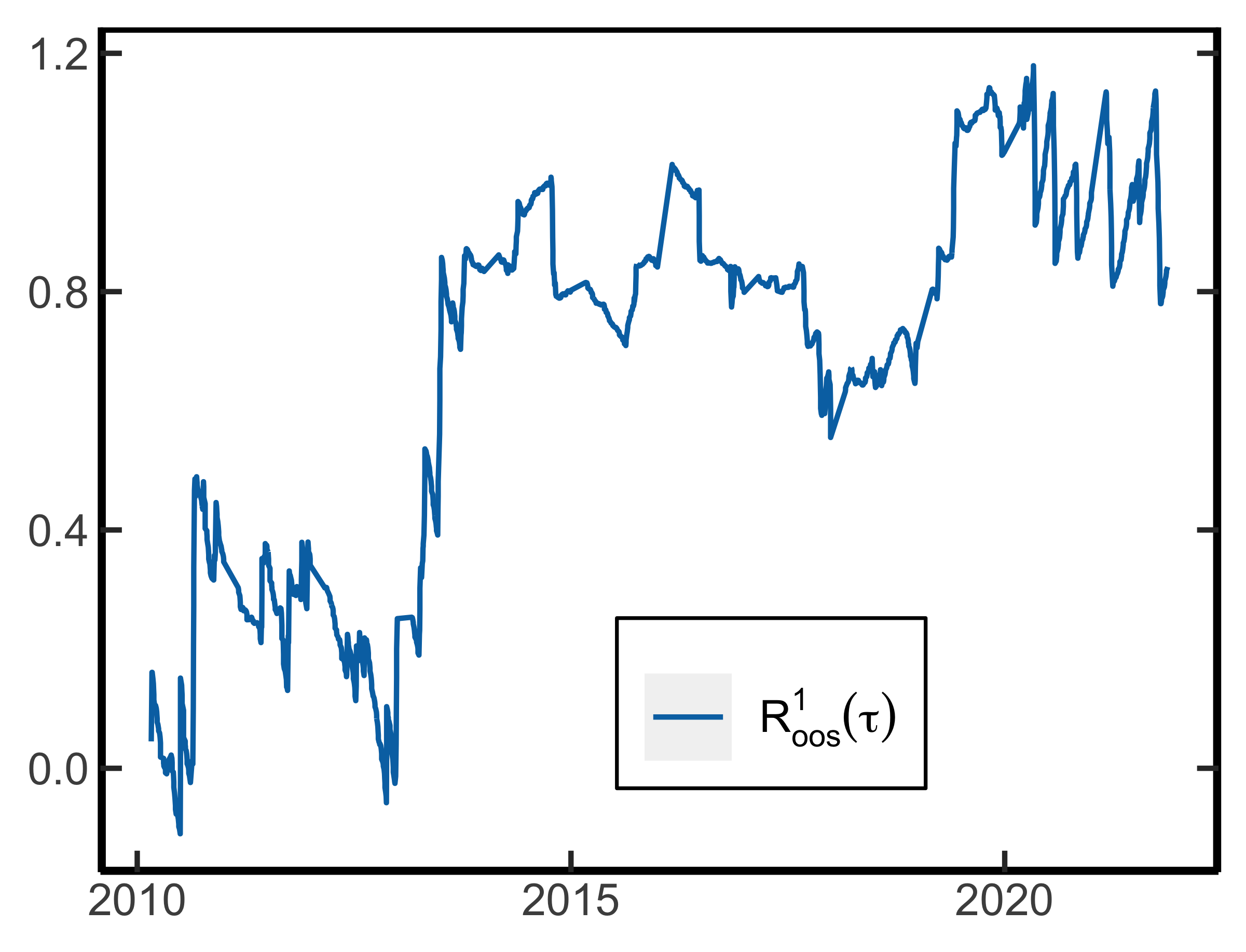}
		\caption{$\tau = 0.9$}
	\end{subfigure}
	\hfill
	\begin{subfigure}[b]{0.49\textwidth}
		\centering
		\includegraphics[width=\textwidth]{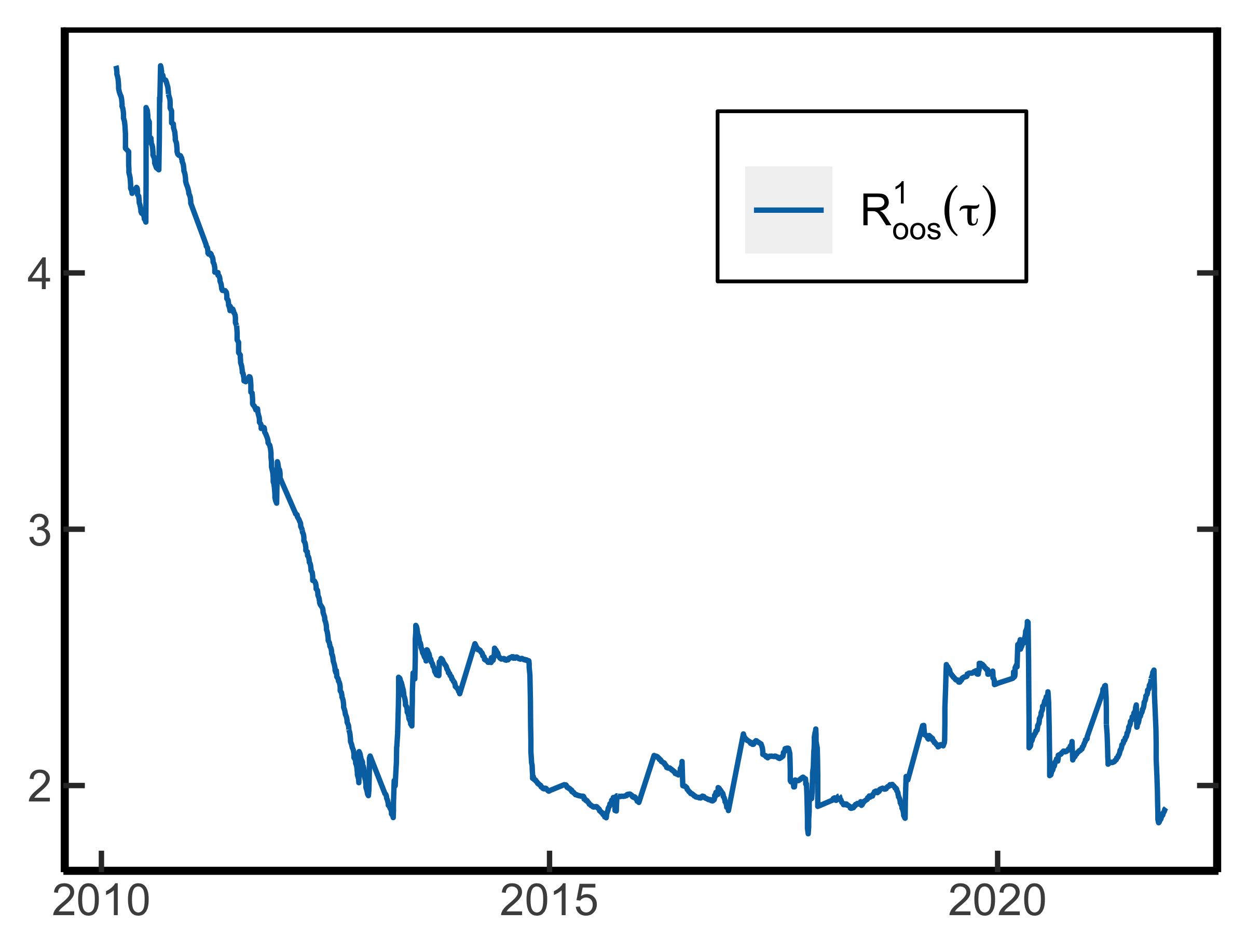}
		\caption{$\tau = 0.95$}
	\end{subfigure}
	\caption{\textbf{Out-of-sample forecast using risk-neutral quantile with VIX benchmark}. \footnotesize This figure shows the cumulative out-of sample $R^1(\tau)$, defined as $R_{oos}^1(\tau) = 1-\sum_{t=500}^T \rho_\tau(\omrkt - \tilde{Q}_{t,\tau})/\sum_{t=500}^T \rho_\tau(\omrkt - \hat{Q}_{t,\tau}^\text{VIX})$, where $\hat{Q}_{t,\tau}^\text{VIX} = \hat{\beta}_0(\tau) + \hat{\beta}_1(\tau) \cdot \text{VIX}_t$, and $\hat{\beta}_0(\tau),\hat{\beta}_1(\tau)$ are the regression estimates from a quantile regression of $\omrkt$ on $\text{VIX}_t$, using data only up to time $t$. The horizon is 30 days and the QR estimates are dynamically updated using an expanding window over the period 2003--2021. The initial sample uses 500 observations.}
	\label{fig:R_oos}
\end{figure}

Figure \ref{fig:R_oos} performs a similar exercise in the right-tail, but using $\cquant$ instead of $\hat{Q}_{t,\tau}$, since Table \ref{tab:only.rn.quantile} shows that the risk-neutral quantile is a good approximation to $Q_{t,\tau}$ in the  right-tail. We see that $\cquant$ outperforms $\hat{Q}_{t,\tau}^\text{VIX}$ at all quantile levels. Hence, the risk-neutral approximation in the  right-tail is more accurate than using the in-sample VIX measure.

\subsection{Lower bound in Black-Scholes model}\label{app:bs_sim_evidence}
This section illustrates the accuracy of the quantile approximation in \eqref{eq:q_first_order} in a discretized version of the Black-Scholes model with time-varying parameters. Specifically, I assume the following DGP
\begin{align}\label{eq:model_bs}
\omrkt &= \exp\lro{(\mu_t- \frac{1}{2}\sigma_t^2)N + \sigma_t \sqrt{N} Z_{t+N}}, \quad Z_{t+N} \sim \mathcal{N}(0,1) \\
\sigma_t &\sim \unif{0.05}{0.35}\nonumber\\
\mu_t &\sim \unif{-0.02}{0.2}\nonumber.
\end{align}
The return distribution under risk-neutral dynamics is given by 
\begin{align}\label{eq:rn_return}
\tilde{R}_{m,t\to N} &= \exp\lro{(r_t- \frac{1}{2}\sigma_t^2)N + \sigma_t \sqrt{N} Z_{t+N}}\\
r_t &\sim \unif{0}{0.03}.
\end{align}
Finally, assume that all parameters are \iid over time and that options are priced according to the Black-Scholes formula, conditional on time $t$. In this setup, it is fruitless to use historical data to predict future quantiles, since parameters change unpredictably over time. We use $N = 30$ to mimic the monthly application in this paper.  It is assumed that the risk-neutral quantile function is known at the start of period $t$, as it is in the real world, by the result of \citet{breeden1978prices}. I use the risk-neutral quantile function to calculate $\lrb$ at time $t$. Then, following the approximation in \eqref{eq:q_first_order}, the physical quantile function is estimated by
\begin{equation}\label{eq:q_predict}
\widehat{Q}_{t,\tau} = \cquant + \lrbdp.
\end{equation}
We take 3,000 return observations that are generated according to \eqref{eq:model_bs}. This exercise is repeated 1,000 times. To assess the accuracy of the approximation in \eqref{eq:q_predict}, I use several metrics. For every sample, I estimate a quantile regression of the form
\begin{equation*}
Q_\tau(\risky) = \beta_0(\tau) + \beta_1(\tau) \widehat{Q}_{t,\tau},
\end{equation*} 
where $\widehat{Q}_{t,\tau}$ comes from \eqref{eq:q_predict}. The first two columns in Table \ref{tab:bs_simulation} report the average values of the QR estimates across the 1,000 simulations. The means are rather close to 0 and 1 respectively for all quantiles. If \eqref{eq:q_risk_adjust} is a good approximation, one expects $\cpquant > \widehat{Q}_{t,\tau}$, since $\lrb \le \tau - F_t(\cquant)$. The third column in Table \ref{tab:bs_simulation} shows this happens for the majority of samples. The fourth column shows the correlation between $\cpquant$ and $\widehat{Q}_{t,\tau}$, which is very close to one, and corroborates the view that the approximation is quite accurate. Columns four and five document the percentage of non rejection of $H_0$, which is indeed quite high. The last column considers non rejection of the joint null hypothesis, which is also high except for the 10th percentile. Overall, Table \ref{tab:bs_simulation} suggests that \eqref{eq:q_predict} is a highly accurate predictor of the physical quantile function.

\begin{table}[!htb]
	\captionsetup{width=12cm}
	\centering
	\caption{Simulation results}
	\label{tab:bs_simulation}
	\begin{adjustbox}{max width=\textwidth}
		\begin{threeparttable}
			\begin{tabular}{lccccccc}
				\toprule
				\midrule
				& $\mathbb{E}{\hat{\beta}_0(\tau)}$ & $\mathbb{E}{\hat{\beta}_1(\tau)}$ & $Q > \hat{Q} $  & $\rho(Q,\hat{Q})$ & $ \hat{\beta}_0(\tau)=0$ &  $ \hat{\beta}_1(\tau)=1$ &  $[\hat{\beta}_0(\tau),\hat{\beta}_1(\tau)] = [0,1]$ \\ 
				\cmidrule(lr){2-8}
			$\tau =    0.01 $ & 0.01 & 0.99 & 0.85 & 1 & 0.94 & 0.96 & 0.80 \\
			$\tau =    0.05 $ & -0.03 & 1.04 & 0.69 & 0.99 & 0.90 & 0.89 & 0.66 \\
			$\tau =    0.1 $ & -0.06 & 1.07 & 0.64 & 0.99 & 0.78 & 0.76 & 0.47 \\  \bottomrule
			\end{tabular}%
			\begin{tablenotes}
				\footnotesize
				\item \textit{Note}: $\mathbb{E}{\hat{\beta}_0(\tau)}$ denotes the average QR estimate of $\hat{\beta}_0(\tau)$ and likewise $\mathbb{E}{\hat{\beta}_0(\tau)}$ shows it for $\hat{\beta}_1(\tau)$. $Q > \hat{Q}$ shows the fraction of times the true physical quantile is larger than our predicted quantile. Columns $H_0: \hat{\beta}_0(\tau)=0$ and $H_0: \hat{\beta}_1(\tau)=1$ report the fraction of times the individual null hypotheses $\beta_0(\tau) = 0 , \beta_1(\tau) = 1$ are not rejected. The last column reports the fraction of times the joint null hypothesis is not rejected.
			\end{tablenotes}
		\end{threeparttable}
	\end{adjustbox}
\end{table}

\begin{figure}[!htb]
	\centering
	\includegraphics[width=0.65\textwidth]{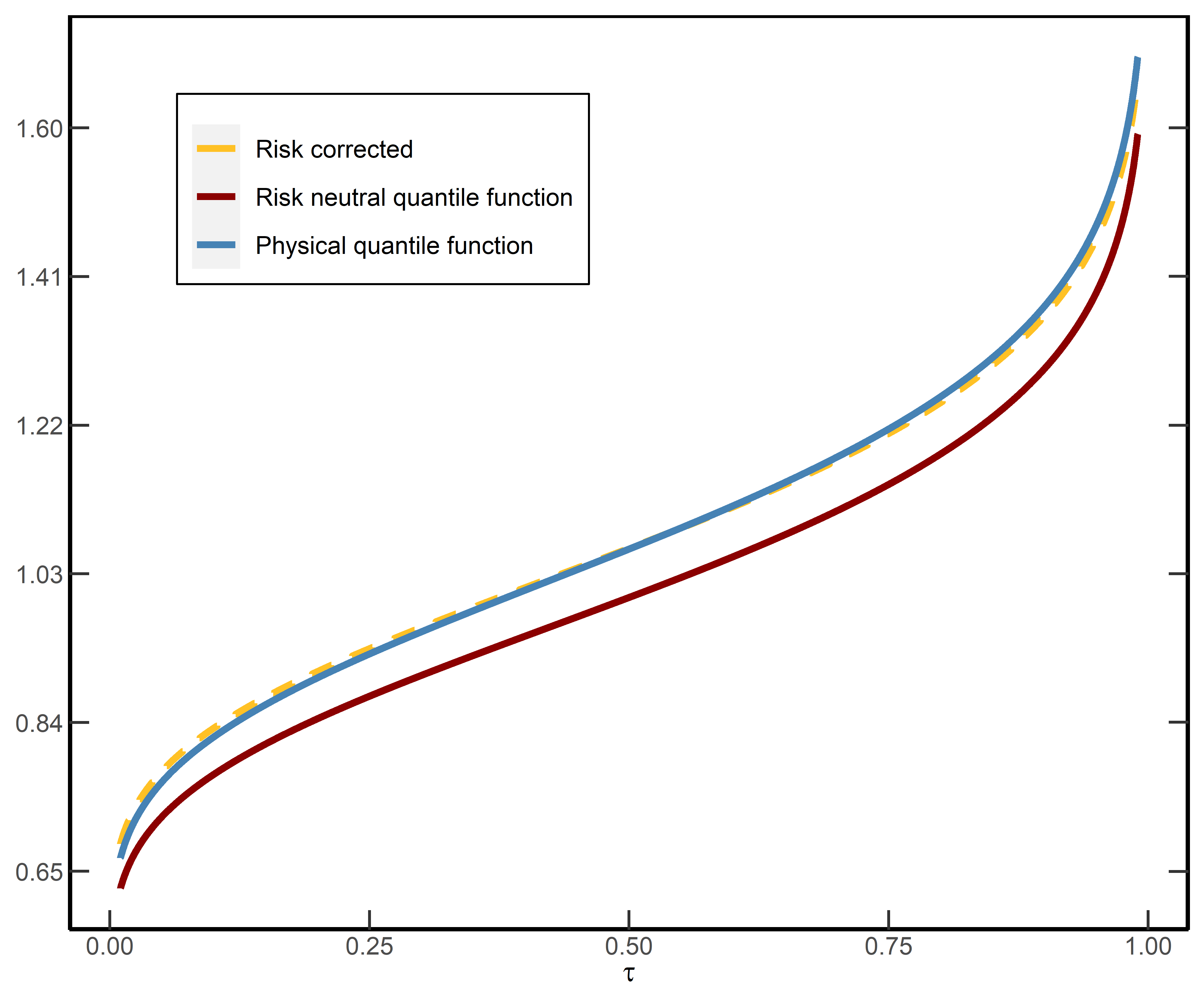}
	\caption{\textbf{Quantile approximation in the Black-Scholes model.} This figure shows that quantile approximation \eqref{eq:q_predict} in the Black-Scholes model with parameters: $\mu = 0.08, r = 0.02, \sigma = 0.2$ over a one-year time horizon.}
	\label{fig:mises}
\end{figure}

\begin{example}
	I illustrate the approximation \eqref{eq:q_risk_adjust} in the Black-Scholes model with fixed parameters: $N=365$ (one year), $\mu = 0.08, r = 0.02, \sigma = 0.2$.\footnote{For illustrative purposes, I use $N = 1$, instead of $N = 1/12$, otherwise the physical quantile function and its approximation are indistinguishable.} We can explicitly calculate $F^{-1}, \tilde{F}^{-1}$ and $\tilde{f}$ owing to the lognormal assumption. Figure \ref{fig:mises} shows the risk-neutral quantile function, the approximation \eqref{eq:q_risk_adjust} and the true physical quantile function.  Observe that the approximation \eqref{eq:q_risk_adjust} is very accurate in this case.
\end{example}

\subsection{Bias in quantile regression}\label{app:bias_quantile}
In the empirical application, we have to estimate $\cquant, \tilde{f}(\cdot)$ and $\lrb$ from market data. Therefore, the estimated coefficients in the quantile regression are biased due to measurement error in the covariate. I present simulation evidence which shows that the bias is small in finite samples. \\

\begin{figure}[!htb]
	\centering
	\includegraphics[width=0.7\textwidth]{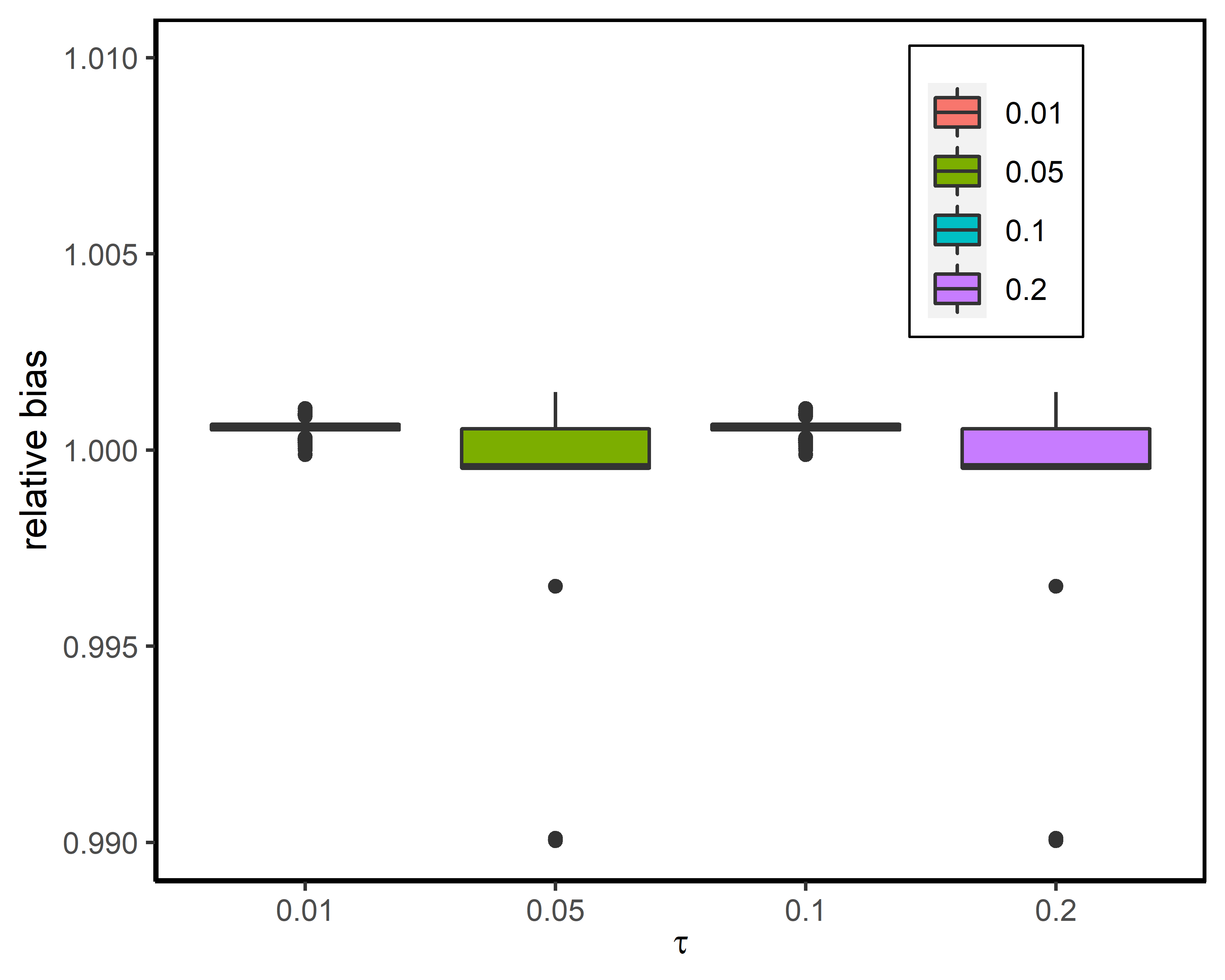}
	\caption{\textbf{Bias in QR resulting from measurement error.} This boxplot shows the relative bias in  the quantile regression estimate as a result of measurement error. }
	\label{fig:bias_box}
\end{figure}

The setup is as follows. I simulate returns according to model \eqref{eq:model_bs} and assume that options are priced according to the \citet{black1973pricing} formula at the start of period $t$. We want to calculate the lower bound for a maturity of 90 days. As in the empirical application, I assume that options with an exact 90 day maturity are not available, but instead we observe options with maturity 85 and 97 days. I generate a total of 1,000 options every time period with maturities randomly sampled from 85 and 97 days.\footnote{So on average there will 500 options with maturity 85 days and 500 with maturity 97 days.} These numbers are roughly consistent with the latter part of the empirical sample. The procedure is repeated for a total of 1,000 time periods. For the entire sample, I compare the estimated and analytical lower bound, which are given by respectively
\begin{align*}
\lrbdp^{e} &\coloneqq \widehat{\tilde{Q}}_{t,\tau} +  \frac{\widehat{\lrb}}{\widehat{\tilde{f_t}(\cquant)}} \\
\lrbdp^{a} &\coloneqq \cquant + \frac{\lrb}{\tilde{f_t}(\cquant)}.
\end{align*}
The hats signify that the risk-neutral quantile, PDF and CDF lower bound are estimated from the available options at time $t$, using the procedure in Appendix \ref{app:risk_neutral_quantile_function}. The terms in $\lrbdp^a$ are obtained from the known analytical expression of the risk-neutral distribution (recall \eqref{eq:rn_return}). I then use QR to estimate the models
\begin{align*}
Q(\risky) &= \hat{\beta}_0(\tau) + \hat{\beta}_{1,e}(\tau) \lrbdp^{e}\\
Q(\risky) &= \hat{\beta}_0(\tau) + \hat{\beta}_{1,a}(\tau) \lrbdp^{a}.
\end{align*}
I use the ratio $\hat{\beta}_{1,e}/\hat{\beta}_{1,a}$ to measure the relative bias in the sample. This experiment is repeated 500 times to get a distribution of the relative bias. Figure \ref{fig:bias_box} shows boxplots of the bias for several quantiles. We see that the relative bias is very small and centered around 1. Hence, the error in measurement problem resulting from estimating the lower bound is limited in this case.  

\section{Robustness of Quantile Regression}

\subsection{Linear versus Non-linear Model: Out-of-Forecasting Accuracy}\label{app:oos_linear}
This section explores alternative specifications to the linear quantile model presented in \eqref{eq:tail}, focusing only on 30-day returns. The findings for longer time horizons are very similar and omitted for parsimony. Specifically, I consider the addition of higher-order terms to the linear model, such as:
\begin{equation}\label{eq:tail_nl}
\cpquant = \beta_0(\tau) + \beta_1(\tau) \cquant + \beta_2(\tau)\cquant^2. 
\end{equation}

 To evaluate the performance of the non-linear model in \eqref{eq:tail_nl} vs.\ the linear model in \eqref{eq:tail}, I recursively estimate the model parameters based on an expending window, starting at January 2, 2003. The first sub-sample ends at August 15, 2012 and I increase the sample size on a monthly basis. For each sub-sample, I calculate the out-of-sample forecasting accuracy  using the formula:
\begin{equation}\label{eq:oos_loss}
\frac{1}{\# t}\sum_t \rho_\tau(\omrkt - \hat{Q}_{t,\tau}),
\end{equation}
where $\hat{Q}_{t,\tau}$ is the predicted physical quantile based on the parameters estimated from the sub-sample. The summation includes all dates that are at least one month ahead of the end of the sub-sample period.\\

\begin{figure}[!htb]
	\centering
	\begin{subfigure}[b]{0.32\textwidth}
		\centering
		\includegraphics[width=\textwidth]{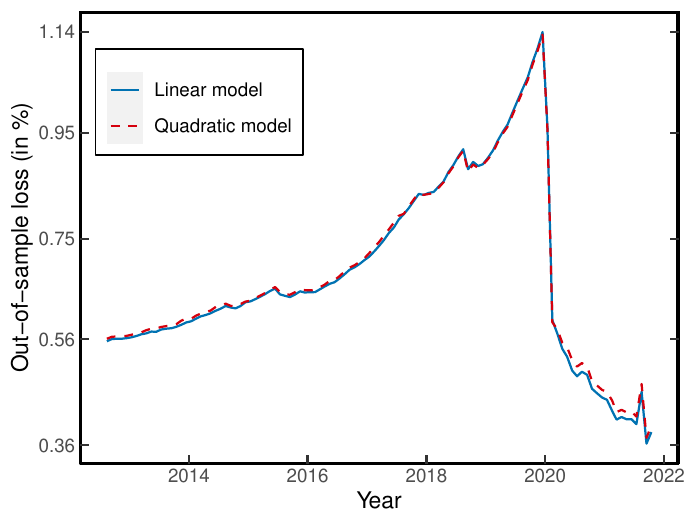}
		\caption{5th percentile}
		\label{fig:loss5}
	\end{subfigure}
	\begin{subfigure}[b]{0.32\textwidth}
	\centering
	\includegraphics[width=\textwidth]{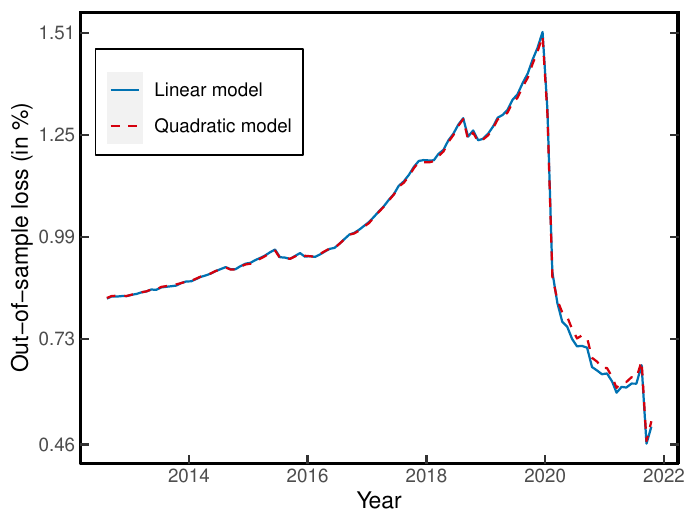}
	\caption{10th percentile}
	\label{fig:loss10}
\end{subfigure}
	\begin{subfigure}[b]{0.32\textwidth}
		\centering
		\includegraphics[width=\textwidth]{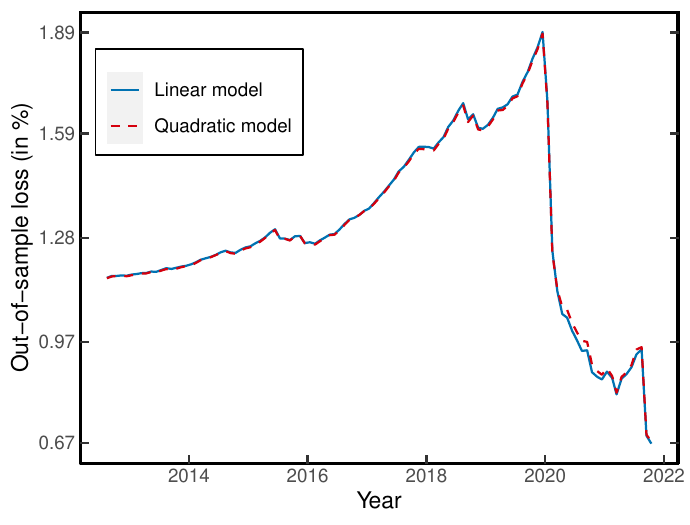}
		\caption{20th percentile}
		\label{fig:loss20}
	\end{subfigure}
	\begin{subfigure}[b]{0.32\textwidth}
	\centering
	\includegraphics[width=\textwidth]{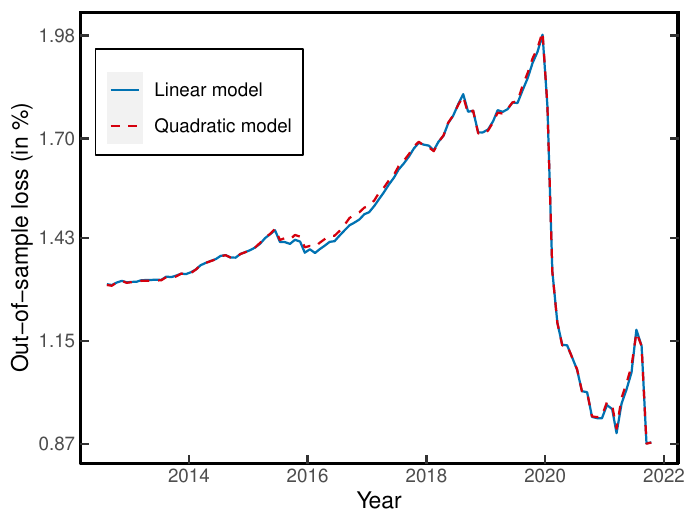}
	\caption{30th percentile}
	\label{fig:loss30}
\end{subfigure}
	\begin{subfigure}[b]{0.32\textwidth}
	\centering
	\includegraphics[width=\textwidth]{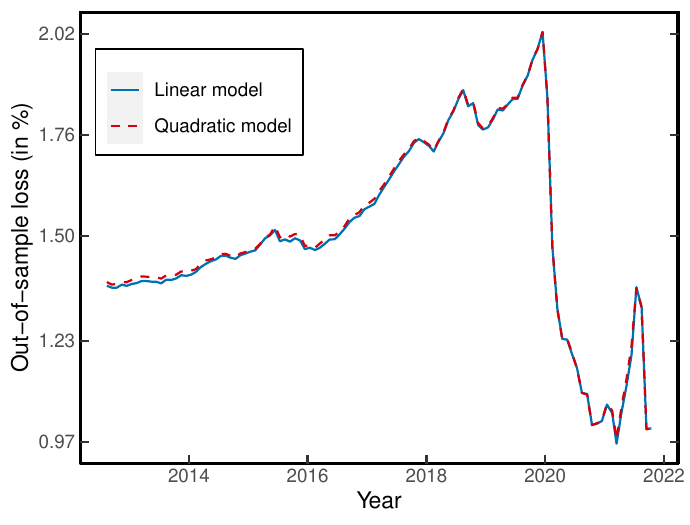}
	\caption{40th percentile}
	\label{fig:loss40}
\end{subfigure}
	\begin{subfigure}[b]{0.32\textwidth}
	\centering
	\includegraphics[width=\textwidth]{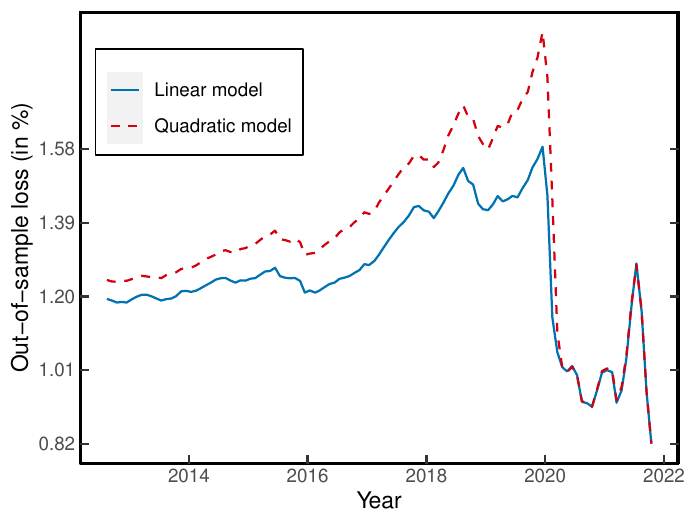}
	\caption{60th percentile}
	\label{fig:loss60}
\end{subfigure}
	\begin{subfigure}[b]{0.32\textwidth}
	\centering
	\includegraphics[width=\textwidth]{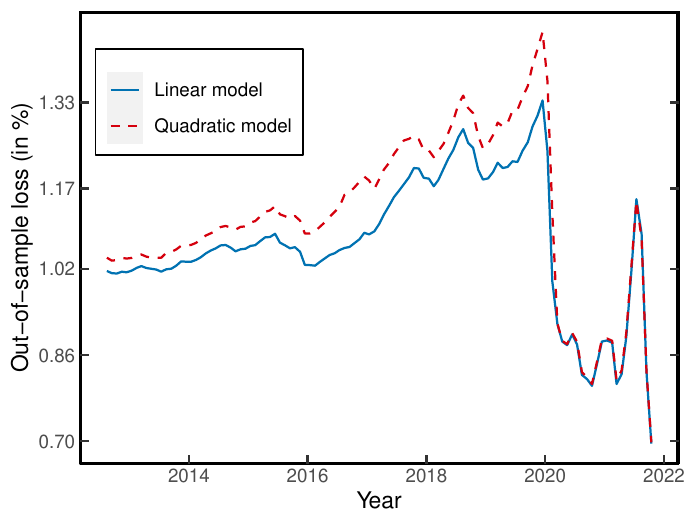}
	\caption{70th percentile}
	\label{fig:loss70}
\end{subfigure}
	\begin{subfigure}[b]{0.32\textwidth}
		\centering
		\includegraphics[width=\textwidth]{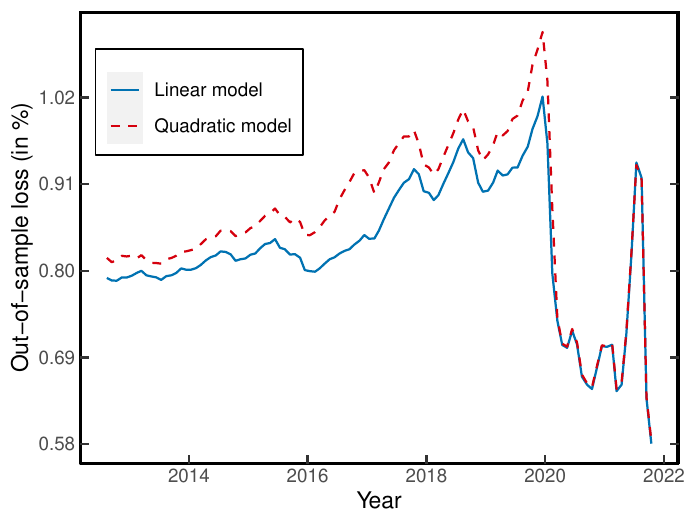}
		\caption{80th percentile}
		\label{fig:loss80}
	\end{subfigure}
	\begin{subfigure}[b]{0.32\textwidth}
		\centering
		\includegraphics[width=\textwidth]{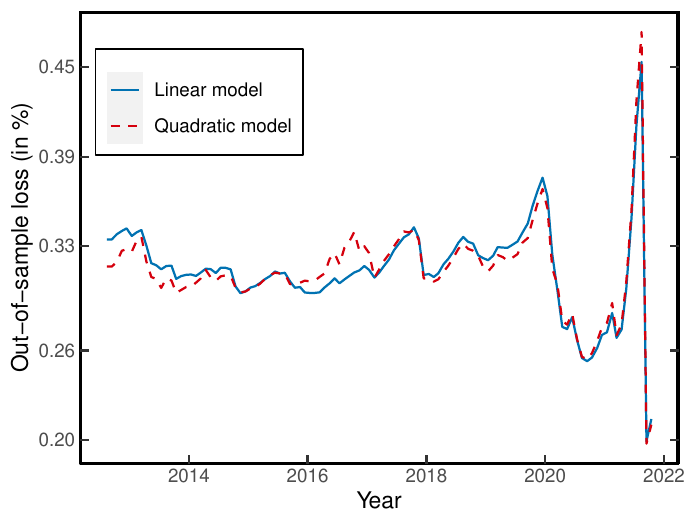}
		\caption{95th percentile}
		\label{fig:loss95}
	\end{subfigure}
	\caption{\textbf{Out-of-sample quantile forecasting loss}. \footnotesize These figures show the out-of-sample loss of forecasting the physical quantile function, based on an expanding window. The loss at different percentiles is calculated by \eqref{eq:oos_loss}.}
	\label{fig:robust_linear}
\end{figure}

Figure \ref{fig:robust_linear} shows the out-of-sample loss at various percentiles. In most cases, the linear model outperforms the quadratic model, with some exceptions observed at the 95th percentile during specific periods. These results continue to hold when adding other non-linear terms, such as cubic, exponential or logarithmic factors. Additionally, I find that the risk-neutral quantile function exhibits a high correlation with higher-order terms. Consequently, the non-linear model tends to produce quantile forecasts that closely resemble those generated by the linear model.

\subsection{Additional Evidence Against the Lognormal Assumption}\label{app:add_logn}

Table \ref{tab:only.rn.quantile} already indicates evidence against the lognormal model since the QR estimates in the left- and right-tail are rather different, in contradiction with \eqref{eq:claim}.  To further assess the implications of the lognormal model, I analyze the accuracy of the physical quantile forecast in \eqref{eq:q_forecast} out-of-sample. Specifically, I use QR based on the first $t_0$ observations to estimate the model
\begin{equation}\label{eq:rq_logn}
	Q_{t,\tau}(\omrkt) = \hat{\beta}_{0,t_0}(\tau) + \hat{\beta}_{1,t_0}(\tau) \cquant,
\end{equation}
where the $t_0$-subscript in $\beta_{\cdot,t_0}$ refers to the fact that the coefficients are estimated using observations up to time $t_0$. Using an expanding window to estimate $\beta_{\cdot,t_0}$, the model produces dynamic quantile forecasts of the form
\begin{equation}\label{eq:dynamic_rnq}
	\qhatln = \hat{\beta}_{0,t}(\tau) + \hat{\beta}_{1,t}(\tau) \cquant.
\end{equation}
In the lognormal case,  Proposition \ref{lemma:bs}(\ref{item:logn2}) suggests that $\cpquant(\omrkt) \approx \qhatln$. This approximation can be tested using the joint restriction
\begin{equation*}
	H_0: \quad [\beta_0(\tau), \beta_1(\tau)] = [0,1],
\end{equation*}
in the quantile regression
\begin{equation*}
	\min_{\beta_0,\beta_1 \in \mathbb{R}} \sum_t \rho_\tau\left(\omrkt - \beta_0- \beta_1 \qhatln \right).
\end{equation*}
The results are summarized in Table \ref{tab:rolling_window} and show that the point estimates are quite far from the $[0,1]$ benchmark. The Wald test on the joint restriction tends to reject $H_0$ far enough in the tail, but for $\tau = 0.2$ the null hypothesis is never rejected due to the large standard errors.  Additionally, the $R^1(\tau)$ statistic shows that the explanatory power is low relative to Table \ref{tab:only.rn.quantile}, even though the sample sizes are different. Hence, the results are incompatible with \eqref{eq:claim} and \eqref{eq:q_forecast} and provide evidence against the conditional lognormal assumption, which is in line with evidence from the literature (see e.g. \citet[Result 4]{martin2017expected}).

\begin{table}[!htb]
	\captionsetup{width=12cm}	
	\centering
	\caption{\textbf{Expanding quantile prediction with risk-neutral quantile}}
	\label{tab:rolling_window}
	\begin{adjustbox}{max width=\textwidth}
		\begin{threeparttable}
			\begin{tabular}{lcccccc}
				\toprule
				\midrule
				Horizon & $\tau $& $\hat{\beta}_0(\tau)$ & $\hat{\beta}_1(\tau)$ & $\underset{(p\text{-value})}{\text{Wald test}}$ & $R^1(\tau)${[}\%{]} & Obs \\ 
				\cmidrule(lr){1-7}
				&  &  &  &  &  &  \\
				\underline{30 days} & 0.05 & $\underset{( 0.185 )}{ 0.54 }$ & $\underset{( 0.193 )}{ 0.42 }$ & 0.00 & 4.36 & 3804 \\
				& 0.1 & $\underset{( 0.205 )}{ 0.59 }$ & $\underset{( 0.212 )}{ 0.39 }$ & 0.01 & 2.39 &  \\
				& 0.2 & $\underset{( 0.332 )}{ 0.83 }$ & $\underset{( 0.338 )}{ 0.15 }$ & 0.04 & 0.28 &  \\
				&  &  &  &  &  &  \\
				\underline{60 days} & 0.05 & $\underset{( 0.310 )}{ 0.55 }$ & $\underset{( 0.329 )}{ 0.39 }$ & 0.06 & 1.84 & 3753 \\
				& 0.1 & $\underset{( 0.339 )}{ 0.80 }$ & $\underset{( 0.352 )}{ 0.16 }$ & 0.03 & 0.22 &  \\
				& 0.2 & $\underset{( 0.416 )}{ 0.87 }$ & $\underset{( 0.425 )}{ 0.11 }$ & 0.10 & 0.25 &  \\
				&  &  &  &  &  &  \\
				\underline{90 days} & 0.05 & $\underset{( 0.335 )}{ 0.78 }$ & $\underset{( 0.358 )}{ 0.11 }$ & 0.01 & 0.88 & 3702 \\
				& 0.1 & $\underset{( 0.376 )}{ 0.74 }$ & $\underset{( 0.395 )}{ 0.21 }$ & 0.08 & 1.34 &  \\
				& 0.2 & $\underset{( 0.481 )}{ 0.73 }$ & $\underset{( 0.491 )}{ 0.26 }$ & 0.31 & 0.52 &  \\
				\bottomrule
			\end{tabular}%
			\begin{tablenotes}
				\footnotesize
				\item \textit{Note}: This table reports the QR estimates of \eqref{eq:dynamic_rnq} using an expanding window based on an initial 500 observations. The sample period is 2003-2021. \emph{Wald test} denotes the $p$-value of the joint restriction $[\beta_0(\tau),\beta_1(\tau)] = [0,1]$. Standard errors are reported in parentheses and calculated using the SETBB with a block length equal to the prediction horizon. $R^1(\tau)$ denotes the goodness of fit measure \eqref{eq:R1tau}.
			\end{tablenotes}
		\end{threeparttable}
	\end{adjustbox}
\end{table}

\section{Additional Figures}
\begin{figure}[!htb]
	\centering
	\includegraphics[width=0.6\textwidth]{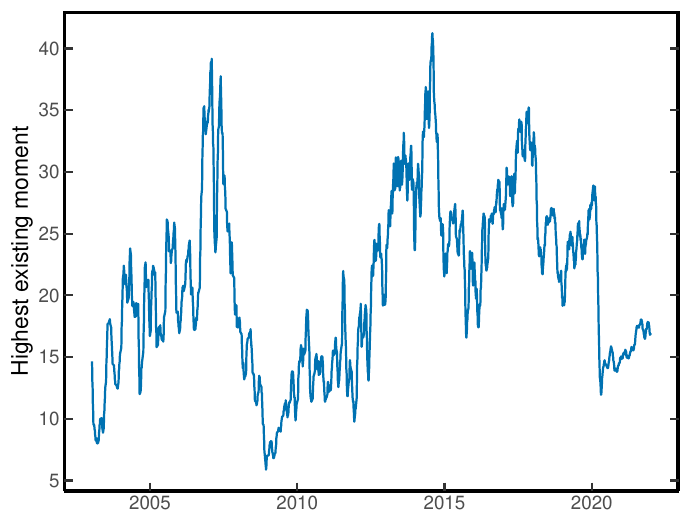}
	\caption{\textbf{Highest existing risk-neutral moment for 30-day returns.}  This figure shows $p_t^* \coloneqq \sup\{p: \texptilde{\omrkt^{p}} < \infty  \}$ over time, where $\omrkt$ represents the 30-day return. $p_t^*$  is calculated from the moment formula of \citet{Lee2004}, $p_t^* = \frac{1}{2\beta_R} + \frac{\beta_R}{8} + \frac{1}{2}$, where $\beta_R = \limsup_{x \to \infty} \frac{\sigma_{\text{IV}}^2(x)}{\abs{x}/N}$, $\sigma_{\text{IV}}(x)$ is the implied volatility at log-moneyness $x = \log(K/(e^{rN}S_0))$, and $N = 30/365$ is the time horizon. $\beta_R$ is estimated from the call option with highest available strike price.  The figure is smoothed using a 30-day moving average. }
	\label{fig:lee_bound}
\end{figure}

\end{document}